\documentclass[12pt]{article}
\usepackage{bm}
\usepackage[numbers]{natbib}
\usepackage{xr}
\usepackage[colorlinks,citecolor=blue,linkcolor=blue,urlcolor=blue]{hyperref}
\usepackage{latexsym, epsfig, epstopdf, amssymb, amsmath, amsthm, mathrsfs, booktabs}
\usepackage{rotating, tabularx, setspace,paralist}
\usepackage{color}
\usepackage{subfig}

\usepackage[OT1]{fontenc}
\usepackage{lscape}
\usepackage{multirow,multicol}
\usepackage{anysize}
\usepackage{amsfonts}
\usepackage{accents}
\usepackage{bbm} 
\usepackage[linesnumbered,ruled,vlined]{algorithm2e} 
\usepackage{algorithmic}
\usepackage{mathtools}

\usepackage{verbatim} 
\usepackage[utf8]{inputenc}
\usepackage[english]{babel}

\usepackage{multibib}

\newcites{SM}{Appendix References}

\newtheorem{defn}{Definition} 
\newtheorem*{defn*}{Definition} 
\newtheorem{exmp}{Example} 
\newtheorem{model}{Model} 
\newtheorem{condi}{Condition} 

\newtheorem{theorem}{Theorem}
\newtheorem{corollary}{Corollary}
\newtheorem{lemma}{Lemma}
\newtheorem*{lemma*}{Lemma}
\newtheorem{proposition}{Proposition}

\theoremstyle{plain}
\newtheorem{remark}{Remark}
\newtheorem{claim}{Claim}
\newtheorem*{claim*}{Claim}
\newtheorem*{rwr*}{Removed When Ready}

\usepackage{enumitem}
\newlist{steps}{enumerate}{1}
\setlist[steps, 1]{label = Step \arabic*:}

\usepackage{mathtools}
\DeclarePairedDelimiter\ceil{\lceil}{\rceil}

\SetKwInput{KwInput}{Input}                
\SetKwInput{KwOutput}{Output}              

\newcommand{\pkg}[1]{{\fontseries{b}\selectfont #1}}

\newcommand\norm[1]{\left\lVert#1\right\rVert} 
\renewcommand{\hat}{\widehat}
\renewcommand{\tilde}{\widetilde}
\renewcommand{\top}{^{T}}
\renewcommand{\dots}{\cdots}

\renewcommand{\uparrow}{\rightarrow}

\newcommand{\independent}{\perp \!\!\! \perp}

\usepackage[margin=0.97in,nohead]{geometry}
\begin{document}
	
	\title{High-Dimensional Knockoffs Inference for Time Series Data%
		\thanks{
			Chien-Ming Chi is Assistant Research Fellow, Institute of Statistical Science, Academia Sinica, Taiwan (E-mail: \textit{xbbchi@stat.sinica.edu.tw}). %
			Yingying Fan is Centennial Chair in Business Administration and Professor, Data Sciences and Operations Department, Marshall School of Business, University of Southern California, Los Angeles, CA 90089 (E-mail: \textit{fanyingy@marshall.usc.edu}). %
			Ching-Kang Ing is Tsing Hua Chair Professor, Institute of Statistics, National Tsing Hua University, Hsinchu 300044, Taiwan (E-mail: \textit{cking@stat.nthu.edu.tw}). %
			Jinchi Lv is Kenneth King Stonier Chair in Business Administration and Professor, Data Sciences and Operations Department, Marshall School of Business, University of Southern California, Los Angeles, CA 90089 (E-mail: \textit{jinchilv@marshall.usc.edu}). %
			This work was supported by NSF Grants DMS-1953356, EF-2125142, DMS-2310981, and DMS-2324490, and by Grants 111-2118-M-001-012-MY2 and 109-2118-M-007-MY3 from the National Science and Technology Council, Taiwan.}
		\date{November 10, 2024}
		\author{Chien-Ming Chi$^{1}$, Yingying Fan$^2$, Ching-Kang Ing$^3$ and Jinchi Lv$^2$
			\medskip\\
			Academia Sinica$^1$, University of Southern California$^2$\\ and National Tsing Hua University$^3$ 
			\\
		} %
	}
	
	\maketitle
	
	\begin{abstract}
		We make some initial attempt to establish the theoretical and methodological foundation for the model-X knockoffs inference for time series data. We suggest the method of time series knockoffs inference (TSKI) by exploiting the ideas of subsampling and e-values to address the difficulty caused by the serial dependence. We also generalize the robust knockoffs inference in \cite{barber2018robust} to the time series setting to relax the assumption of known covariate distribution required by model-X knockoffs, since such an assumption is overly stringent for time series data. We establish sufficient conditions under which TSKI achieves the asymptotic false discovery rate (FDR) control. Our technical analysis reveals the effects of serial dependence and unknown covariate distribution on the FDR control. We conduct a power analysis of TSKI using the Lasso coefficient difference knockoff statistic under the generalized linear time series models. The finite-sample performance of TSKI is illustrated with several simulation examples and an economic inflation study.
	\end{abstract}
	
	\textit{Running title}: TSKI
	
	\textit{Key words}: Model-X knockoffs; Time series; High dimensionality; FDR control; Power analysis; Sparsity; Interpretable forecasting; E-values

	\section{Introduction} \label{Sec1}

	Identifying key economic factors among a large number of potential variables (e.g., numerous types of consumer price indices, unemployment rates, and housing prices) that can influence inflation is a long-standing research pursuit~\citep{king1995temporal, stock1999forecasting, crump2022unemployment} that remains crucial due to inflation's significance. However, statistical inference for economic time series such as inflation is challenging due to the serial dependence, large number of potentially important covariates (including time series covariates, their lags, and non-time series covariates), regime shifts~\cite{hamilton1989new, tong1980threshold}, and possible nonlinear relationships.

\begin{figure}[t!]
		\centering
		\includegraphics[width=11cm]{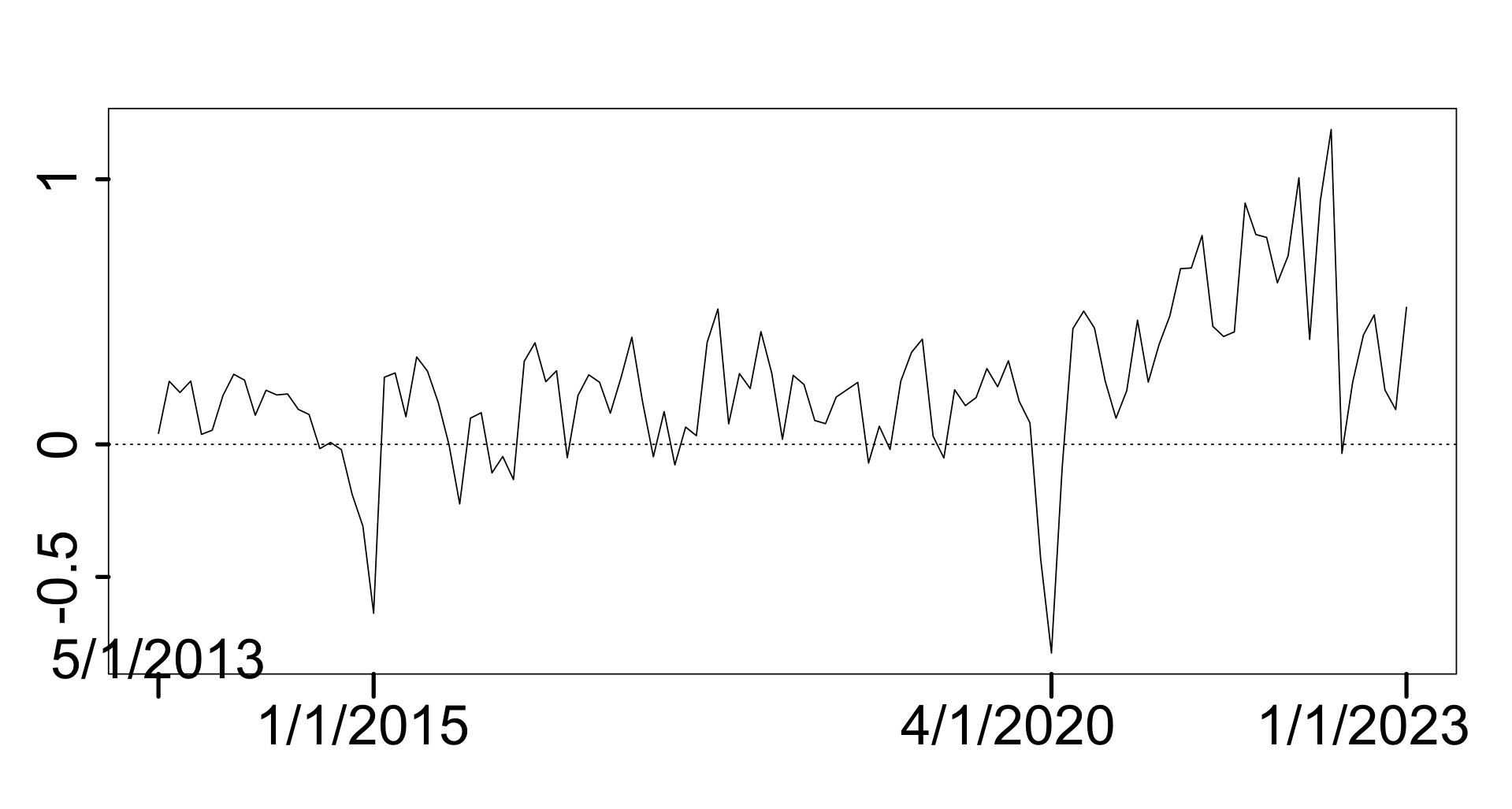}
		\caption{The U.S. inflation from May 2013 to January 2023.}
		\label{fig:inflation}
	\end{figure}

	Let us exemplify these challenges with Figure~\ref{fig:inflation}, which depicts the monthly inflation rate (hereafter referred to as inflation) time series of the U.S. economy from May 2013 to January 2023, sourced from the FRED-MD database~\citep{mccracken2016fred} and the U.S. Bureau of Labor Statistics. In addition to the inflation series, the FRED-MD database includes $126$ other time series variables that can be used to predict the inflation for the following month. Here, inflation is calculated as the first-order difference of the Consumer Price Index (CPI) \citep{mccracken2016fred} divided by the CPI of the previous month; see Section \ref{real.data.1} for details. Figure~\ref{fig:inflation} shows that during the period from 2020 to late 2022, inflation displayed an upward mean drift due to the U.S. economy's post-COVID-19 recovery and the impacts of the Russia–Ukraine conflict, along with some possible stationary phase-changing behaviors~\citep{hamilton1989new, tong1980threshold}. To address concerns on potential nonstationarity over the entire time span caused by, say, the mean drift, the rolling-window method is commonly employed in time series analysis. The intuition behind this method is that the time series, including both the response and predictors, are more likely to be stationary within a small time window\footnote{We also obtain numerical evidence supporting the stationarity of the inflation within each window through standard unit root tests; details are provided in Section~\ref{unit_root_sec}.}. However, the use of rolling windows further complicates the problem in that the sample size in each window is usually small; for example, there are only 60 data points if sampled monthly over a 
	five-year period. The presence of serial dependence, small sample size, high-dimensional covariates, and possibly nonlinear relationships makes statistical inference for time series regression highly challenging. Variable selection has been a popular solution to address such challenges, under the assumption that only a small subset of covariates contribute to the response of interest. Correctly selecting these important covariates can help simplify the model and improve interpretability and prediction accuracy. Our goal in this paper is to develop a reliable variable selection approach that accounts for these unique challenges in time series regression.

	Popularly used measures for evaluating the performance of high-dimensional variable selection include the false discovery rate (FDR)~\cite{benjamini1995controlling}, model selection consistency~\cite{zhang2011adaptive, medeiros2016l1, ing2020model},  and feature importance ranking~\citep{Breiman2001, medeiros2021forecasting}. Our paper focuses on controlling the FDR, whose formal definition is in \eqref{def:FDR}.
	Most existing works address the FDR control by building procedures based on the p-values constructed for assessing the importance of individual variables; see, e.g., the seminal works of \cite{benjamini1995controlling, benjamini2001control}.  Yet, the high dimensionality in covariates and the possibly complicated nonlinear model structure make many conventional ways of p-value calculations inapplicable or even completely fail \cite{FanDemirkayaLv2019}.  To overcome such difficulties, the framework of knockoffs inference was proposed in~\cite{Barber2015, CandesFanJansonLv2018} to achieve the goal of exact FDR control in variable selection in finite samples, completely bypassing the use of the conventional p-values in high-dimensional regression models.  It allows for an arbitrary dependence structure of the response on covariates and an arbitrary dimensionality of covariates at the cost of assuming the known joint covariate distribution. See Section \ref{Sec2.1} for a brief review of the model-X knockoffs framework.

	Two critical assumptions in model-X knockoffs \cite{CandesFanJansonLv2018} are: 1) the observations across time are independent and identically distributed (i.i.d.) and 2) the joint distribution of covariate vector is known for generating the  knockoff variables.  
	Both assumptions are unreasonably strong for time series data. Time series data exhibit serial dependence, and in some applications where covariates are lagged response variables with a stationarity assumption, assuming a known covariate distribution directly reveals the set of important variables, thus invalidating the problem of variable selection. We will address these challenges by relaxing the two aforementioned assumptions.

	To relax the first assumption, we adopt the popular idea of subsampling. To relax the second assumption, we generalize the robust knockoffs inference in \cite{barber2018robust} to the time series setting, where the robust knockoffs inference \cite{barber2018robust} is a recent innovation 
	that allows for approximate covariate distribution (as opposed to known covariate distribution) developed for i.i.d. data. We name the knockoff variables generated using approximate covariate distribution as \textit{approximate knockoffs} to ease the presentation. Our analyses reveal that with approximate knockoffs generated in a rowwise fashion ignoring the serial dependence, the FDR inflation has an upper bound depending on the Kullback–Leibler (KL) divergence between the distributions of data matrices corresponding to the approximate and exact model-X knockoff variables. Our theoretical results are comparable to Theorem 1 of \citep{barber2018robust}, with the difference that we do not need the i.i.d. observations assumption. 
	Our theory shows that there is generally no guarantee that such KL divergences asymptotically vanish in the existence of serial dependence, suggesting that the corresponding FDR could be uncontrolled. We also show that subsampling ~\citep{yu1994rates} can successfully address such difficulty and warrant asymptotically vanishing KL divergence if the subsampling rate is appropriately chosen, provided that the time series are $\beta$-mixing. 
	We then apply the robust knockoffs inference to each of the subsampled data, resulting in multiple sets of selected variables. 
	We aggregate these sets via the e-value method~\citep{wang2022false, ren2022derandomized}. The complete framework is presented in Section \ref{Sec2.new} and named as the \textit{time series knockoffs inference} (TSKI). We provide a rigorous characterization of how the serial dependence and the accuracy of the approximate knockoffs affect the FDR control and prove that the TSKI procedure can achieve asymptotic FDR control when the subsampling is done appropriately and the approximate knockoffs are accurate enough.

	It is well-known that FDR and power are two sides of the same coin. We then turn to the power analysis of TSKI in Section~\ref{Sec3.new}. Assuming the generalized linear time series models and some regularity conditions, we show that 
	for TSKI with subsampling and e-value aggregation, the set of selected variables is either empty or enjoys the sure screening property  with asymptotic probability one. 
	These results are formally summarized in Theorem \ref{theorem4.new} and discussed after it. 
	
	We test the empirical performance of TSKI on both simulated and real data. Our simulation results in Section \ref{Sec6} demonstrate that the TSKI with subsampling controls the FDR when the data are generated from some popular $\beta$-mixing processes. The selection power is generally satisfactory with a sufficient sample size. Furthermore, in Section \ref{real.data.1} we apply the TSKI to study the temporal relations between inflation and other macroeconomic time series from the U.S. economy
	over the past ten years.

	

	\subsection{Related work}

	Established methods such as the BH and BY~\citep{benjamini1995controlling, benjamini2001control} are commonly used in biology applications with non-time series data. Some new developments such as~\citep{ramdas2019unified,  bogdan2015slope} either assume independent test statistics, require the availability of p-values, or rely on specific model structures, making them unsuitable for time series applications due to incompatible assumptions.
	
	Among these existing methods,  BY achieves the FDR control in variable selection based on a set of valid p-values with no requirements on the dependence structure among p-values. Similarly, e-BH~\citep{wang2022false} uses e-values without requirements on their dependence structure. While these methods offer the potential for valid FDR control for time series data, obtaining valid p-values or e-values remains an unresolved challenge in many applications. 
	
	Regularized regression~\citep{medeiros2016l1} and information criterion-based model selection~\cite{ing2020model} are also popularly used for selecting important variables in time series regressions. Yet, they often assume some specific model structure (e.g., the linear model) to prove variable/model selection consistency, posing uncertainty about their performance when deployed on real data with nonlinear dependency. In addition, they are not specifically designed for controlling the variable selection error rate, and hence may not be suitable for certain applications when the error rate control is a concern.

	\subsection{Notation}
	Let $(\Omega, \mathcal{F}, \mathbb{P})$ and $\mathcal{R}$ be the underlying probability space and the Borel $\sigma$-algebra on the real line $\mathbb{R}$, respectively. We use the boldface for random vectors and matrices,  the tilde for the knockoff variables, and the vector notation $\vec{x}$ to denote vectors in the Euclidean space. For random vector $\boldsymbol{x}$, define $\boldsymbol{x}_{-j}$ as the subvector by removing the $j$th coordinate. We use parentheses for matrix concatenation. For any real sequences $\{a_{n}\}$ and $\{b_{n}\}$, $a_{n} =  O(b_{n})$ means $\limsup_{n\to \infty} \left|\frac{a_{n}}{b_{n}} \right|< \infty$, and $a_{n} =  o(b_{n})$ means $\limsup_{n\to \infty}\left|\frac{a_{n}}{b_{n}}\right| = 0 $. We use $\# S$ to denote the cardinality of a given set $S$. Moreover, a transition kernel is defined as a map $p : (\mathbb{R}^{k_{1}}, \mathcal{R}^{k_{2}}) \longrightarrow [0, 1]$ for some positive integers $k_{1}$ and $k_{2}$ satisfying that (i) for each $\mathcal{D} \in\mathcal{R}^{k_{2}}$, $p(\cdot, \mathcal{D})$ is a measurable function and (ii) for each $x \in \mathbb{R}^{k_{1}}$, $p(x, \cdot)$ is a probability measure.

	\section{Robust time series knockoffs inference with TSKI} \label{Sec2.new}
	
	Given an observed stationary time series $\{Y_{t}, \boldsymbol{x}_{t}\}_{t=1}^{n}$ with $Y_t\in \mathbb R$ a scalar response and $\boldsymbol{x}_{t}\in \mathbb R^p$ a high-dimensional covariate vector, we are interested in accurately selecting relevant covariates (i.e., non-null features) in  $\boldsymbol{x}_{t}$, where the definition of the null feature is given below.

	\begin{defn}(Null feature)\label{null.features} Consider the response $Y$ and  the covariate vector $\boldsymbol{x} = (X_{1}, \dots, X_{p})^{\top}$. 
		Covariate $X_{j}$ with $j\in \{1,\cdots, p\}$ is said to be null with respect to response $Y$ if and only if $X_{j}\independent Y \vert (X_{1}, \dots, X_{j-1}, X_{j+1}, \dots, X_{p})^T$.
	\end{defn}
	
	We denote the set of null features according to Definition \ref{null.features} above as $\mathcal{H}_{0} \subset \{1, \dots, p\}$ for each $t$, where $\mathcal H_0$ is independent of $t$ because $(Y_{t}, \boldsymbol{x}_{t})$'s have the same distribution due to the stationarity assumption. Our goal is to estimate the important variable set $\mathcal H_1 = \mathcal \{1,\dots, p\}\backslash \mathcal H_0$ using data $\{Y_{t}, \boldsymbol{x}_{t}\}_{t=1}^{n}$.  For an estimated set $\widehat{S}\subset\{1, \dots ,p\}$ returned by some algorithm, we measure the accuracy by evaluating the False Discovery Rate (FDR)  defined as
	\begin{equation}\label{def:FDR}
		\textnormal{FDR} \coloneqq 
		\mathbb{E}\left(\frac{ \#(\widehat{S}\cap \mathcal{H}_{0}) }{ (\# \widehat{S} ) \vee 1}\right).
	\end{equation}
	
	In this paper, we focus on time series data with serial dependency and high dimensionality where covariate dimensionality $p$ can be much larger than sample size $n$. We \textit{do not} assume any specific dependence structure of $Y_t$ on $\boldsymbol{x}_t$ other than the one in Definition \ref{null.features} and that $Y_{t}$ is $\boldsymbol{x}_{t+1}$-measurable (see Corollary~\ref{collary.2}) for our FDR analysis. As a result, our proposed method can accommodate an unknown relationship between $Y_t$ and $\boldsymbol{x}_t$ including both linear and nonlinear ones.     Our method builds on the recent work of model-X knockoffs \cite{CandesFanJansonLv2018} and its robust extension \cite{barber2018robust} proposed for the non-time-series data, where we briefly review the former in the next section to set the stage.  
	
	\subsection{A brief review of the model-X knockoffs framework}\label{Sec2.1.newb}
	
	Let $\boldsymbol{y}=(Y_1,\cdots, Y_n)^T\in \mathbb R^n$  and $\boldsymbol{X}=(\boldsymbol{x}_{ 1}, \cdots, \boldsymbol{x}_{ n})^T\in \mathbb R^{n\times p}$ be the response vector and design matrix collecting the $n$ observations. The model-X knockoffs~\citep{CandesFanJansonLv2018} was proposed for the setting where the rows of the augmented matrix $(\boldsymbol{y},\boldsymbol{X})\in \mathbb R^{n\times (p+1)}$ are i.i.d. random vectors with known distribution for $\boldsymbol{x}_1$. The knockoffs inference aims at estimating $\mathcal H_1 $ while keeping the FDR \eqref{def:FDR} under control.  To this end,  it constructs an $n\times p$ matrix $\tilde{\boldsymbol{X}}$ in a rowwise fashion independently using the known joint distribution of $\boldsymbol{x}_1$  such that 
	\begin{equation}\label{MX2}
		\tilde{\boldsymbol{X}} \independent \boldsymbol{y}| \boldsymbol{X} \ \text{ and } \ 		(\boldsymbol{X}, \tilde{\boldsymbol{X}} )_{\textnormal{swap}(S)} \overset{d}{=} (\boldsymbol{X}, \tilde{\boldsymbol{X}})
	\end{equation}
	for each $S\subset\{1, \dots, p\}$,	where  \textnormal{swap($S$)} denotes the swapping operation meaning that for each $j \in S$, columns $j$ and $j + p$ are swapped, and $\overset{d}{=}$ stands for equal in distribution.   
	At a high level,  the model-X knockoffs create a ``fake" covariate matrix $\tilde{\boldsymbol{X}}$ which perfectly mimics the ``behavior" of the original covariate matrix. By using these fake covariates as controls, the importance of original covariates can be inferred.

	As an example, when the covariate distribution for  $\boldsymbol{x}_t$ is known to be $N({\bf 0}, {\Sigma})$ with $\Sigma$ the $p\times p$ covariance matrix, a valid way to construct the corresponding ideal knockoff vector $\widetilde{\boldsymbol{x}}_t$ is to sample from the conditional multivariate Gaussian distribution
	\begin{equation}\label{ex-Gaussian-knockoffs}
		\widetilde{\boldsymbol{x}}_t|\boldsymbol{x}_t \overset{d}{\sim} N(\boldsymbol{x}_t - \textnormal{diag}(\vec{s}) {\Sigma^{-1}} \boldsymbol{x}_t, 2\textnormal{diag}(\vec{s}) - \textnormal{diag}(\vec{s})\Sigma^{-1} \textnormal{diag}(\vec{s})),
	\end{equation}
	where diag$(\vec{s})$ is a diagonal matrix of tuning parameters with positive diagonal entries. Larger components of $\vec{s}$ imply that the resulting knockoff variables are more independent of the original variables, thereby providing higher power in distinguishing them. For the general covariate distribution, a conditional distribution for generating knockoff variables can also be constructed following the same high-level idea above.  Further details about knockoff variable sampling procedure for general distributions can be found in~\citep{CandesFanJansonLv2018, FanDemirkayaLiLv2020}. 
	
	
	With the knockoff variable matrix $\tilde{\boldsymbol{X}}$, the knockoff statistics $W_{j}$'s, measuring the importance of original covariates, are constructed such that the sign-flip property~\cite{Barber2015, CandesFanJansonLv2018} is satisfied: for each $S\subset \{1, \dots, p\}$ and each $1\le j\le p$,
	{\small \begin{equation}
			\label{sign-flip.2}
			W_{j}(\boldsymbol{y}, [\boldsymbol{X}, \widetilde{\boldsymbol{X}}]_{\textnormal{swap}(S)}) =\begin{cases}
				-W_{j}(\boldsymbol{y}, \boldsymbol{X}, \widetilde{\boldsymbol{X}}) \quad  \textnormal{ if } j\in S,\\
				W_{j}(\boldsymbol{y}, \boldsymbol{X}, \widetilde{\boldsymbol{X}}) \quad \textnormal{ otherwise.}
			\end{cases}
	\end{equation}}%

	As variable importance measures, high-quality knockoff statistics $W_j$'s should have the desired properties that 1) $W_j$'s have large positive values for $j\in \mathcal H_1$ and 2) for null features $j\in \mathcal H_0$, $W_j$'s have small magnitude and are symmetric around zero. See \cite{CandesFanJansonLv2018} for the formal characterization. 
	Examples~\ref{knockoff.exmp.2}--\ref{knockoff.exmp.1} in the next subsection are two important instances of the knockoff statistics that satisfy the sign-flip property. 
	
	Model-X knockoffs \cite{CandesFanJansonLv2018} estimates $\mathcal H_1$ as $\widehat{S}=\{1\leq j\leq p: W_j\geq T\}$ with  the so-called knockoff threshold
	$ T   = \min\left\{t>0:  \frac{1+ \#\{ j : W_{j}\le -t \}}{\#\{ j : W_{j} \ge t\}\vee 1}\le \tau \right\}$, where $\tau \in (0,1)$ is some pre-specified  target level for FDR control. It has been shown in \cite{CandesFanJansonLv2018} that the model-X knockoffs framework achieves FDR control in finite samples with arbitrary dimensionality of $\boldsymbol{x}_t$ and arbitrary (unknown) dependence structure of $Y_t$ on $\boldsymbol{x}_t$. 

As discussed in the Introduction, the i.i.d. row assumption for $(\boldsymbol{y},\boldsymbol{X})$ and the known distribution assumption for $\boldsymbol{x}_t$ are too stringent for time series data. 
We will develop a new framework for time series knockoffs inference, which relaxes these assumptions. It is important to note that the remaining part of the paper does \textit{not} assume i.i.d. rows in data matrix $(\boldsymbol{y},\boldsymbol{X})$.

\subsection{Outline of the TSKI framework}\label{Sec2.1}
In this subsection, we provide an outline of the TSKI framework. Some technical details will be presented in the next subsection.  
TSKI has three key ingredients: subsmapling, robust knockoffs inference on each subsample, and e-value aggregation of the selected sets from different subsamples. Algorithm \ref{Algorithm1} provides a detailed implementation of TSKI. Our presentation may not strictly adhere to the order of the three ingredients stated above.  

To relax the known covariate distribution assumption, we adopt the robust knockoffs framework in \cite{barber2018robust} and introduce the following Definition \ref{knockoff.generator.1}. In what follows, the knockoff generator, conditional distribution, and transition kernel are examples of regular conditional probability (r.c.p.) in probability theory.

\begin{defn}[Knockoff generator]
	\label{knockoff.generator.1}
	$\kappa:\mathbb{R}^{p}\times \mathcal{R}^{p}\longmapsto\mathbb{R}^{p}$ is said to be a knockoff generator if \textnormal{1)} $\kappa(\vec{z}, \cdot)$ is a probability measure for each $\vec{z}\in\mathbb{R}^{p}$ and \textnormal{2)} $\kappa( \cdot, \mathcal{A})$ is a measurable function for each  $\mathcal{A}\in\mathcal{R}^{p}$, where $\mathcal{R}$ is the smallest $\sigma$-algebra of $\mathbb{R}$ that contains all open sets. 
\end{defn}

For each observed covariate vector $\boldsymbol{x}_{t}$ with $t\in [n] := \{1, \cdots, n\}$, we generate its knockoff vector $\widetilde{\boldsymbol{x}}_{t}$ from $\kappa(\boldsymbol{x}_{t},\cdot)$. When $\boldsymbol{x}_{t}$ has known distribution $N({\bf 0},\Sigma)$ with known parameters $\Sigma$, the conditional Gaussian distribution in \eqref{ex-Gaussian-knockoffs} is an instance of the knockoff generator which yields \textit{ideal} knockoff variables satisfying \eqref{MX2}. If $\Sigma$ in \eqref{ex-Gaussian-knockoffs} is unknown and replaced with an estimated version (see \eqref{ko.sampling.1} for details), the resulting knockoff generator produces only \textit{approximate} knockoff variables that violate the exchangeability condition in \eqref{MX2}. Additional examples of the knockoff generator can be found in~\citep{CandesFanJansonLv2018, FanDemirkayaLiLv2020}.
To achieve asymptotic FDR control using knockoff variables generated from $\kappa(\boldsymbol{x},\cdot)$, we need some additional conditions on 
$\kappa$ which will be presented in Condition \ref{knockoff.generator.2} in the next section. Note that because of the rowwise generation of knockoff variables, although $\{\boldsymbol{x}_t\}_{t=1}^n$ have serial dependence across $t$, $\{\tilde{\boldsymbol{x}}_t\}_{t=1}^n$ are independent across $t$ conditional on $\{\boldsymbol{x}_t\}_{t=1}^n$; this violates the second property in \eqref{MX2} and thus the FDR control result in \cite{CandesFanJansonLv2018} or \cite{barber2018robust} cannot be applied directly. From now on, we work with  the inference sample $\{Y_{t}, \boldsymbol{x}_{t}, \widetilde{\boldsymbol{x}}_{t}\}_{t=1}^{n}$.

To overcome such difficulty, we consider subsamples each with index set $H_{k} = \{k+ s(q+1) : s = 0, 1,\cdots,  \lfloor \frac{n-k}{q+1} \rfloor \}$ for $k\in\{1, \dots ,q+1\}$ with some integer $q\ge 0$. 
To simplify the technical presentation, let $\{V_{t}, \boldsymbol{u}_{t}, \widetilde{\boldsymbol{u}}_{t}\}_{t=1}^{N}$ be a generic subsample that can be any of the $q+1$ subsamples. Denote by \sloppy$\boldsymbol{v} \coloneqq (V_{1}, \dots, V_{N})^{\top}$, $\boldsymbol{U}\coloneqq (\boldsymbol{u}_{1}, \dots, \boldsymbol{u}_{N})^{\top}$, and $\widetilde{\boldsymbol{U}}\coloneqq (\widetilde{\boldsymbol{u}}_{1}, \dots, \widetilde{\boldsymbol{u}}_{N})^{\top}$. As in the robust knockoffs inference \cite{barber2018robust}, we construct knockoff statistics $W_{j}(\boldsymbol{v}, \boldsymbol{U}, \widetilde{\boldsymbol{U}})$'s based on $(\boldsymbol{v}, \boldsymbol{U}, \widetilde{\boldsymbol{U}})$ and select the set of important variables using these knockoff statistics following the identical procedure as reviewed in the last subsection.  Thus, we end up with $q+1$ sets of selected variables  $\{j:W_j^k \geq T^k\}$ with each corresponding to a subsample $k$. Here, $W_j^k$'s and $T^k$ are the correspondingly constructed knockoff statistics and the knockoff threshold as specified in \eqref{tau.2}, respectively.

Below are two examples of the knockoff statistics. The random forests model in Example~\ref{knockoff.exmp.1} can be replaced with other 
learning models such as the deep learning model.

\begin{exmp}[Lasso coefficient difference (LCD)]\label{knockoff.exmp.2}
	For a given sample $(\boldsymbol{v}, \boldsymbol{U}, \widetilde{\boldsymbol{U}})$ and a tuning parameter $\lambda\ge 0$, we define  
	$ W_{j} = W_{j}(\boldsymbol{v}, \boldsymbol{U}, \widetilde{\boldsymbol{U}}) = |\widehat{\beta}_{j}| - |\widehat{\beta}_{j+p}|, $
	where 
	$(\widehat{\beta}_{1}, \dots , \widehat{\beta}_{2p})^T$ is given by either the Lasso estimate $ \arg\min_{\boldsymbol{\beta}\in\mathbb{R}^{2p}} \left\{n^{-1}\sum_{t=1}^{n} (V_{t} - (\boldsymbol{u}_{t}^{\top}, \widetilde{\boldsymbol{u}}_{t}^{\top}) \boldsymbol{\beta} )^2 + \lambda \sum_{j=1}^{2p}|\beta_{j}|\right\}$ with $\boldsymbol{\beta} = (\beta_1, \cdots, \beta_{2p})^T$, or the generalized linear model (GLM) Lasso estimate defined in \eqref{ing1}.
\end{exmp}

\begin{exmp}[Random forests mean decrease accuracy (MDA)]\label{knockoff.exmp.1}
	For a given sample $(\boldsymbol{v}, \boldsymbol{U}, \widetilde{\boldsymbol{U}})$, we define  
	$ W_{j} = W_{j}(\boldsymbol{v}, \boldsymbol{U}, \widetilde{\boldsymbol{U}}) = N^{-1}\sum_{t=1}^{N}\big\{[V_{t} - \widehat{m}(\boldsymbol{u}_{t}^{(j)}, \widetilde{\boldsymbol{u}}_{t})]^2 - [V_{t} - \widehat{m}(\boldsymbol{u}_{t}, \widetilde{\boldsymbol{u}}_{t}^{(j)})]^2\big\}$
	for each $j\in \{1, \dots, p\} $, where $(\boldsymbol{u}_{t}^{(j)}, \widetilde{\boldsymbol{u}}_{t}^{(j)})=[\boldsymbol{u}_{t}, \widetilde{\boldsymbol{u}}_{t}]_{\textnormal{swap}(\{j\})}$ and $\widehat{m}:\mathbb{R}^{2p}\longmapsto\mathbb{R}$ is the random forests regression function trained by regressing $V_{t}$'s on $(\boldsymbol{u}_{t}, \widetilde{\boldsymbol{u}}_{t})$'s.     
\end{exmp}

The LCD uses the linear model as a working model, while the MDA does not assume any explicit model structure. When the underlying true model is nonlinear, the LCD is based on the misspecified model, whereas the MDA is free of such issues. Yet, MDA demands a large sample size, which is a common drawback for all nonparametric regression models. Our simulation study will provide additional insights into the performance of LCD and MDA in various model settings. 

When $q>0$, subsampling yields more than one set of selected variables. Naively taking the intersection or union over these sets would not guarantee FDR control. The TSKI uses the e-BH procedure ~\citep{wang2022false} to overcome such a difficulty; see Step 3 in Algorithm \ref{Algorithm1} for the calculation of e-values and how it forms the final set of selected variables $\widehat{S}$. The e-BH procedure works similarly to the BH procedure~\cite{benjamini1995controlling} with the difference that e-values \cite{shafer2021} are used in place of the p-values. Given a null hypothesis, we call a non-negative random variable $E$ an ``e-value" if $\mathbb E[E] \leq 1$ under the null. To test a hypothesis at significance level $\alpha$, we can reject the null hypothesis when $E \geq 1/\alpha$, noting that $\mathbb P(E\geq 1/\alpha)\leq \alpha \mathbb E[E] \leq \alpha$ under the null. One appealing property is that the average of multiple e-values is still a valid e-value regardless of their dependence structure. This motivates us to use it to aggregate results from different subsamples. We acknowledge that the e-value idea has been used in the literature for aggregating knockoffs inference results ~\citep{wang2022false, ren2022derandomized}. We note that tuning parameter $\tau_{1}$ used for individual subsample in Step 3 should be set smaller than the overall target level $\tau^*$ in Step 4 in order to achieve high selection power; this is formally stated in Theorem~\ref{theorem4.new} in Section~\ref{Sec3.new}.         

\bigskip
{\small
	\begin{algorithm}\setstretch{1.35}
		\SetAlgoLined
		Let $0< \tau_{1}<1$ be a constant and $0<\tau^{*}<1$ the target FDR level.
		
		For each $k\in\{1, \dots ,q+1\}$, calculate the knockoff statistics $W_{1}^{k}, \dots, W_{p}^{k}$ satisfying \eqref{sign-flip.2} using sample $\{\boldsymbol{x}_{i}, \widetilde{\boldsymbol{x}}_{i}, Y_{i}\}_{i\in H_{k}}$.
		
		Calculate the e-value statistics $e_{j} = (q+1)^{-1}\sum_{k=1}^{q+1}e_{j}^{k}$, where\footnote{$\min \emptyset$ and $\max \emptyset$ are defined as  $\infty$ and $0$, respectively.} 
		{\small \begin{equation}
				\begin{split}\label{tau.2}
					e_{j}^{k} & = \frac{p\times \boldsymbol{1}_{\{W_{j}^{k} \ge T^{k}\} }}{ 1+\sum_{s=1}^{p} \boldsymbol{1}_{\{W_{s}^{k} \le - T^{k}\} } },  \ \ \ T^{k}   = \min\left\{t \in\mathcal{W}_{+}^k:  \frac{1+ \#\{ j : W_{j}^{k}\le -t \}}{\#\{ j : W_{j}^{k} \ge t\}\vee 1}\le \tau_{1} \right\},
				\end{split}
		\end{equation}}%
		and  $\mathcal{W}_{+}^{k} = \{|W_{s}^k|:|W_{s}^k|>0\}$ for each $k\in \{1, \dots, q+1\}$. Here, $\boldsymbol{1}_{\{\cdot\}}$ is the indicator function.
		
		Let $\widehat{S} = \{j: e_{j} \ge p(\tau^* \times \widehat{k})^{-1}\}$ with $\widehat{k} = \max\{k:e_{(k)} \ge p(\tau^* \times k)^{{-1}}\}$, where  $e_{(j)}$'s are the ordered statistics of $e_{j}$'s such that $e_{(1)}\ge \dots \ge e_{(p)}$.

		\caption{{Robust time series knockoffs inference (TSKI) via e-values} }\label{Algorithm1}
\end{algorithm}}
\vspace{-1cm}

\subsection{FDR control by TSKI}\label{Sec2.2}

We need some technical conditions to bound the FDR of TSKI.

\begin{condi}\label{knockoff.generator.3}
	The density function of  $(\boldsymbol{X}, \widetilde{\boldsymbol{X}}, \boldsymbol{Y})$ exists and $(Y_{t}, \boldsymbol{x}_{t})$'s are identically distributed across $t$. In addition, the supports of $(\boldsymbol{X}, \widetilde{\boldsymbol{X}}, \boldsymbol{Y})$ and $[\boldsymbol{X}, \widetilde{\boldsymbol{X}}, \boldsymbol{Y}]_{\textnormal{swap}(\{j\})}$ are the same for each $j\in \{1, \dots, p\}$.
\end{condi}

\begin{condi}\label{knockoff.generator.4}
	The knockoff generator $\kappa(\cdot, \cdot)$ is constructed independently of observed time series $\{\boldsymbol{x}_{t}, Y_{t}\}_{t=1}^{n}$.  
\end{condi}

Condition~\ref{knockoff.generator.3} is a basic regularity condition. Condition~\ref{knockoff.generator.4} may be relaxed if we use sample splitting to obtain an asymptotically independent training subsample for estimating the unknown covariate distribution and constructing the knockoff generator.

To ease the presentation, let $(Y, \boldsymbol{x}, \widetilde{\boldsymbol{x}})$ 
be an independent copy of $(Y_{1}, \boldsymbol{x}_{1}, \widetilde{\boldsymbol{x}}_{1})$. As outlined in the last subsection, we allow certain deviation of $\kappa(\boldsymbol x,\cdot)$ from the one derived from the true covariate distribution of $\boldsymbol{x}$. Consequently, $\tilde{\boldsymbol{x}}$ generated from $\kappa(\boldsymbol x,\cdot)$ is only an \textit{approximate} knockoff vector, which may violate the exchangeability condition in \eqref{MX2}. The use of approximate knockoffs instead of ideal model-X knockoffs may incur a potential FDR inflation. Our Theorem~\ref{theorem1.new}, which imposes mild conditions on $\kappa(\cdot, \cdot)$,  demonstrates that such FDR inflation can be controlled in time series applications. These results generalize the robust knockoffs inference results for i.i.d. observations in \citep{barber2018robust} to the time series setting.

Let $\{\boldsymbol{x}_{t}^{\pi}, \widetilde{\boldsymbol{x}}_{t}^{\pi}, Y_{t}^{\pi}\}_{t=1}^{n}$ be a sequence of i.i.d. random vectors such that $(\boldsymbol{x}_{1}^{\pi}, \widetilde{\boldsymbol{x}}_{1}^{\pi}, Y_{1}^{\pi})$ and $( \boldsymbol{x}_{1}, \widetilde{\boldsymbol{x}}_{1}, Y_{1})$ have the same distribution. Denote by $\mathcal{X}_{k} = \{\boldsymbol{x}_{i}, \widetilde{\boldsymbol{x}}_{i}, Y_{i}\}_{i\in H_{k}}$ and  $\mathcal{X}_{k}^{\pi} = \{\boldsymbol{x}_{i}^{\pi}, \widetilde{\boldsymbol{x}}_{i}^{\pi}, Y_{i}^{\pi}\}_{i\in  H_{k}}$ for each $k\in \{1, \dots,  q+1\}$. Let $f_{\boldsymbol{z}}(\cdot)$ be the density function of random vector $\boldsymbol{z}$. 
\begin{theorem} \label{theorem1.new}
	Let $\widehat{S}$ be the set of variables selected by TSKI with  Algorithm~\ref{Algorithm1}. Then under Conditions~\ref{knockoff.generator.3}--\ref{knockoff.generator.4}  and the assumption of positive $T^{k}$'s in \eqref{tau.2}, we have
	{\small\begin{equation}\label{eq: FDR-bound-general}
			\begin{split}
				\textnormal{FDR}  & \le \inf_{\varepsilon> 0} \Big[\tau^{*} e^{\varepsilon} + \sum_{k=1}^{q+1}\mathbb{P}(\max_{1\le j\le p} \widehat{\textnormal{KL}}_{j}^{k\pi} >\varepsilon )\Big]  + \sum_{k=1}^{q+1}\sup_{\mathcal{D}\in\mathcal{R}^{\#H_{k}\times (2p+1)}}|\mathbb{P}(\mathcal{X}_{k}\in\mathcal{D})-\mathbb{P} (\mathcal{X}_{k}^{\pi} \in\mathcal{D})|,
			\end{split}
	\end{equation}}
	where $0<\tau^*<1$ is the target FDR level and for each $1\le k \le q+1$ and $1\le j\le p$,
	\begin{equation}
		\begin{split}\label{kl.1}
			\widehat{\textnormal{KL}}_{j}^{k\pi} & = \sum_{i\in H_{k}}\log{\left(\frac{f_{X_{j}, \widetilde{X}_{j}, \boldsymbol{x}_{-j}, \widetilde{\boldsymbol{x}}_{-j}} ( X_{ij}^{\pi}, \widetilde{X}_{ij}^{\pi}, \boldsymbol{x}_{-ij}^{\pi}, \widetilde{\boldsymbol{x}}_{-ij}^{\pi})}{f_{X_{j}, \widetilde{X}_{j}, \boldsymbol{x}_{-j}, \widetilde{\boldsymbol{x}}_{-j}} (\widetilde{X}_{ij}^{\pi}, X_{ij}^{\pi}, \boldsymbol{x}_{-ij}^{\pi}, \widetilde{\boldsymbol{x}}_{-ij}^{\pi} )}\right)}.
		\end{split}
	\end{equation}
\end{theorem}

An example is provided in Section \ref{SecA.1} of the Supplementary Material, where the KL divergence term on the right-hand side of \eqref{eq: FDR-bound-general} asymptotically vanishes. The KL divergence term can be further simplified provided that Condition~\ref{knockoff.generator.2} below, adapted from Definition 1 in \cite{barber2018robust}, is satisfied.  This condition concerns the knockoff generator $\kappa(\cdot, \cdot)$ and $p$ additional coordinate-wise knockoff generators $\kappa_{j} : \mathbb{R}^{p-1} \times \mathcal{R} \mapsto \mathbb{R}$ for $j \in \{1, \dots,  p\}$, where each $\kappa_{j}(\boldsymbol{x}_{-j}, \cdot)$ approximates the conditional distribution of $X_{j}$ given $\boldsymbol{x}_{-j}$.

\begin{condi}[Definition 1 in \cite{barber2018robust}]\label{knockoff.generator.2}
	For each $1\le j\le p$, \textnormal{1)} if $\widetilde{\boldsymbol{z}}=(\widetilde Z_1,\cdots, \widetilde Z_p)^T$ is sampled from the conditional distribution \sloppy$\kappa((X_{1}, \dots, X_{j-1}, \widetilde{X}_{j}^{\dagger}, X_{j+1}, \dots, X_{p}), \cdot)$, then $(\widetilde{X}_{j}^{\dagger}, \widetilde{Z}_{j}, \boldsymbol{x}_{-j}, \widetilde{\boldsymbol{z}}_{-j})$ and $(\widetilde{Z}_{j}, \widetilde{X}_{j}^{\dagger}, \boldsymbol{x}_{-j}, \widetilde{\boldsymbol{z}}_{-j})$ have the same distribution, where  $\widetilde{X}_{j}^{\dagger}$ is sampled from $\kappa_{j}(\boldsymbol{x}_{-j},\cdot)$; and \textnormal{2)} the density function of the distribution of $(\widetilde{X}_{j}^{\dagger}, \widetilde{Z}_{j}, \boldsymbol{x}_{-j}, \widetilde{\boldsymbol{z}}_{-j})$ exists. 
\end{condi}

\begin{corollary} \label{collary.1} 
	Assume that all the conditions of Theorem \ref{theorem1.new} hold.  If further  Condition~\ref{knockoff.generator.2} is satisfied, then \eqref{eq: FDR-bound-general} holds with  
	\begin{equation*}
		\begin{split}
			\widehat{\textnormal{KL}}_{j}^{k\pi} 
	& = \sum_{i\in H_{k}}\log{\left(\frac{f_{X_{j}| \boldsymbol{x}_{-j}}( X_{ij}^{\pi}| \boldsymbol{x}_{-ij}^{\pi}) 
			f_{\widetilde{X}_{j}^{\dagger}| \boldsymbol{x}_{-j}}(\widetilde{X}_{ij}^{\pi}| \boldsymbol{x}_{-ij}^{\pi}) 
		}{f_{X_{j}| \boldsymbol{x}_{-j}}(\widetilde{X}_{ij}^{\pi}|  \boldsymbol{x}_{-ij}^{\pi}) 
			f_{\widetilde{X}_{j}^{\dagger}| \boldsymbol{x}_{-j}}( X_{ij}^{\pi}| \boldsymbol{x}_{-ij}^{\pi}) } \right)},
\end{split}
\end{equation*}
where $\widetilde{X}_{j}^{\dagger}$'s are given in Condition~\ref{knockoff.generator.2} and $f_{\boldsymbol{z}_{1}| \boldsymbol{z}_{2}} ( \boldsymbol{z}_{1}| \boldsymbol{z}_{2}) = f_{\boldsymbol{z}_{1}, \boldsymbol{z}_{2}} ( \boldsymbol{z}_{1}, \boldsymbol{z}_{2}) [f_{\boldsymbol{z}_{2}} ( \boldsymbol{z}_{2})]^{-1}$ is the conditional probability density function of $\boldsymbol{z}_{1}$ given $\boldsymbol{z}_{2}$. 
\end{corollary}

It is seen that when Condition~\ref{knockoff.generator.2} is met, the FDR inflation can be measured by the Kullback--Leibler (KL) divergence between the conditional distributions of $\widetilde{X}_{j}^{\dagger}\vert \boldsymbol{x}_{-j}$ and  $X_{j}\vert \boldsymbol{x}_{-j}$, where the former is described by the coordinate-wise knockoff generator $\kappa_j(\boldsymbol{x}_{-j},\cdot)$. When these two conditional distributions are identical, the knockoff generator $\kappa(\cdot, \cdot)$ satisfying Condition~\ref{knockoff.generator.2} reduces to the ideal knockoff generator in model-X knockoffs \cite{CandesFanJansonLv2018}. In this sense, Condition~\ref{knockoff.generator.2} relaxes the requirement of knowing the exact covariate distribution. 

We borrow the Gaussian example discussed in \cite{barber2018robust} to help understand Condition \ref{knockoff.generator.2}. Consider the Gaussian knockoff generator described around \eqref{ko.sampling.1}.
Lemma~4 of \citep{barber2018robust} demonstrates that $\kappa(\cdot, \cdot)$ and $\kappa_{j}(\cdot, \cdot)$ satisfy Condition~\ref{knockoff.generator.2} if the former is chosen as
{\small\begin{equation}\label{ko.sampling.1}
\kappa(\boldsymbol{x}_{t}, \cdot) \overset{d}{\sim} N(\boldsymbol{x}_{t} - \textnormal{diag}(\vec{s}) \widehat{\Theta} \boldsymbol{x}_{t}, 2\textnormal{diag}(\vec{s}) - \textnormal{diag}(\vec{s})\widehat{\Theta} \textnormal{diag}(\vec{s})),
\end{equation}}%
and each $\kappa_{j}(\boldsymbol{x}_{-j}, \cdot)$ follows the distribution $N( \boldsymbol{x}_{-j}^{\top} \widehat{\Theta}_{-j, j}  \widehat{\Theta}_{jj}^{-1} , \widehat{\Theta}_{jj}^{-1})$, where $\widehat{\Theta}$ is a postive definite estimate of $ [\mathbb{E}(\boldsymbol{x}\boldsymbol{x}^{\top})]^{-1}$ constructed independently from the observed data $(\boldsymbol{y}, \boldsymbol{X})$, $\widehat{\Theta}_{-j, j}$ represents the $j$th column of $\widehat{\Theta}$ with the $j$th component removed, and $\widehat{\Theta}_{jj}$ is the $j$th diagonal entry of $\widehat{\Theta}$. Here, for simplicity, we assume that  $\boldsymbol{x}_t$ has mean zero. 
Corollary \ref{collary.1} indicates that the asymptotic FDR control depends on the estimation accuracy of $\widehat{\Theta}$. See Section~\ref{SecA1} for how we estimate the precision matrix in practice. Meanwhile, see \citep{barber2018robust} for other examples of knockoffs generators satisfying Condition~\ref{knockoff.generator.2}.

Under Condition \ref{ge4} below, a simple upper bound for the second term on the right-hand side of \eqref{eq: FDR-bound-general} can be derived; this result is formally summarized in  Corollary \ref{collary.2} below. An example satisfying Condition \ref{ge4} is provided in Section~\ref{ge2b}.

\begin{condi}[$h$-step $\beta$-mixing with exponential decay]\label{ge4}
The covariate process $\{\boldsymbol{x}_{t}\}$ is a $p$-dimensional stationary Markov chain with a transition kernel $p: \mathbb{R}^{p}\times \mathcal{R}^{p} \longmapsto \mathbb{R}$ and a stationary distribution $\pi$. There exist a positive integer $h$, a measurable function $V: \mathbb{R}^{p}\longrightarrow[0, \infty)$, and some constants $0 \le \rho< 1$ and $0 < C_{0} < \infty$ such that for each $\vec{x}\in\mathbb{R}^{p}$,
$\norm{p^{h}(\vec{x}, \cdot) - \pi(\cdot)}_{TV} \le C \rho^{h}V(\vec{x})$,
where $C>0$ is some constant with $C_{0} \ge C \int_{\mathbb{R}^{p}} V(\vec{x})d\pi(\vec{x})$ and $\norm{\cdot}_{TV}$ denotes the total variation (TV) norm associated with measures. Moreover, for each $\vec{x}\in\mathbb{R}^{p}$, $p(\vec{x}, \cdot)$ is absolutely continuous with respect to the Lebesgue measure.
\end{condi}

\begin{corollary} \label{collary.2} 
Assume that all the conditions of Theorem \ref{theorem1.new} hold.
If further  $\{\boldsymbol{x}_{t}\}_{i\ge1}$ satisfies Condition~\ref{ge4} with $q$-step and constants $C_{0}>0$ and $0 \le \rho<1$, and $Y_{i}$ is $\boldsymbol{x}_{t+1}$-measurable, then \eqref{eq: FDR-bound-general} holds with 
{\small \begin{equation}\label{asym.1}
	\sum_{k=1}^{q+1}\sup_{\mathcal{D}\in\mathcal{R}^{\#H_{k}\times (2p+1)}}|\mathbb{P}(\mathcal{X}_{k}\in\mathcal{D})-\mathbb{P} (\mathcal{X}_{k}^{\pi} \in\mathcal{D})| \le C_{0} \times \rho^{q}\times n.
	\end{equation}}%
	Moreover, when $(Y_{t}, \boldsymbol{x}_{t})$'s are i.i.d., \eqref{asym.1} holds with $\rho=0$.
\end{corollary}

As shown in Corollary \ref{collary.2},  in the i.i.d. data setting where $\rho=0$, the FDR upper bound in \eqref{eq: FDR-bound-general} replicates the result in \cite{barber2018robust} when $q=0$ (i.e., no subsampling). With serial dependence and general $q$, from \eqref{eq: FDR-bound-general} and \eqref{asym.1}, we observe an interesting tradeoff between the KL divergence and the TV upper bound as $q$ changes. First, in view of the expression in Corollary \ref{collary.1},  $\mathbb{P}(\max_{1\le j\le p} \widehat{\textnormal{KL}}_{j}^{k\pi} >\varepsilon )$ depends on $q$ in a complicated way via the subsample size $\lfloor n/(q+1)\rfloor$. In addition, since the probability is upper bounded by 1, the term $\sum_{k=1}^{q+1}\mathbb{P}(\max_{1\le j\le p} \widehat{\textnormal{KL}}_{j}^{k\pi} >\varepsilon )$ increases at most linearly with $q$. On the other hand, Corollary \ref{collary.2} suggests that the TV upper bound decreases exponentially with $q$. Despite this tradeoff, it is unreasonable to determine the optimal choice of $q$ by directly analyzing these upper bounds since they can be conservative. We leave the theoretical investigation of the optimal $q$ for future research. Our empirical study suggests that $q = 1$ or $2$ often yields satisfactory finite-sample control of the FDR, as long as each subsample is reasonably large.


To provide a concrete time series example where our theory in this section applies, we analyze the Gaussian ARX model (Example~\ref{ARX_example} in the next subsection) with $\Sigma = \mathbb{E}(\boldsymbol{x} \boldsymbol{x}^{\top})$ assumed to be known, and $q  = \lceil (\log{n})^{1+\delta} \rceil$ and $p = O(n^{K_{0}})$ for some constants $\delta>0$ and $K_{0}>0$. We show in Section~\ref{SecA.1} that the KL divergence is zero and prove in Section~\ref{ge2b} that Condition~\ref{ge4} is satisfied. Thus, the FDR upper bound in Theorem~\ref{theorem1.new} becomes $\tau^*+ C_0n\rho^q$.   When $\Sigma$ is unknown and needs to be estimated, it is possible to obtain an upper bound using an estimated covariance matrix formed from a large independent learning sample using the proof idea for Lemma 5 of \citep{barber2018robust}.

When moving beyond the Gaussian time series covariates, for example, Model~\ref{SETAR1} in Section~\ref{Sec6}, we assume the high-level condition $\sum_{k=1}^{q+1}\mathbb{P}(\max_{1\le j\le p} \widehat{\textnormal{KL}}_{j}^{k\pi} >\varepsilon ) = o(1)$ and the $\beta$-mixing condition as in Corollary~\ref{collary.2}. Under these assumptions, our theory applies. Despite these technical assumptions, our simulation results suggest that the FDR can be controlled in broad model settings with in-sample estimated covariate distribution. Theoretical justification of such empirical results for time series data is highly challenging and is left for future investigation.

\subsection{Stationary processes satisfying Condition \ref{ge4}} \label{ge2b}

We provide a linear time series example that satisfies Condition \ref{ge4}.   Additional examples, including nonlinear ones, are provided in Section \ref{ge2} of the Supplementary Material. 
\begin{exmp}[Autoregressive models with exogenous variables (ARX)]\label{ARX_example} For each  $t$, define 
\[Y_{t} = \sum_{j = 1}^{k_{1}} \alpha_{j}Y_{t-j} + \sum_{l = 1}^{k_{2}}\sum_{j=1}^{k_{3l}} \beta_{j}^{(l)}H_{t - j + 1}^{(l)}+ \varepsilon_{t} \]
with $H_{t}^{(l)} = \epsilon_{t}^{(l)} + \sum_{j=1}^{k_{3l}}b_{j}^{(l)} H_{t-j}^{(l)}$, where for some positive constants $L_{0}, L_{1}$, and $L_{2}$, it satisfies that $1 - \sum_{j=1}^{k_{3l}} b_{j}^{(l)}z^{j} \not = 0$ and $1 - \sum_{j=1}^{k_{1}}\alpha_{j}z^{j} \not = 0$ for each $|z|\le 1 + L_{0}$ and each $l\in\{1, \dots, k_{2}\}$. Additionally, $\sum_{l=1}^{k_{2}}\sum_{j=1}^{k_{3l}}| \beta_{j}^{(l)}| < L_{1}$ and  $k_{1}, k_{31}, \dots, k_{3k_{2}} < L_{2}$. Here,  $k_{1}$, $k_{2}$, and $k_{31}, \dots, k_{3k_{2}}$ are all positive integers. Moreover, 
$(\epsilon_{t}^{(1)}, \dots$, $\epsilon_{t}^{(k_{2})}, \varepsilon_{t})$'s are $(k_{2} + 1)$-dimensional i.i.d. Gaussian random vectors with zero mean and positive definite covariance matrix $\Sigma_{0}$ that satisfy $\mathbb{E}(\varepsilon_{t} \epsilon_{t}^{(l)}) = 0$ for each $l$.
\end{exmp}
Example~\ref{ARX_example}  is a benchmark model for describing the behavior of macroeconomic variables.  It is seen that the covariate vector with respect to response $Y_t$ is  $\boldsymbol{x}_{t} = (Y_{t-1}, \dots, Y_{t-k_{1}}, \boldsymbol{h}_{t}^{\top})^{\top}$ with $\boldsymbol{h}_{t}  = (H_{t}^{(1)}, \dots, H_{t - k_{31} + 1}^{(1)}, \dots, H_{t}^{(k_{2})}, \dots, H_{t-k_{3 k_{2}} + 1}^{(k_{2})})^{\top}$. It has been shown (e.g., \citep{an1996geometrical}) that the stationary process $\{\boldsymbol{x}_{t}\}$ in Example~\ref{ARX_example} with fixed dimensionality satisfies Condition~\ref{ge4} with $h$-step and some constants $\rho, C_{0}$ for each positive integer $h$.
When the dimensionality of $\boldsymbol{x}_t$ increases, certain growth conditions such as \eqref{hds1} below on the value of $h$ and the dimensionality of $\boldsymbol{x}_t$ are needed for Condition~\ref{ge4} to hold. For this reason, we make the dependence of the stationary processes on $h$ explicit: hereafter  $\{\boldsymbol{x}_{t}^{(h)}\}$ denotes a $p_{h}$-dimensional stationary process satisfying Condition~\ref{ge4} for $h \geq 1$, as guaranteed by Proposition~\ref{glp1} below.

\begin{proposition}\label{glp1}
Let $\{\boldsymbol{x}_{t}^{(h)}\}$ be a sequence of $p_{h}$-dimensional linear process in  Example \ref{ARX_example} with constant $L_{i}$'s  and uniformly positive definite $\Sigma_{0}^{(h)}$'s. Assume that for some constant $C_{2} > 0$ and sufficiently small $s_{2} > 0$, 
{\small \begin{equation}\label{hds1}\sup_{h>0} \{p_{h}\exp{(-s_{2}h)}\} \le C_{2}.
\end{equation}}%
Then for some constants $0 \le \rho < 1$ and $C_{0} > 0$ and all large $h$, $\{\boldsymbol{x}_{t}^{(h)}\}$ satisfies Condition~\ref{ge4}. 
\end{proposition}
In view of Theorem~\ref{theorem1.new} and Proposition~\ref{glp1}, if we assume Example~\ref{ARX_example} with $p = p_{q_{n}} = O(n^{K_{0}})$ and  subsample with $q=q_{n}=\ceil{(\log{n})^{1+\delta}}$, where  $K_{0}>0$ and $\delta>0$ are some constants, then the $\beta$-mixing convergence rate required by Theorem~\ref{theorem1.new} 
is satisfied by the subsampled data for all large $n$.

\section{Power analysis under generalized linear time series models}\label{Sec3.new}

Since the selection power of any procedure depends on the signal strength, we showcase the power analysis using the GLM time series model where the signal strength can be measured conveniently by the regression coefficients. Correspondingly, we consider the LCD knockoff statistic in Example~\ref{knockoff.exmp.2} because Lasso is popularly used in high-dimensional GLM regression.   

The canonical GLM has the conditional mean function
{\small \begin{equation}
\label{ing2}
\mathbb{E}(Y_t|\boldsymbol{x}_t) =
g(\boldsymbol{x}_t^T \vec{\beta}^{o}),
\end{equation}}%
where $\vec{\beta}^{o}$
is a vector of unknown coefficients
and $g(\cdot)$ is the derivative of a 
differentiable function $r(\cdot)$.
The inverse function of 
$r'(\cdot)$, denoted as $g^{-1}(\cdot)$, is referred to as the canonical link function.
Commonly used canonical link functions include:
(1) the identity link $g^{-1}(\mu) = \mu$ for the linear model, (2) the logit link $g^{-1}(\mu) = \log (\mu/(1-\mu))$ with $0<\mu<1$ for the logistic model,
and (3) the log link $g^{-1}(\mu) = \log \mu$ with $\mu>0$ for the Poisson model.

To form the LCD knockoff statistics, we first fit the following GLM Lasso regression
{\small \begin{eqnarray}
\label{ing1}
(\widehat{\beta}_{1}, \dots , \widehat{\beta}_{2p})^T = \arg\min_{\vec{\beta}\in\mathbb{R}^{2p}} \left\{n^{-1}\sum_{t=1}^{n} 2 \left(-Y_{t} \times  (\boldsymbol{x}_{t}^{\top}, \widetilde{\boldsymbol{x}}_{t}^{\top}) \vec{\beta} + r\left((\boldsymbol{x}_{t}^{\top}, \widetilde{\boldsymbol{x}}_{t}^{\top}) \vec{\beta}\right)\right) + \lambda_{n} \sum_{j=1}^{2p}|\beta_{j}|\right\},
\end{eqnarray}}%
where $\lambda_n\geq 0$ is the regularization parameter. Define $\vec{\beta}^*:=(\beta_{1}^{*}, \dots, \beta_{2p}^{*})^T $ with $(\beta_1^*, \ldots, \beta_p^*)^T=\vec{\beta}^{o}$
and $\beta_{p+1}^*= \dots = \beta_{2p}^*= 0$.  
By the conditional independence 
$Y_t\independent\tilde{\boldsymbol{x}}_t\vert\boldsymbol{x}_t$, we have 
$\mathbb{E}(Y_t|\boldsymbol{x}_t,\tilde{\boldsymbol{x}}_t)
= \mathbb{E}(Y_t|\boldsymbol{x}_t)=
g(\boldsymbol{x}^T_t \vec{\beta}^{o}) = g((\boldsymbol{x}_t^{\top}, \widetilde{\boldsymbol{x}}_{t}^{\top}) \vec{\beta}^{*})$. Thus,  \eqref{ing1} estimates the population parameter vector $\vec{\beta}^*$.  When the link function is the identity, the GLM Lasso estimate becomes the linear Lasso estimate.

Extensive research has been conducted to study the performance of Lasso with time series data. See, for example,
~\cite{
adamek2022lasso,
basu2015regularized,
han2023high,
kock2015oracle,  
medeiros2016l1, wong2020lasso}.
Specifically, error bounds for the linear Lasso
estimates in ARX models with conditionally heteroskedastic errors have been considered in \citep{ adamek2022lasso, medeiros2016l1, wong2020lasso}. 
Moreover, \citep{han2023high}
established error bounds and support recovery guarantees of the GLM
Lasso when both the response and covariate vector
are stationary time series
with dependence measures (as defined therein)
satisfying certain summability conditions. Our power analysis needs the following technical conditions.

\begin{condi}    
\label{theorem4.new.4c}
For some constant $c_{0}>0$ and sequence $k_{3n}>0$ with $\lim_{n\rightarrow \infty}k_{3n}=0$, it holds that
$\mathbb{P}\left(\sum_{j=1}^{2p}|\widehat{\beta}_{j} - \beta_{j}^{*}|\le c_{0}(\# S^*)\lambda_{n}\right) \ge 1- k_{3n}$, where
$S^* = \{j: |\beta_{j}^*|>0\}$. 
\end{condi}

\begin{condi}\label{power.new.1}
There exists some sequence $k_{1n}>0$ with $k_{1n} q^{-1}\uparrow\infty$ as $n\uparrow\infty$ such that $\min_{j\in S^*}|\beta_{j}^{*}|> k_{1n} \lambda_{n}$. 
\end{condi}

\begin{condi} \label{power.new.2}
For some constant $c_{1}\in(0,1)$ and sequence $\{k_{2n}\}$ with  $\lim_{n\uparrow\infty}k_{2n} = 0$, it holds that $2(\tau_{1} \# S^*)^{-1}<c_{1}$ and $\mathbb{P}(\# \{j:W_{j}^{k} \ge T^k\} \ge c_{1}(\#S^*)) \ge  1-k_{2n}$ for $k\in\{1, \dots, q+1\}$. 
\end{condi}

We note that  
Condition~\ref{theorem4.new.4c} and \eqref{theorem4.new.4b} below
are the $L_1$- and $L_2$-estimation error bounds for the linear Lasso, which have been established in previous studies
\citep{
adamek2022lasso, basu2015regularized, 
kock2015oracle, medeiros2016l1, wong2020lasso} in time series settings under various different model assumptions with model-specific choice of $\lambda_n$. In these studies, $k_{1n}, k_{2n}$, and $k_{3n}$ typically decrease at
polynomial rates as $n$ increases. Additionally,  the GLM Lasso estimation error bounds have been established in  \citep{han2023high}.
As a concrete example, for the model in Example~\ref{ARX_example} with dimensionality $p$ increasing at most exponentially with sample size $n$, we can prove that Condition~\ref{theorem4.new.4c} and \eqref{theorem4.new.4b}  hold with $\lambda_n = O([\log{p}]^{3/2} n^{- 1/2 + \alpha})$ for some $\alpha \in (0,  1/2)$ following the same proof as that for Theorem 3.1 in \citep{han2023high} if, say, the Gaussian knockoff generator \eqref{ex-Gaussian-knockoffs} is used. Since the proof is a straightforward extension, we opt to omit it to save space. 


Condition~\ref{power.new.1}
shares similarities with the often-imposed beta-min assumption $\min_{j\in S^*}|\beta_{j}^{*}|>\sqrt{\#S^*} \times  \lambda_{n}$
for ensuring Lasso's model selection consistency.
If $q$ is chosen as
$O((\log{n})^{1+\delta})$ as suggested by Proposition \ref{glp1} and $\sqrt{\#S^*}  \gg (\log{n})^{1+\delta}$, then
Condition~\ref{power.new.1} 
becomes less restrictive than the beta-min condition.
Condition~\ref{power.new.2} is a technical assumption
that can be proved using the same derivations as in \citep{FanDemirkayaLiLv2020} under Condition~\ref{theorem4.new.4c}   and the following $L_2$-estimation error bound condition
{\small \begin{equation}
\label{theorem4.new.4b}
\mathbb{P}\big(\big[\sum_{j=1}^{2p}(\widehat{\beta}_{j} - \beta_{j}^{*})^2\big]^{\frac{1}{2}}\le c_{0}\sqrt{\# S^*}\lambda_{n}\big) \ge 1- k_{2n}.
\end{equation} }%
For technical simplicity in power analysis, we assume that there are no ties in the magnitude of nonzero knockoff statistics and Lasso solutions. Furthermore, we assume that
$T^k$'s in \eqref{tau.2} satisfy
$\max_{1\le k\le q+1}\{T^k\}<\infty$ almost surely.

\begin{theorem}\label{theorem4.new}
Assume $S^*\subset\{1, \dots, p\}$ with $\# S^*>0$. Let $\widehat{S}$ be returned by Algorithm~\ref{Algorithm1} with $\tau_{1}, \tau_{*}\in(0, 1)$ and the LCD knockoff statistics (Cf. Example~\ref{knockoff.exmp.2}). Assume that Conditions~\ref{power.new.1}--\ref{power.new.2} are satisfied and Condition \ref{theorem4.new.4c} holds for the Lasso estimates applied to each subsample $H_k$ in Algorithm~\ref{Algorithm1}. Then it holds that for all large $n$, 
{small\begin{equation}
	\label{theorem4.eq.2}
	\mathbb{P}\left(\{\widehat S = \emptyset\}\cup \left\{\frac{\#(S^*\cap \widehat{S}) }{\# S^*} \ge 1 - 4c_0(1+q)k_{1n}^{-1}\right\}\right)\ge 1-(q+1)\times (k_{2n}+k_{3n}).
	\end{equation}}%
	If we further assume that $\tau_{1} = \tau^* \times K^{-1}\times (1- 4(q+1)c_{0}k_{1n}^{-1})$ with some $K>1$, then for all large $n$, 
	{\small \begin{equation}
	\label{theorem4.new.2}
	\mathbb{E} \left(\frac{\#(S^*\cap \widehat{S}) }{\# S^*} \right)\ge \left(1 - \frac{(q+1)(\tau_{1}+\theta_{\varepsilon})K}{K-1} - (q+1)\times (k_{2n}+k_{3n})\right)\times k_{4n},
	\end{equation}}%
	where $\{k_{4n}\}$ is some increasing sequence with $\lim_{n\rightarrow\infty}k_{4n} = 1$ and 
	{\small\begin{equation}
	\begin{split}\label{power.new.10}
		\theta_{\varepsilon}= \inf \left\{ \theta\ge 0: \max_{1\le k\le q+1}\mathbb{E}\left(\frac{\# (\{j: W_{j}^k\ge T^k\} \cap (S^*)^{c}) }{ \#\{j: W_{j}^k\ge T^k\} \vee 1 }\right)\le \tau_{1} + \theta \right\}.
	\end{split}
	\end{equation}}%

\end{theorem}

As previously discussed,  $k_{2n}$ and $k_{3n}$ generally converge to zero at polynomial rates as $n$ increases. Hence, setting
$q = O((\log{n})^{1+\delta})$ makes the right-hand side of \eqref{theorem4.eq.2} asymptotically approaching one. Thus, with asymptotic probability one, Algorithm \ref{Algorithm1} either makes no discovery or has the percentage of correct discovery $\frac{\#(S^*\cap \widehat{S}) }{\# S^*}$ approaching one. The event of no discovery occurs when most individual knockoff filters from different subsamples select an abundantly large number of false positives so that the resulting e-values for all variables become too small to be selected. To further exclude the event $\{\widehat S = \emptyset\}$,  it is essential for $(\tau_1 + \theta_{\varepsilon})q$ to be asymptotically vanishing (see \eqref{theorem4.new.2}). To understand this, note that when
$(\tau_{1} + \theta_{\varepsilon})q$
is sufficiently small, it effectively controls the false discovery proportion for each knockoff filter, consequently providing an upper bound on $\sum_{s=1}^p\mathbf 1_{\{W_s^k\leq -T^k\}}$ for each $k=1,\cdots, q+1$.
This further entails that the $j$th e-value statistic given in Algorithm~\ref{Algorithm1} is sufficiently large for each $j\in S^*$ so that it can be selected by the e-value procedure and hence  $\widehat S \not =\emptyset$. 
In practical implementation, in light of the definition of $\tau_1$, we can make $\tau_1q$ asymptotically negligible by choosing $K\gg q =O((\log{n})^{1+\delta})$. 


\section{Simulation studies}\label{Sec6}

We now investigate the finite-sample performance of the TSKI procedure with Algorithm \ref{Algorithm1}. 
Two selection methods considered in this section are TSKI with LCD in Example~\ref{knockoff.exmp.2} and TSKI with random forests MDA in Example~\ref{knockoff.exmp.1}; we abbreviate them as TSKI-LCD and TSKI-MDA, respectively. 
We generate the approximate knockoff variables using the idea of the second-order approximation~\cite{CandesFanJansonLv2018, FanDemirkayaLiLv2020} via the knockoff generator  \eqref{ko.sampling.1}; 
See 
Section~\ref{SecA1} for the estimation of $\widehat\Theta$ and the choice of $\vec{s}$.

We compare with two benchmark methods. The first method selects features using the Benjamini--Yekutieli (BY) method~\citep{benjamini2001control}. We choose not to use the Benjamini--Hochberg (BH) method~\citep{benjamini1995controlling} because it assumes independence among the underlying p-values, which is not expected to hold true for time series applications. Due to the limited results on obtaining p-values for high-dimensional time series regression models, for the BY method, we calculated p-values by utilizing the ordinary least squares method in our simulations when $n>p$. When $n<p$, the BY method is not applicable due to the lack of an effective p-value calculation method. The second benchmark method is the adaptive Lasso~\cite{zou2006adaptive, medeiros2016l1}, which was designed for feature selection without the aim of FDR control. These two approaches are abbreviated as LS-BY and adaLasso, respectively. 


\subsection{Simulation settings} \label{Sec6.1}

We consider the self-exciting threshold autoregressive model with exogenous variables (SETARX~\cite{tong1980threshold}), which is one of the most commonly used models for regime changes in time series data. 
Regime changes are frequently observed in economic time series (e.g., time series in Figure \ref{fig:inflation}).   It is well-known that SETARX with fixed dimensionality $p$ satisfies the $\beta$-mixing Condition~\ref{ge4} and is stationary; see, e.g., \citep{an1996geometrical}.
We present additional simulation experiments in Section~\ref{SecA1} of the Supplementary Material, where some other popular time series models such as the linear autoregressive model with exogenous variables and the autoregressive conditional heteroskedasticity model with exogenous variables~\cite{engle1982autoregressive} are considered.

\begin{model}[SETARX model]\label{SETAR1} For each integer $t$ and $\iota\in \{0, 5\}$, we define 
{\small	\[Y_{t} = \begin{cases} \sum_{j=1}^{2} (-0.5)^{j-1} \beta Y_{t-j} + 0.6(\sum_{j=1}^{\iota}  H_{t, j} + \sum_{j=\iota+1}^{15} H_{t, j})+ \varepsilon_{t}, \quad \textnormal{ if } Y_{t-d} > r,\\
\sum_{j=1}^{2} -(-0.5)^{j-1} \beta Y_{t-j} +0.6( -\sum_{j=1}^{\iota}  H_{t, j} + \sum_{j=\iota+1}^{15} H_{t, j}) + \varepsilon_{t}, \quad \textnormal{ otherwise,}
\end{cases}
\]}%
where we choose $r = 0.7$ and $ d = 1$ as the threshold value and the threshold lag, respectively.
\end{model}
We set the autoregressive coefficient $\beta=0.7$, and choose the model errors $\varepsilon_{t}$'s as i.i.d. $N(0,1)$. 
The time series covariates $H_{t,j}$'s are generated as $H_{t, j} =  \eta\times  H_{t-1, j} + \epsilon_{t, j}$ with $j\in\{1, \dots, 50\}$ and $\eta=0.2$, where $(\epsilon_{t, 1}, \dots, \epsilon_{t, 50})$'s are i.i.d. Gaussian random vectors across $t$ with zero mean and $\mathbb{E}(\epsilon_{t, k}\epsilon_{t, l}) = (0.2)^{|k-l|}$ for all $k, l$. We formulate the covariate vector with respect to response $Y_t$ as $\boldsymbol{x}_{t} = (Y_{t-1}, \dots, Y_{t-20}, \boldsymbol{h}_{t}, \boldsymbol{h}_{t-1}, \boldsymbol{h}_{t-2}, \boldsymbol{h}_{t-3}, \boldsymbol{h}_{t-4})$ with   $\boldsymbol{h}_{t} = (H_{t, 1},\dots, H_{t, 50})$, giving rise to $p =270$. Throughout the simulation, we keep $p=270$ while varying the sample size $n$ across experiments with $n\in\{200, 300, 500, 1000\}$. Due to the space constraint, we move the results for $n=300$ and $1000$ to Section~\ref{SecA1}. 

\begin{figure}[tp]
\centering
\includegraphics[width=7cm]{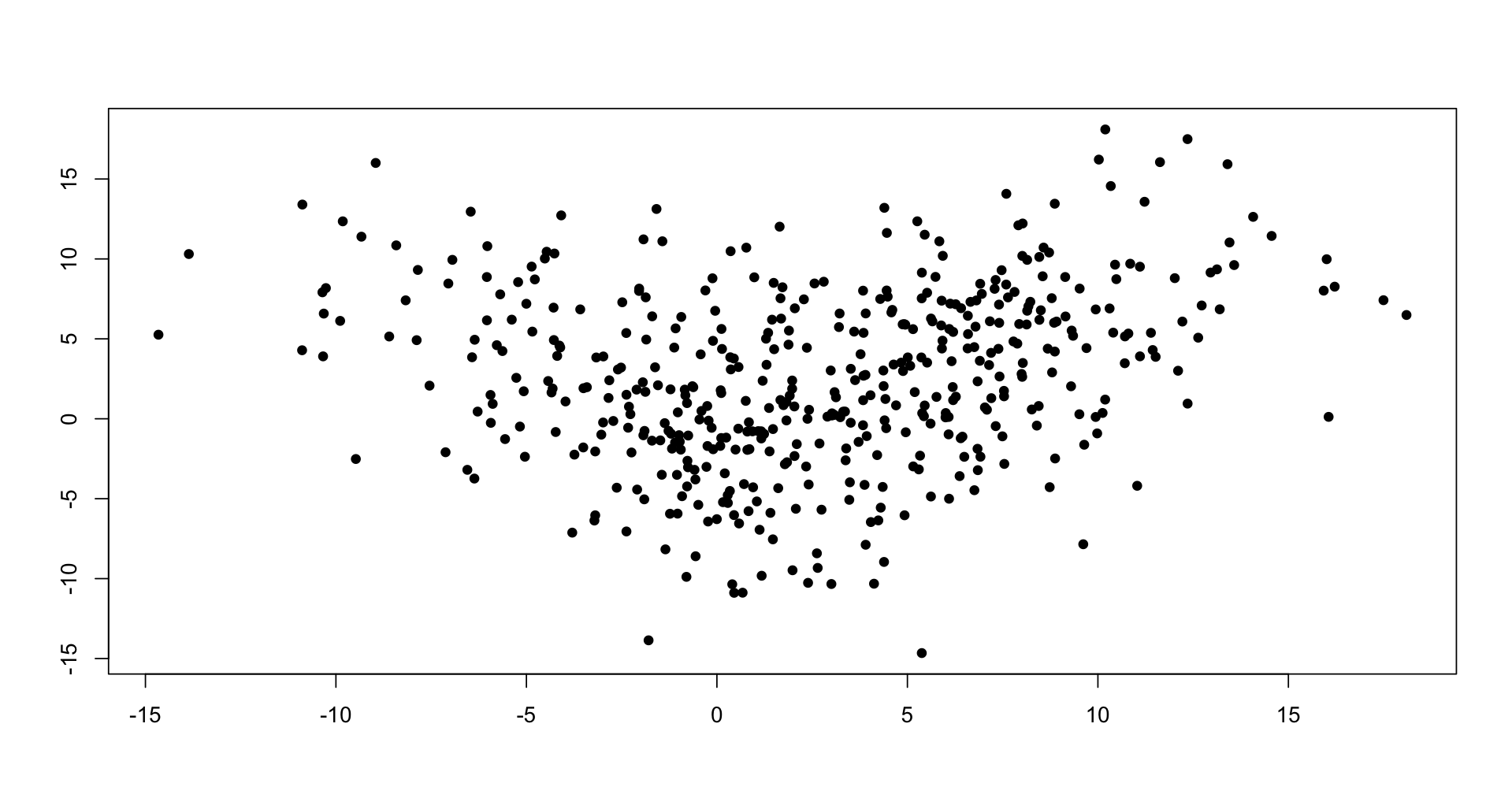}
\caption{A graphical representation depicting $Y_{t-1}$ on the $x$-axis and $Y_{t}$ on the $y$-axis is provided under Model~\ref{SETAR1} with $(\eta, \iota) = (0.2, 5)$.}
\label{fig:nonlinear}
\end{figure}

It is seen that the mean function is comprised of two components: a piecewise linear function of the lags and the first $\iota$ time series covariates, along with a linear function involving $(H_{t, \iota+1}, \dots,  H_{t, 15})$ for $\iota \in \{0, 5\}$. The setup with $\iota = 5$ mimics a realistic and straightforward scenario where certain variables $H_{t,j}$, but not all, together with lagged variables $Y_{t-j}$, undergo changes in regime. The mean regression function of Model~\ref{SETAR1} is therefore nonlinear, as illustrated graphically in Figure~\ref{fig:nonlinear}. The relevant index set according to Definition~\ref{null.features} is $S_0 = \{1, 2, 21, \dots, 35\}$, while the null set $\mathcal{H}_{0}= \{3,\dots,20, 36, \dots, 270\}$.

For implementation, the target FDR level is set as $\tau^* = 0.2$, and the R packages \texttt{glmnet} and \texttt{randomForest} are used for calculating the Lasso estimates and the random forests MDA, respectively.  The TSKI Algorithm~\ref{Algorithm1} 
with parameters $q \in \{{0, 1, 2}\}$, $\tau^* = 0.2$, and $\tau_{1} = \tau^*/(q+1)$ is used in our simulation. The empirical versions of the FDR and power are calculated as the sample averages of the false discovery proportion and true discovery proportion across 100 repetitions, respectively.  
The values of $(p, q, n)$ are included in both Tables~\ref{fig:setarx2}--\ref{fig:adalasso2}, and the R codes are available in the Supplementary Material.

\subsection{Empirical performance of TSKI}

\begin{table}[t!]
\footnotesize
\centering
\begin{tabular}{c | cccc |cccc}
\hline
Method & $n/p/\eta/ \iota$   &  $q$ & FDR & Power  & $n/p/\eta/ \iota$   & $q$ & FDR & Power   \\   \hline
TSKI-LCD & \multirow{4}{*}{$200$/$270$/0.2/0} 
&0  & 0.157 &0.698 & \multirow{4}{*}{$500$/$270$/0.2/0} &0  & 0.176 & 0.939 \\    
TSKI-LCD & & 1  & 0.026 & 0.051 & &1  &  0.099 & 0.872 \\ 
TSKI-MDA & & 0 & 0.173 &  0.456 & &0 &  0.181 &  0.922 \\ 
TSKI-MDA & & 1 &  0.026 &  0.028 & &1 &  0.092 & 0.550 \\ 
\hline     
TSKI-LCD & \multirow{4}{*}{$200$/$270$/0.2/5} 
& 0  & 0.139 & 0.287 & \multirow{4}{*}{$500$/$270$/0.2/5} 
&0  & 0.141 & 0.634 \\
TSKI-LCD & &1  & 0.023 & 0.019 & &1  &  0.086 & 0.267 \\ 
TSKI-MDA & & 0 & 0.138 &  0.215 & &0 &  0.166 &  0.679 \\ 
TSKI-MDA & & 1 & 0.012 & 0.011 &  &1 &  0.084 & 0.216 \\
\hline
\end{tabular}

\caption{The simulation results on the empirical FDR and power for the TSKI with $\tau_{1} = \tau^* / (q+1)$ and $\tau^* = 0.2$ under Model \ref{SETAR1} in Section \ref{Sec6.1}. The results for $n\in \{300, 1000\}$ and $q=2$ are given in Section~\ref{SecA1}}.\label{fig:setarx2}
\end{table}

\begin{table}[t!]
\footnotesize
\centering

\begin{tabular}{cccc}
\multicolumn{3}{c}{Adaptive Lasso}\\
\hline
{\footnotesize $n/p/\eta/\iota$}   &   FDR & Power  \\ \hline
{\footnotesize$200$/270/0.2/0} & 0.520 & 0.964 \\
{\footnotesize$300$/270/0.2/0} & 0.468 & 0.997 \\
{\footnotesize$500$/270/0.2/0} & 0.657 & 1.000 \\
{\footnotesize$200$/270/0.2/5} & 0.604 & 0.705 \\
{\footnotesize$300$/270/0.2/5} & 0.563 & 0.786 \\
{\footnotesize$500$/270/0.2/5} & 0.677 & 0.891 \\
\hline
\end{tabular}		
\begin{tabular}{cccc}
\multicolumn{3}{c}{LS +  BY}\\
\hline
{\footnotesize$n/p/\eta/\iota$ }  &   FDR & Power  \\ \hline
{\footnotesize$200$/270/0.2/0 }& -- & -- \\
{\footnotesize$300$/270/0.2/0 }& 0.000 & 0.001 \\
{\footnotesize$500$/270/0.2/0} & 0.027 & 0.763 \\  
{\footnotesize$200$/270/0.2/5 }& -- & -- \\
{\footnotesize$300$/270/0.2/5} & 0.018 & 0.006 \\
{\footnotesize$500$/270/0.2/5 }& 0.026 & 0.276 \\

\hline

\end{tabular}
\caption{Left panel: the simulation results on the empirical FDR and power for the adaptive Lasso~\cite{medeiros2016l1, zou2006adaptive}; there is no target FDR level for the adaptive Lasso. 
Right panel: the simulation results on the empirical FDR and power for the ordinary least squares +  Benjamini--Yekutieli (BY~\citep{benjamini2001control}) with the target FDR level at $0.2$; this approach does not apply to high-dimensional scenarios when $n<p$.
}\label{fig:adalasso2}
\end{table}

Table~\ref{fig:setarx2} presents the results for TSKI-LCD and TSKI-MDA with $q=0$ and $q=1$. The complete results, which also include $q=2$ and a larger sample size $n=1000$, are provided in Section~\ref{SecA1}.  It is seen from Table~\ref{fig:setarx2} that both methods control the FDR in finite samples below the target level of $\tau^*=0.2$, while larger $q$ gives lower selection power compared to that of $q=0$ (i.e., no subsampling). The lower power for larger $q$ is reasonable and caused by the small sample size in each subsample when fitting the time series regression model. The knockoffs method is empirically known to be conservative when the sample size is overly small. This conservative nature offsets the FDR inflation due to serial dependence in the small sample size scenario, which explains why FDR is still controlled even without subsampling. The extended Model~\ref{SETAR1} simulation presented in Section~\ref{SecA1} shows that for a larger sample size $n=1000$, we start observing FDR inflation when $q=0$. In addition, in the additional simulation examples presented in Section~\ref{SecA1}, we observe severe FDR inflation when $q=0$ in various scenarios (see Table \ref{fig:arx}).   

From Table~\ref{fig:setarx2}, we see that for Model~\ref{SETAR1} with $\iota=0$ (i.e., lower nonlinearity), the LCD-based method demonstrates superior performance over the MDA-based method in terms of power. Meanwhile, when $\iota=5$ (i.e., higher nonlinearity), we observe from Table~\ref{fig:setarx2} that MDA outperforms LCD in power when the sample size is large (i.e., $n=500$ and $q=0$), an intuitive observation considering the nonparametric nature of the MDA measure. Indeed, the empirical performance of MDA decreases drastically in all settings as $q$ increases, because of the smaller sample size when calculating the MDA measures.  

The results from Table~\ref{fig:setarx2} highlight that TSKI-MDA may not be suitable for some real data applications, such as our study where only $60$ observations are accessible for each inference window. They underline the need to explore ways to enhance selection power for nonlinear time series, especially with limited samples. 
On the other hand, Table~\ref{fig:adalasso2} highlights that adaLasso demonstrates one of the highest selection powers among all four methods. However, adaLasso fails to control error rates under Model~\ref{SETAR1}. This is expected because adaLasso is designed for the linear model and focuses on consistent model selection, lacking an error rate tuning parameter, such as the target FDR level.  LS-BY  is inapplicable to high-dimensional time series applications when $n<p$, such as our real data example in Section~\ref{real.data.1}, due to the lack of a reliable p-value calculation method. 
It is important to emphasize that obtaining valid p-values under high-dimensional linear or nonlinear time series models presents a highly challenging and currently unresolved issue; such challenge is also reflected in the poor performance of LS-BY when $p$ becomes comparable to $n$ (i.e., $(n,p)=(300, 270)$). 

It is worth pointing out that our additional simulation experiments in Section~\ref{SecA1} of the Supplementary Material reveal that: 1) TSKI-LCD and TSKI-MDA with $q=1$ consistently control FDR below the target level, whereas setting $q=0$ yields results occasionally exceeding the target level; and 2) the LS-BY method can have FDR exceeding the target level when data is simulated from autoregressive conditional heteroskedasticity models with exogenous variables and higher value of $\eta$ (as in Model~\ref{SETAR1}). See the Supplementary Material for full details.


Overall, our simulation experiments show that the time series FDR is controlled stably by the TSKI procedure with subsampling parameter $q=1$ in finite samples, whereas the selection power depends on the sample size, the subsampling parameter $q\ge0$, and the choice of the feature importance measure. Importantly, our theoretical results (Theorem~\ref{theorem1.new} and Corollary~\ref{collary.2}) and simulation results show that TSKI is among the first approaches with theoretically justified FDR control for dependent data under flexible $\beta$-mixing condition in Condition~\ref{ge4}. 
For implementation, we suggest that practitioners working on time series inference with limited sample sizes initiate their diagnosis by using TSKI-LCD with $q$ set to $1$, and generate knockoff variables using knockoff generator~\eqref{ko.sampling.1}.


\section{Real data application}\label{real.data.1}

\begin{figure}[t!]
\includegraphics[width=7.4cm]{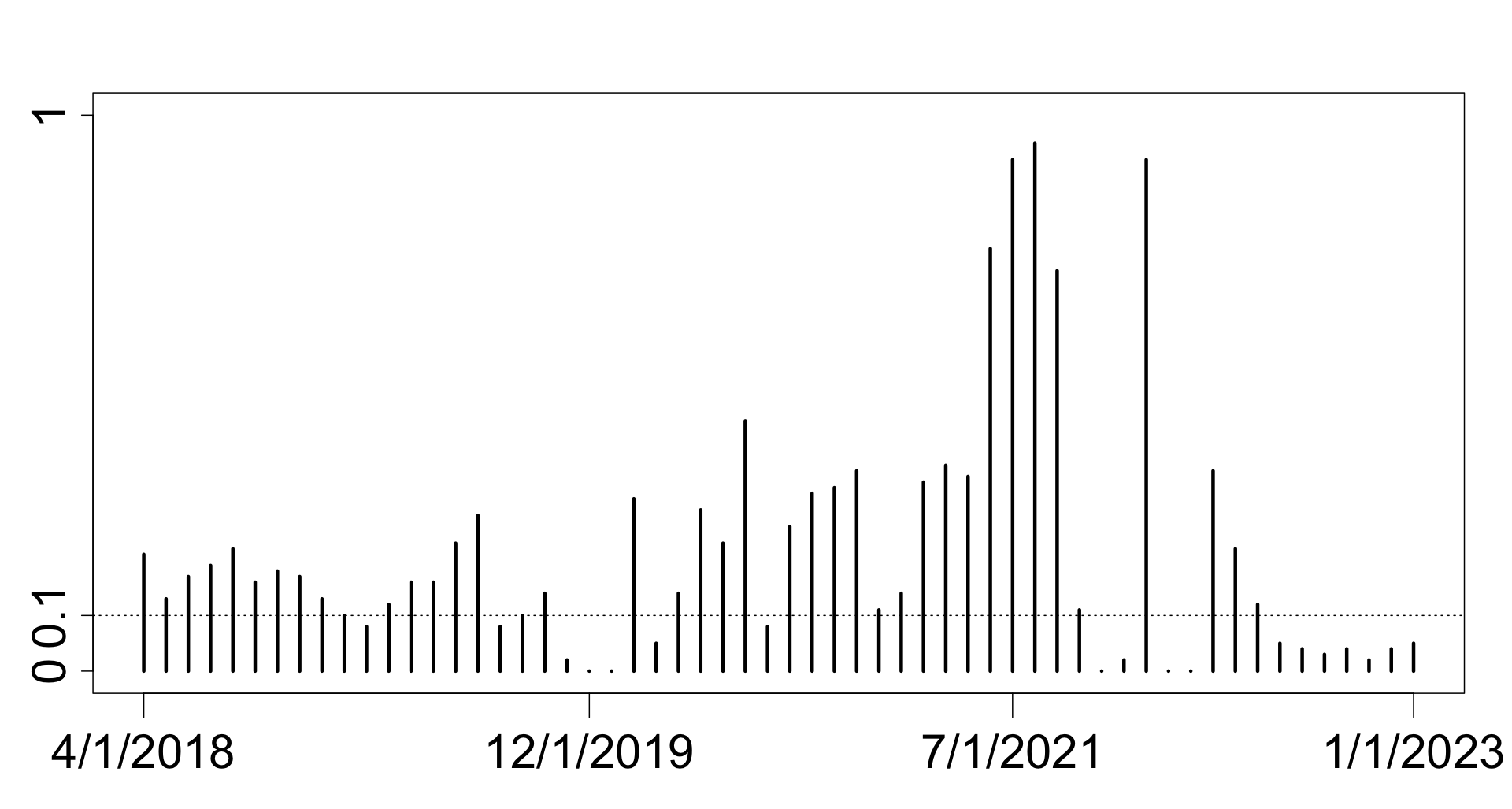}  
\includegraphics[width=7.9cm]{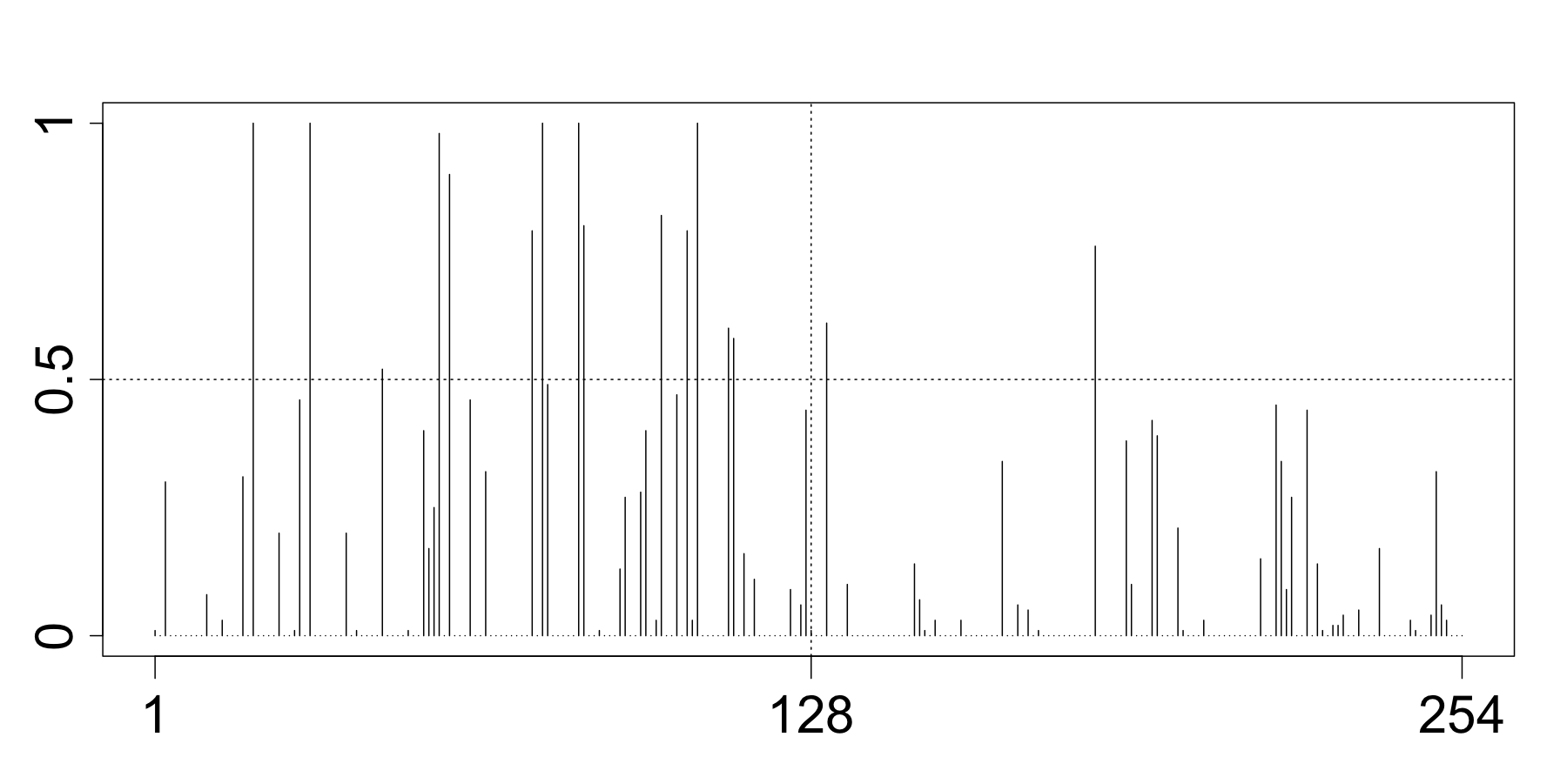}      
\captionsetup{width=1.0\linewidth}
\caption{ The left panel displays the averages of ``having any selections'' indicators over 100 repetitions, where the indicator at each rolling window is one if and only if any covariates are selected, and the $x$-axis indicates the ending time of each rolling window. The right panel shows the analogous results, but the indicator is one at each covariate index if that covariate is selected at any rolling window. The first 127 covariates are current time covariates, and the $128$th to $254$th covariates are one-month lag covariates in the AR(2) model.  Covariates measuring similar economic values are clustered closer (see~\cite{mccracken2016fred} for detailed definitions of these covariates). The selection method here is the TSKI-LCD without subsampling ($q=0$).}
\label{fig:frequency2}
\end{figure}

\begin{figure}[t!]
\centering
\includegraphics[width=7.4cm]{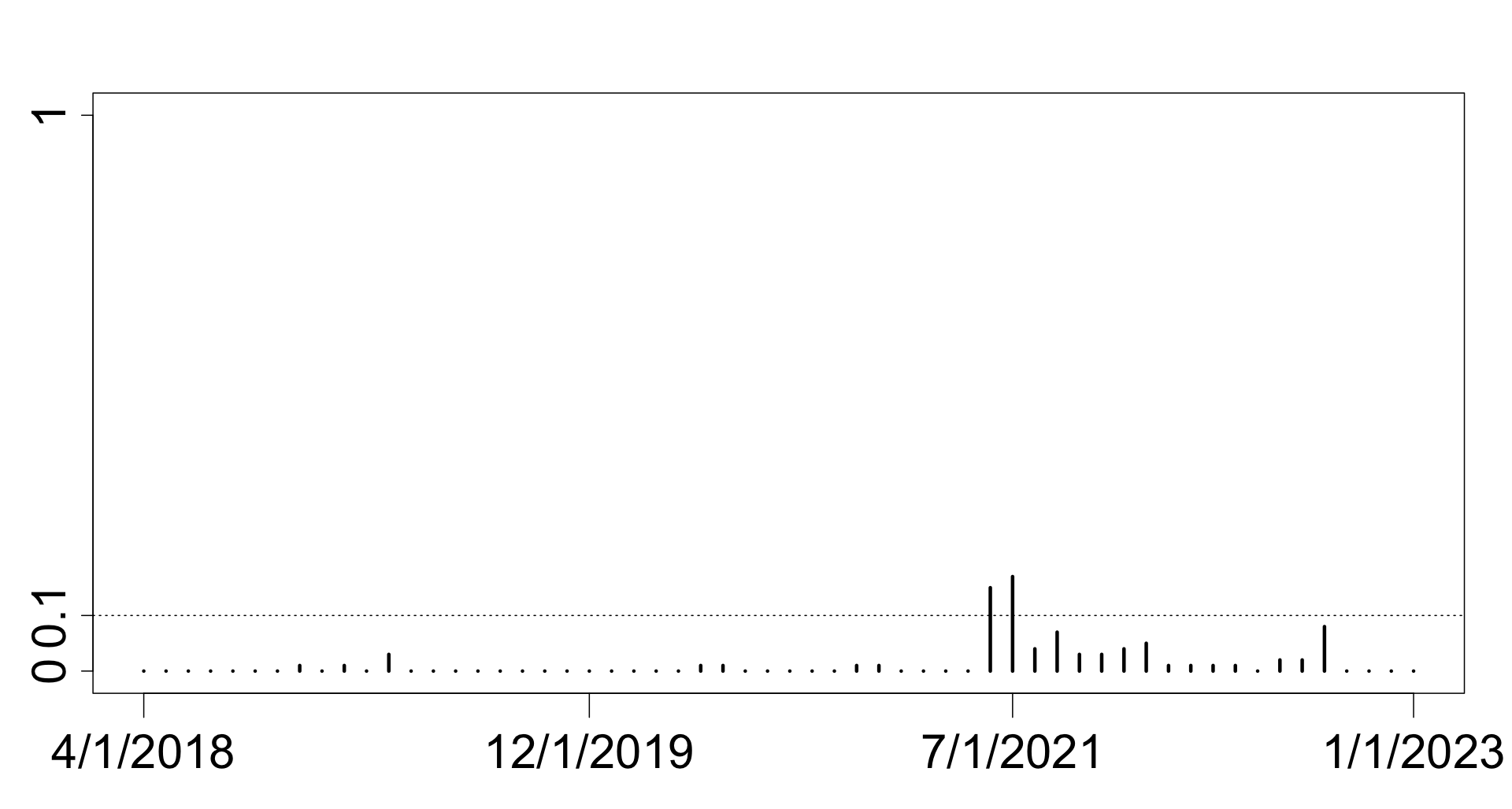}        
\includegraphics[width=7.9cm]{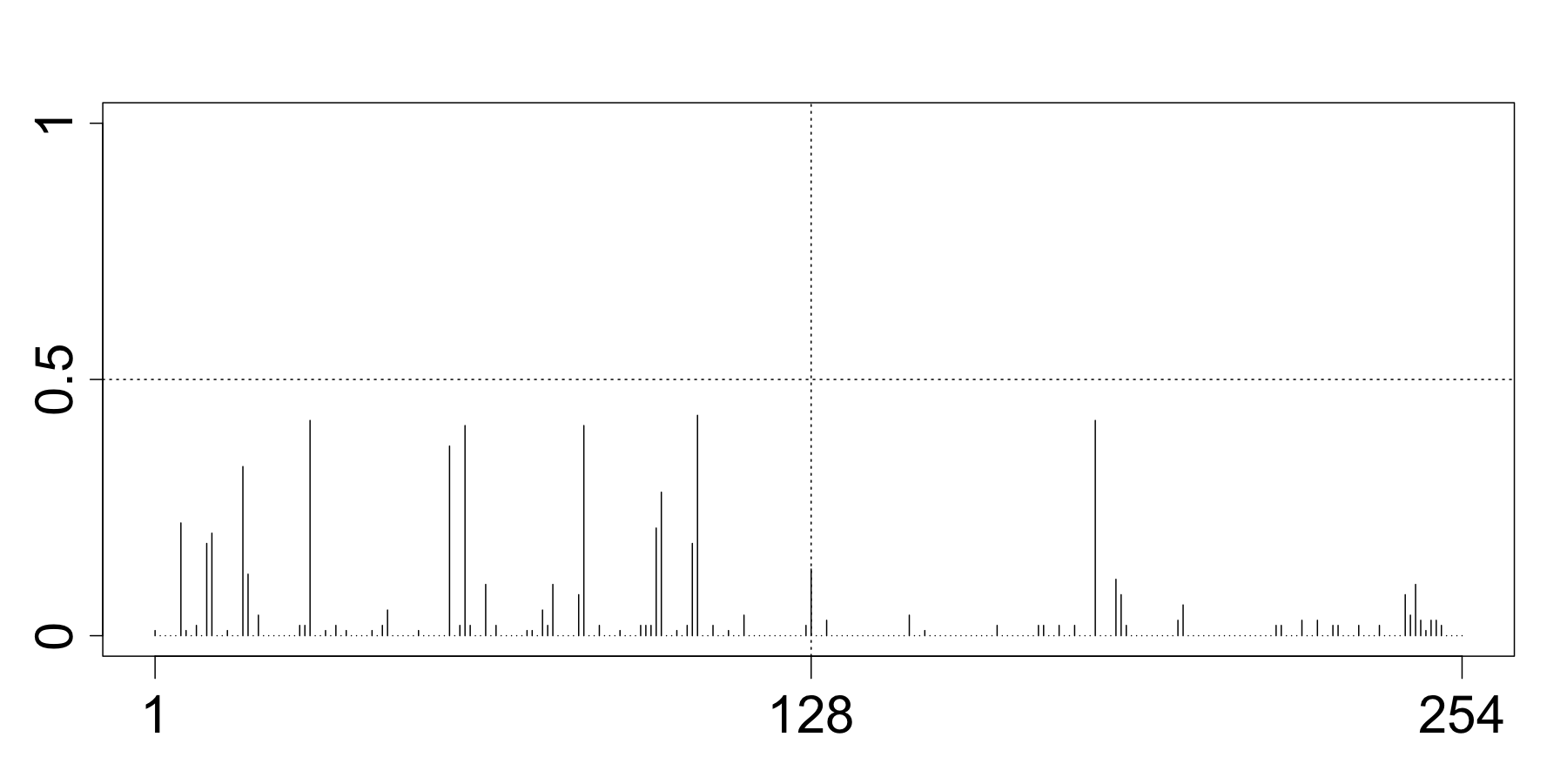}      
\captionsetup{width=1.0\linewidth}
\caption{ These two panels are analogous to those in Figure~\ref{fig:frequency2} but with $q=1$ for the TSKI-LCD procedure.}
\label{fig:tski_mda}
\end{figure}

\begin{figure}[t!]
\centering
\subfloat[The red curve is ACOGNO]{
\includegraphics[width=9cm]{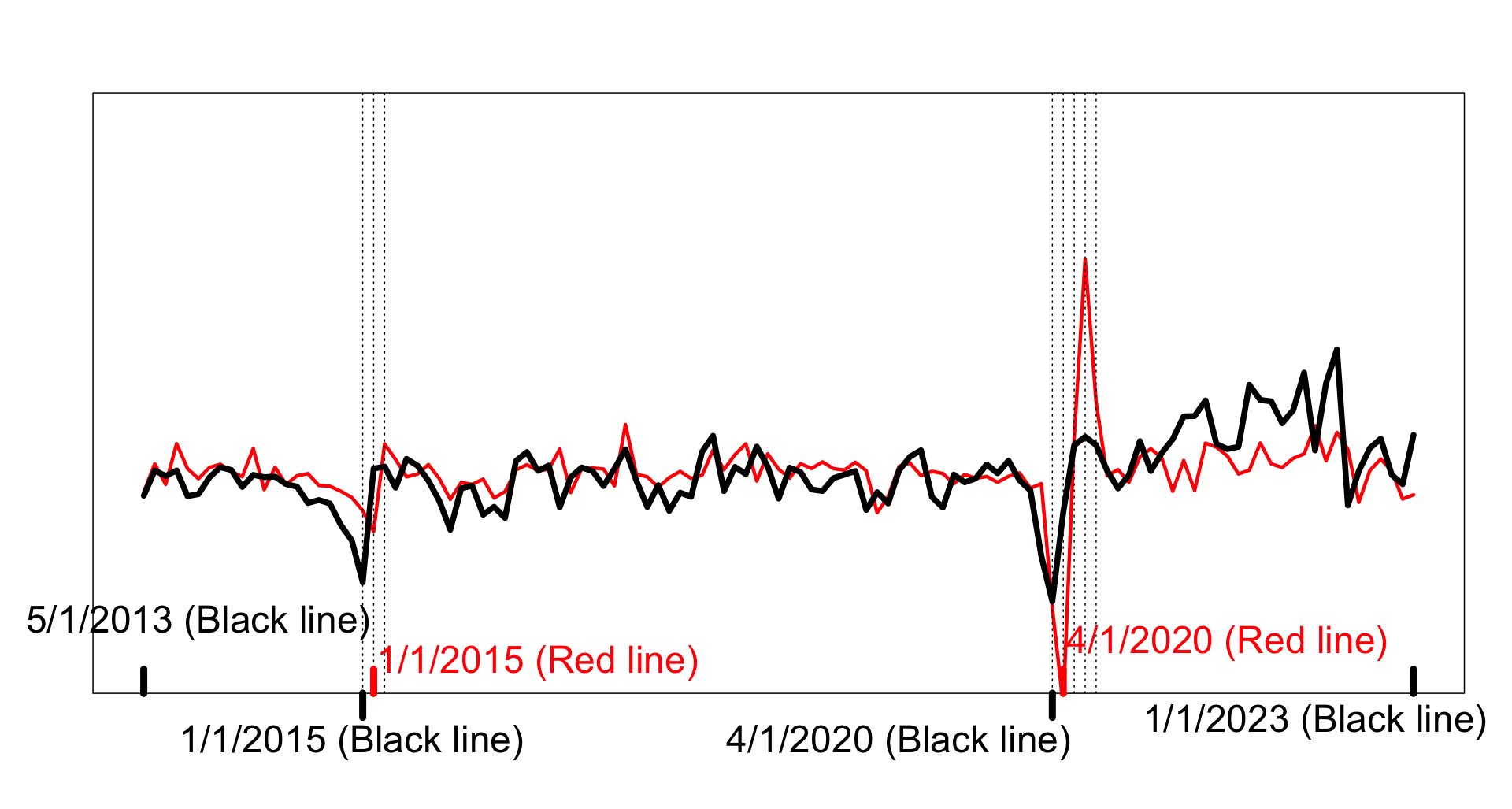}}%


\subfloat[The red curve is EXCAUSx]{\includegraphics[width=9cm]{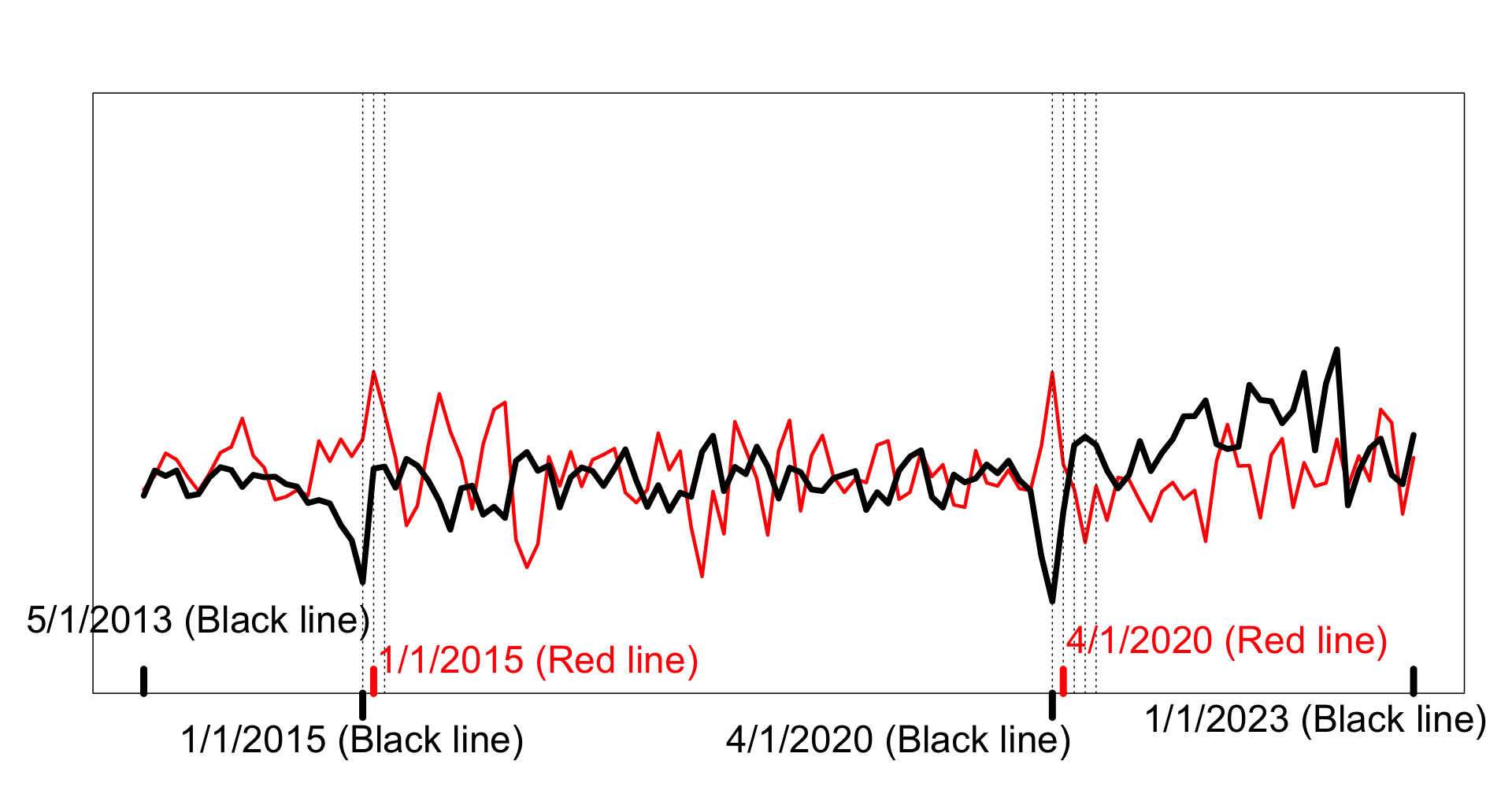}}%

\subfloat[The red curve is CLAIMSx]{\includegraphics[width=9cm]{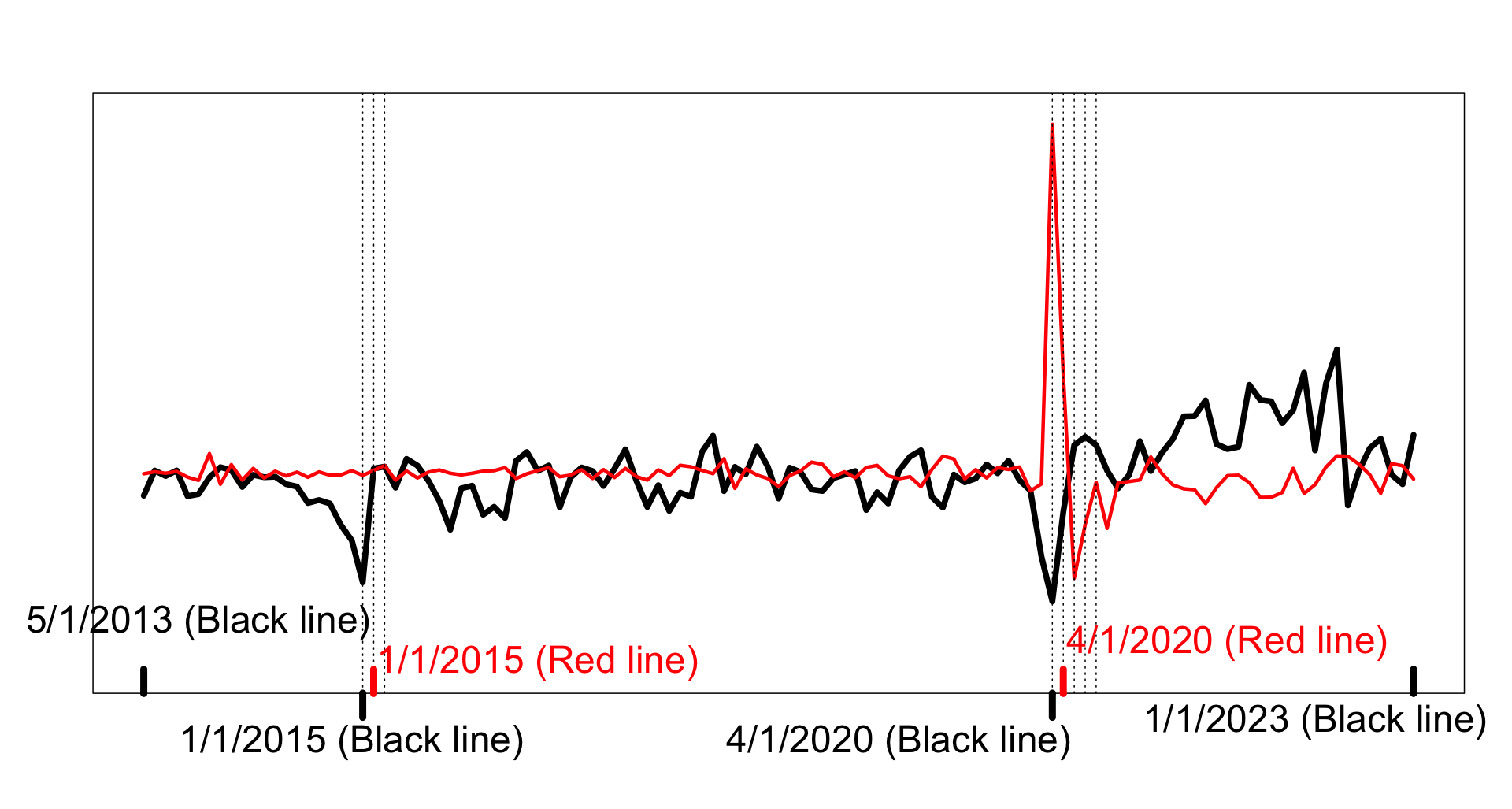}}%
\caption{The black curves in the three panels are the inflation series at time $t+1$. The red curve in panel (a) is the number of new orders for consumer goods at time $t$, the red curve in panel (b) indicates the U.S./Canada exchange rate at time $t$, and the red curve in panel (c) is the U.S. initial claims for unemployment benefits at time $t$. All curves here are standardized and adjusted for visual comparison, and hence the values of these time series are not reported on the $y$-axis.}
\label{fig:relations}
\end{figure}

We analyze the U.S. inflation series data described in the Introduction. The monthly economic time series, including numerous types of consumer price indices, unemployment rates, and housing prices, can be obtained from the FRED-MD database~\citep{mccracken2016fred}\footnote{The website URL: \url{https://research.stlouisfed.org/econ/mccracken/fred-databases/}.} and the U.S. Bureau of Labor Statistics. These time series have been pre-processed following the instructions of the FRED-MD database to make them more stationary \citep{mccracken2016fred}. 
To address the concern on potential nonstationarity over the entire time span, we break it into 58 rolling windows, each covering a five-year period.
Motivated by our simulation study, we apply TSKI-LCD with $q\in\{0, 1\}$ described in Section~\ref{Sec6.1} (with knockoff generator \eqref{ko.sampling.1}) to each five-year rolling window to investigate the temporal relations between the inflation and other time series variables.  The inflation at time $t$ is defined as the adjusted consumer price index for all goods: $\textnormal{Inflation}_{t} \coloneqq \left(\frac{\textnormal{CPI}_{t} - \textnormal{CPI}_{t-1}}{\textnormal{CPI}_{t-1}}  \times 100 \right) \%, $
where CPI$_{t}$ is the consumer price index for all goods at time (month) $t$. Each time series has a FRED-MD code. For example, CPIAUCSL is the FRED-MD code of the inflation series. To reduce randomness resulting from the use of the knockoffs construction, we repeat the inference procedure $100$ times, and report the average results in Figure~\ref{fig:frequency2} with $q = 0$, where the left panel displays whether any significant covariates have been found at each rolling window period, and the right panel shows the selection frequency over $58$ rolling window periods of the 127 time series covariates and their one month lags. That is, we model the one month ahead inflation series $\text{Inflation}_{t+1}$ (response) using an AR(2) model with $127$ time series covariates at current time $t$ and $127$ one month lags of these time series covariates. Consequently, the total covariate dimensionality is $p=254$.

As can be seen in Figure~\ref{fig:frequency2}, the TSKI-LCD with $q=0$ identifies some active windows around the year 2021 (COVID) and 2022 (Russia-Ukraine conflict), with the selection frequency concentrating on a sparse set of covariates. 
The majority of the selected variables are covariates at the current time $t$. The top $10$ most frequently selected covariates by the TSKI-LCD with $q=0$ are employment-related series (HWI, CLAIMSx), consumption-related indices (ACOGNO, CPIAUCSL), housing-related series (PERMITS), U.S. bond yields (GS5, GS10), stock market indices (S.P.div.yield, S.P.500), and exchange rates (EXCAUSx) at their current time $t$, with their FRED-MD codes given in the parentheses.

The simulation results in Section~\ref{Sec6} suggest that the choice of $q=1$ has better FDR control especially when the sample size is limited. Motivated by our simulation results, we next apply the TSKI-LCD with subsampling $q=1$ in Figure~\ref{fig:tski_mda} with the expectation of better FDR control. The results of Figure~\ref{fig:tski_mda} are more conservative in comparison to those in Figure~\ref{fig:frequency2}. Despite being conservative, these new results also suggest some active windows around the same periods as $q=0$, and the selected variables are also mostly covariates at the current time $t$. In addition,  most covariates selected by the TSKI-LCD with $q=1$ belong to the set selected by the TSKI-LCD with $q=0$. In particular, the top 10 selected covariates are CPIAUCSL, CLAIMSx, PERMITS, AMDMUOx, GS10, ACOGNO, EXCAUSx, EXUSUKx, INDPRO, and IPFUELS, where only PERMITS is one-month lag covariate at $t-1$ in the AR(2) model.  Among them, EXUSUKx, AMDMUOx, INDPRO, and IPFUELS are new in comparison to the list of the selected set when $q=0$, where the first two are the U.S./U.K. exchange rate and another type of consumption-related index (the number of unfilled orders for durable goods), respectively, and the last two are industrial production indices that are related to consumption price indices. The difference in the selected sets of covariates is attributed to the fact that some economic covariates are designed to track similar economic factors and tend to be highly correlated. 

The recent literature~\cite{stock2007has} suggests that inflation foresting is a difficult task in the sense that AR models with additional time series covariates rarely outperform simple AR models with only inflation lags. In other words, conditional on the lagged inflation series, additional covariates do not carry strong signals in inflation forecasting. The fact that the TSKI-LCD with $q=1$ selects only a few time series covariates indeed supports such an argument. It is also interesting to notice that the stock market indices are not considered as important covariates by the TSKI-LCD with subsampling (i.e., $q=1$), but are selected as important covariates when $q=0$. In particular, the selection frequencies of S\&P  dividend yields and S\&P 500 (both at lag $t-1$) are $100\%$ and $79\%$, respectively, in Figure~\ref{fig:frequency2}, while only  $5\%$ and $1\%$, respectively, in Figure~\ref{fig:tski_mda}, suggesting that stock market indices could be spurious findings.

The selection results by the TSKI-LCD motivate us to further investigate the dependency of inflation on a few time series covariates. 
In Figure~\ref{fig:relations}, we 
plot three selected series, namely ACOGNO, EXCAUSx, and CLAIMSx, which are among the top 10 lists both when $q=0$ and $q=1$. These three selected covariates are all at their current time $t$ in the AR(2) model. ACOGNO is the number of new orders for consumer goods, which is an important consumption index; EXCAUSx is the exchange rate from the U.S. dollar to the Canadian dollar, and CLAIMSx is the initial claim for unemployment benefits. We also include the inflation series at time $t+1$ in the same period (the black curve in each panel). For better visual comparison, all curves in Figure~\ref{fig:relations} have been standardized and adjusted.

A visual inspection of Figure \ref{fig:relations} shows that the impacts of the COVID-19 pandemic in April 2020 on the U.S. economy are stronger than those of the gasoline price shock in January 2015. This potentially explains why our empirical findings of significant covariates concentrate mostly on this period. From Figure~\ref{fig:relations}, we see that the gasoline price shock in January 2015 affects the consumption index series ACOGNO more mildly compared to the impact of COVID-19 in April 2020. Also, although there is some variation in the exchange rate EXCAUSx after January 2015, it is unclear whether such variation was caused by the gasoline price shock.
In contrast, most of the U.S. economy's time series clearly responded to the pandemic to an unignorable degree.
Particularly, the exchange rate and the number of initial claims dropped in March 2020, suggesting that these covariates were leading indicators of the inflation drop in April 2020.
In summary, we have applied the newly suggested tool of TSKI to study the U.S. economy. 
Our empirical results illustrate the potential of the TSKI to obtain more instructive findings in real data applications.

\bibliographystyle{chicago}
\bibliography{references}	


\newpage
\appendix
\setcounter{page}{1}
\setcounter{section}{0}
\renewcommand{\theequation}{A.\arabic{equation}}
\renewcommand{\thesubsection}{A.\arabic{subsection}}
\setcounter{equation}{0}

\begin{center}{\bf \Large Supplementary Material to ``High-Dimensional Knockoffs Inference for Time Series Data"}

\bigskip

Chien-Ming Chi, Yingying Fan, Ching-Kang Ing and Jinchi Lv
\end{center}

\noindent This Supplementary Material contains the appendix of Section~\ref{Sec2.new}, additional simulation examples, the proofs of all main results and technical lemmas, and some additional technical details. All the notation is the same as defined in the main body of the paper. Additionally, we introduce some technical notation below.  For $\vec{x} \coloneqq (x_{1}, \dots, x_{n})^{\top}\in\mathbb{R}^{n}$, we define  $\norm{\vec{x}}_{k} \coloneqq (\sum_{i=1}^{n}|x_{i}|^{k})^{1/k}$, $\norm{\vec{x}}_{\infty}\coloneqq \max_{1 \le i\le n }|x_{i}|$, and  $\norm{\vec{x}}_{0}\coloneqq \sum_{i=1}^{n} \boldsymbol{1}_{\{x_{i} \neq 0\}}$ with $\boldsymbol{1}_{\{\cdot\}}$ being the indicator function. The distribution of a random mapping $\boldsymbol{X}$ is denoted as $\mu_{\boldsymbol{X}}$. For a matrix $\boldsymbol{X}$ and an index subset $S$, $\boldsymbol{X}(S)$ represents a submatrix of $\boldsymbol{X}$ containing only columns with indices in  $S$. 
The total variation (TV) norm for any measures $\mu_{1}$ and $\mu_{2}$ on $(\Omega, \mathcal{F})$ is defined as $ \norm{\mu_{1} - \mu_{2}}_{TV} \coloneqq 2 \sup_{\mathcal{D}\in\mathcal{F}} |\mu_{1}(\mathcal{D}) - \mu_{2}(\mathcal{D})|$. 

\renewcommand{\thesubsection}{A.\arabic{subsection}}

\section{Appendix of Section~\ref{Sec2.new}} \label{SecA0}

\subsection{Example for Theorem~\ref{theorem1.new}}\label{SecA.1}

We begin with providing an example with asymptotically vanishing KL divergence.
Assume that $\{\boldsymbol{x}_{t}\}$ follows a stationary linear Gaussian process as in Example~\ref{ARX_example} in  Section~\ref{ge2b} with zero mean and precision matrix (i.e., the inverse of the covariance matrix) $\Theta = [\mathbb{E}(\boldsymbol{x}_{t}\boldsymbol{x}_{t}^{\top})]^{-1}$, and the knockoff generator is such that $\kappa(\vec{z}, \cdot)$ follows a Gaussian distribution with mean $(\boldsymbol{I}_{p} - D\widehat{\Theta})\vec{z}$ and variance $2D - D\widehat{\Theta}D $ for each $\vec{z} \in \mathbb{R}^p$, where $\boldsymbol{I}_{p}$ is the $p$-dimensional identity matrix, $\widehat{\Theta}$ is the estimated covariance matrix constructed from an independent learning sample, and $D$ is a diagonal matrix with nonnegative entries such that $2D - D\widehat{\Theta}D$ is positive semidefinite. It has been shown in Lemma 5 of \citep{barber2018robust} that when the data consists of i.i.d. observations without serial dependency,  it holds that for each $\varepsilon > 0$, 
\begin{equation}
\label{rate.1}    \lim_{n\rightarrow\infty}\sum_{k=1}^{q+1}\mathbb{P}(\max_{1\le j\le p} \widehat{\textnormal{KL}}_{j}^{k\pi} >\varepsilon ) = 0
\end{equation}
as long as Condition~\ref{knockoff.generator.2} is satisfied,  $p\gg q$, and
\begin{equation}\label{rate.2}
\max_{1\le j \le p}\Theta_{jj}^{-\frac{1}{2}} \norm{\Theta^{-\frac{1}{2}} (\widehat{\Theta}_{j} - \Theta_{j}) }_{2} = o_p\left\{\frac{1}{\sqrt{n\log{p}}}\right\}, 
\end{equation}
where $\Theta_{jj}$ denotes the $j$th diagonal entry of $\Theta$ and $\Theta_{j}$ represents the $j$th column of $\Theta$. We omit the dependence of the parameters on sample size $n$ here for simplicity. 
More examples on asymptotically vanishing KL divergence for non-time series data can be found in the same paper above. Similar results can be proved for our applications of time series data using the proof techniques in \citep{barber2018robust} by replacing the concentration inequalities for i.i.d. observations with those for $\beta$-mixing data; since the extension is straightforward, we omit the details for simplicity.

We provide two remarks here. First, when $\Theta$ is known, it is obvious that the KL divergence is zero. Second, by Lemma 5 in \citep{barber2018robust}, an independent learning sample of size $\tilde n\gg n\log p$ is needed for \eqref{rate.2} to hold. 
However, our simulation results indicate that using the full sample for both $\Theta$ estimation and TSKI inference can still control the FDR empirically, which suggests that the theoretical requirement may be unnecessarily strong. How to relax such an assumption in a time series setting is left for future investigation.

\subsection{Additional stationary processes satisfying Condition \ref{ge4}} \label{ge2}

\subsubsection{Various nonlinear AR-type processes}\label{Sec2.4.2}

Many time-homogeneous Markov chains satisfy Condition~\ref{ge4}. To name a few, \citep{tjostheim1990non, an1996geometrical} showed that with some additional mild regularity conditions, $\{(Y_{t-1}, \dots, Y_{t-k_{1}})\}$ in  Example~\ref{nonlinear.AR.1} below satisfies Condition~\ref{ge4} for all $h>0$ with some constants $C_{0}$ and $0 \le \rho<1$. 

\begin{exmp}[Nonlinear AR~\citep{tjostheim1990non, an1996geometrical}]\label{nonlinear.AR.1}Let a measurable function $G:\mathbb{R}^{k_{1}}\longrightarrow\mathbb{R}$ for some constant integer $k_1>0$ be given such that $\sup_{\vec{z}\in\mathbb{R}^{k_{1}}}|G(\vec{z})|<\infty$, and $\{\varepsilon_{t}\}$ a sequence of i.i.d. model errors. For each $t$, we define 
$$Y_{t} = G(Y_{t-1}, \dots, Y_{t-k_{1}}) + \varepsilon_{t}.$$
\end{exmp}

The self-exciting threshold autoregressive models (SETAR)~\citep{tong1980threshold, hansen2011threshold} also satisfies Condition~\ref{ge4} for all $h>0$ according to~\citep{an1996geometrical}. For more examples such as the exponential AR models, see~\citep{ozaki19852, an1996geometrical} and the references therein.

\subsubsection{ARCH-type process}\label{garch}

\begin{exmp}[AR($k_{1}$)-X-ARCH($k_{3}$)~\citep{cline2004stability, meitz2010note}] \label{AR.GARCH}
Let $ \varepsilon_{t}$'s be   i.i.d. random variables with zero mean and $\mathbb{E} \varepsilon_{1}^2 =1$,  and  $\{\boldsymbol{h}_{t} \coloneqq(H_{t, 1}, \dots, H_{t, k_{2}})\}$ be a sequence of $k_{2}$-dimensional time series covariates that is independent of $\varepsilon_{t}$'s. Let measurable functions $G_{1}: \mathbb{R}^{k_{3}}\rightarrow (0,\infty)$, $G_{2}: \mathbb{R}^{k_{2}}\rightarrow \mathbb{R}$, and $\gamma_{j}: \mathbb{R}^{k_{1}} \rightarrow \mathbb{R}$ with $1\le j\le  k_{1}$ be given such that $\sum_{j=1}^{k_{1}}\sup_{\vec{z}\in \mathbb{R}^{k_{1}}}|\gamma_{j}(\vec{z})| < 1$. The functional-coefficient  ARX-ARCH model \citep{chen1993functional} is given by 
$$Y_{t}= \sum_{j=1}^{k_{1}}\gamma_{j}(Y_{t-1}, \dots, Y_{t-k_{1}})Y_{t-j}+G_{2}(\boldsymbol{h}_{t})+ \sigma_{t}\varepsilon_{t}
$$
with $\sigma_{t} = G_{1}(\sigma_{t-1}\varepsilon_{t-1}, \dots, \sigma_{t- k_{3}}\varepsilon_{t- k_{3}})$.
\end{exmp}

Example~\ref{AR.GARCH} above is an ARCH model with the mean function consisting of an AR component and exogenous covariates. The covariate vector is $\boldsymbol{x}_{t} = (Y_{t-1} , \dots, Y_{t - k_{1} - k_{3}}, \boldsymbol{h}_{t}, \dots, \boldsymbol{h}_{t-k_{3}})^{\top}$ for response $Y_{t}$. In particular, Model~\ref{ARXGARCH1} in Section~\ref{Sec6} is an example of the ARX-ARCH model above when $\gamma_{j}$’s are constants and the model error follows a standard ARCH process~\citep{engle1982autoregressive}. 

The AR component in Example \ref{AR.GARCH} above can be a general functional-coefficient autoregressive model~\citep{chen1993functional}, and the ARCH component can take the form of a smooth transition ARCH model~\citep{lanne2005non}. With $G_{2}=0$ and additional mild regularity conditions, the Markov chain $\{Y_{t}, \dots, Y_{t-k_{1}-k_{3}+1}\}_{t}$ satisfies Condition~\ref{ge4} for each $h>0$ with  constants $C_{0}>0$ and $0\le \rho<1$ according to \citep{cline2004stability, meitz2010note}. In the presence of exogenous covariates $\boldsymbol{h}_t$, additional regularity conditions on $\boldsymbol{h}_{t}$'s are needed for Condition~\ref{ge4} to hold for the Markov chain; such study is beyond the scope of the current paper and is left for future investigation. 

One challenge in the variable selection problem with time series data is that there does not always exist an obvious definition of the covariate vector. Taking  Example \ref{AR.GARCH} for instance,
the existence of the ARCH component requires us to take into account both the mean function and variance function when selecting the set of non-null variables in the broad sense according to Definition \ref{null.features}.
To better understand this, let us consider an ARX-ARCH(1) model with a standard ARCH component such that $\sigma_{t} = \sqrt{0.1 + 0.9(\sigma_{t-1}\varepsilon_{t-1})^2}$, and write
\begin{equation}\label{arch.component}
\sigma_{t-1}\varepsilon_{t-1} = Y_{t-1} - \sum_{j=1}^{k_{1}}\gamma_{j}(Y_{t-2}, \dots, Y_{t-1-k_{1}})Y_{t-1-j} - G_{2}(\boldsymbol{h}_{t-1}).
\end{equation}
In this example, in addition to variables $(Y_{t-1}, \dots, Y_{t-k_{1}}, \boldsymbol{h}_{t})$ which affect the mean regression function, we should also take into account the lagged covariates in the ARCH component $(Y_{t-1}, \dots, Y_{t-k_{1}-1}, \boldsymbol{h}_{t-1})$ when conducting variable selection in the broad sense according to Definition \ref{null.features}. That is, one sensible choice of the covariate vector is $\boldsymbol{x}_{t} = (Y_{t-1} , \dots, Y_{t - k_{1} - 1}, \boldsymbol{h}_{t}, \boldsymbol{h}_{t-1})^{\top}$. Omitting variables in the variance function (i.e., the ARCH component) and defining the covariate vector as $(Y_{t-1}, \dots, Y_{t-k_{1}}, \boldsymbol{h}_{t})$ may give us a nonsparse set of non-null variables according to Definition \ref{null.features}. Nevertheless, the actual variable selection performance of the TSKI depends on the specific choice of the knockoff statistics, as shown in our simulation section. If the knockoff statistics are constructed based on the mean regression function alone (e.g., the LCD and MDA discussed earlier), then the corresponding TSKI cannot be expected to have power in selecting variables that affect only the variance function. In this sense,  our results in such a scenario should be interpreted as selecting the important variables contributing to the mean regression function alone.

\begin{remark}\label{garch.remark}
We have left out the GARCH-type process~\citep{bollerslev1986generalized} in our discussion because it can be challenging to formulate meaningful covariates for variable selection purposes in such a setting.  Note that a GARCH-type process can be represented as an ARCH-type process with infinite order. Thus, if the covariates vector is not well-formulated such that some active covariates are not included,  the resulting regression model may no longer be sparse, rendering the FDR control problem invalid.  We shall leave the variable selection problem for the GARCH-type process in future work.
\end{remark}

\subsection{Robust TSKI without subsampling}\label{Sec2.4}

In this section, we consider a special case of Algorithm~\ref{Algorithm1} when $q=0$, that is, no subsampling. 
For ease of reference, we provide a full description of the corresponding algorithm in Algorithm~\ref{Algorithm2} below. Our theoretical study here has two major contributions: 1) extending the theory of robust knockoffs inference~\citep{barber2018robust} to its e-value analog, where the non-robust version was first introduced and studied by \citep{ren2022derandomized} for i.i.d. data, and 2) further extending the results to time series applications. By similar analysis as in Theorem~\ref{theorem2.new} below, we can show that the robust knockoffs inference~\citep{barber2018robust} (without using the e-values) can also be extended to time series applications, but the details are omitted here for simplicity. We emphasize that our results \eqref{robust.eq.14}--\eqref{theorem2.new.3} in Theorem~\ref{theorem2.new} below assume \textit{neither} i.i.d. observations \textit{nor} the pairwise exchangibility Condition~\ref{knockoff.generator.2}.

{\small
\begin{algorithm}[h]
\SetAlgoLined
Let $0< \tau_{1}<1$ be a constant and $0<\tau^{*}<1$ the target FDR level.

Calculate the knockoff statistics $W_{1}, \dots, W_{p}$ satisfying \eqref{sign-flip.2} with the full sample $\{Y_{t}, \boldsymbol{x}_{t}, \widetilde{\boldsymbol{x}}_{t}\}_{t=1}^{n}$.

Let $\mathcal{W}_{+} = \{|W_{s}|:|W_{s}|>0\}$. Calculate the e-value statistics $e_{j}$'s such that
\begin{equation}
\begin{split}\label{tau.1}
	e_{j} = \frac{p\times \boldsymbol{1}_{\{W_{j} \ge T\} }}{ 1+\sum_{s=1}^{p} \boldsymbol{1}_{\{W_{s} \le - T\} } }, \ \ 
	T  = \min\left\{t \in \mathcal{W}_{+}:  \frac{1+ \#\{ j : W_{j}\le -t \}}{\#\{ j : W_{j} \ge t\}\vee 1}\le \tau_{1} \right\}.
\end{split}
\end{equation}

Let $\widehat{S} = \{j: e_{j} \ge p(\tau^* \times \widehat{k})^{-1}\}$ with $\widehat{k} = \max\{k:e_{(k)} \ge p(\tau^* \times k)^{{-1}}\}$, where $e_{(j)}$'s are the ordered statistics of $e_{j}$'s such that $e_{(1)}\ge \dots \ge e_{(p)}$.

\caption{{ \footnotesize Time series knockoffs inference (TSKI) via e-values without subsampling} } \label{Algorithm2}
\end{algorithm}
}


Let $\boldsymbol{X}_{-j}$ be the submatrix of $\boldsymbol{X}$ with the $j$th column removed, and $\boldsymbol{X}_{j}$ and $\widetilde{\boldsymbol{X}}_{j}$   the $j$th columns of $\boldsymbol{X}$ and $\widetilde{\boldsymbol{X}}$, respectively. Recall that $(Y, \boldsymbol{x}, \widetilde{\boldsymbol{x}})$ is an independent copy of $(Y_{1}, \boldsymbol{x}_{1}, \widetilde{\boldsymbol{x}}_{1})$ and  $\widetilde{X}_{j}^{\dagger}$ is given in Condition~\ref{knockoff.generator.2} by the $j$th coordinatewise knockoff generator.

\begin{theorem} \label{theorem2.new}
Let $\widehat{S}$ be the set of variables selected by the TSKI Algorithm~\ref{Algorithm2} and $0<\tau^*<1$ the target FDR level. Assume that  Condition~\ref{knockoff.generator.3} holds and $T$ in \eqref{tau.1} is positive. Then we have 
\begin{equation}
\begin{split}\label{robust.eq.14}
\textnormal{FDR}  & \le \inf_{\varepsilon> 0} \Big[\tau^{*} \times e^{\varepsilon} + \mathbb{P}(\max_{1\le j\le p} \widehat{\textnormal{KL}}_{j} >\varepsilon )\Big] ,
\end{split}
\end{equation}
where for each $1\le j\le p$,
\begin{equation}\label{theorem2.new.3}
\widehat{\textnormal{KL}}_{j} = \log{\left(\frac{f_{\boldsymbol{X}_{j}, \widetilde{\boldsymbol{X}}_{j}, \boldsymbol{X}_{-j}, \widetilde{\boldsymbol{X}}_{-j}, \boldsymbol{Y}} ( \boldsymbol{X}_{j}, \widetilde{\boldsymbol{X}}_{j}, \boldsymbol{X}_{-j}, \widetilde{\boldsymbol{X}}_{-j}, \boldsymbol{Y})}{f_{\boldsymbol{X}_{j}, \widetilde{\boldsymbol{X}}_{j}, \boldsymbol{X}_{-j}, \widetilde{\boldsymbol{X}}_{-j}, \boldsymbol{Y}} (\widetilde{\boldsymbol{X}}_{j}, \boldsymbol{X}_{j}, \boldsymbol{X}_{-j}, \widetilde{\boldsymbol{X}}_{-j}, \boldsymbol{Y})}\right)}.
\end{equation}
If we further assume that  Condition~\ref{knockoff.generator.4} is satisfied and $\boldsymbol{X}_{j}$ is independent of $\boldsymbol{Y}$ conditional on $\boldsymbol{X}_{-j}$ for each $j\in \mathcal{H}_{0}$, then we have 
\begin{equation}\label{theorem2.new.2}
\widehat{\textnormal{KL}}_{j} = \log{\left(\frac{f_{\boldsymbol{X}_{j}, \widetilde{\boldsymbol{X}}_{j}, \boldsymbol{X}_{-j}, \widetilde{\boldsymbol{X}}_{-j}} ( \boldsymbol{X}_{j}, \widetilde{\boldsymbol{X}}_{j}, \boldsymbol{X}_{-j}, \widetilde{\boldsymbol{X}}_{-j})}{f_{\boldsymbol{X}_{j}, \widetilde{\boldsymbol{X}}_{j}, \boldsymbol{X}_{-j}, \widetilde{\boldsymbol{X}}_{-j}} (\widetilde{\boldsymbol{X}}_{j}, \boldsymbol{X}_{j}, \boldsymbol{X}_{-j}, \widetilde{\boldsymbol{X}}_{-j})}\right)}.
\end{equation}
Moreover, if $(\boldsymbol{x}, \widetilde{\boldsymbol{x}}), (\boldsymbol{x}_{1}, \widetilde{\boldsymbol{x}}_{1}), \dots, (\boldsymbol{x}_{n}, \widetilde{\boldsymbol{x}}_{n})$ are further assumed to be i.i.d., then we have 
\begin{equation}
\begin{split}\label{robust.eq.6}
\widehat{\textnormal{KL}}_{j} & = \sum_{t=1}^{n}\log{\left(\frac{f_{X_{j}, \widetilde{X}_{j}, \boldsymbol{x}_{-j}, \widetilde{\boldsymbol{x}}_{-j}} ( X_{tj}, \widetilde{X}_{tj}, \boldsymbol{x}_{-tj}, \widetilde{\boldsymbol{x}}_{-tj})}{f_{X_{j}, \widetilde{X}_{j}, \boldsymbol{x}_{-j}, \widetilde{\boldsymbol{x}}_{-j}} (\widetilde{X}_{tj}, X_{tj}, \boldsymbol{x}_{-tj}, \widetilde{\boldsymbol{x}}_{-tj})}\right)}.
\end{split}
\end{equation}
If further Condition~\ref{knockoff.generator.2} is satisfied, then we have 
\begin{equation}
\begin{split}\label{robust.eq.7}
\widehat{\textnormal{KL}}_{j} & = \sum_{t=1}^{n}\log{\left(\frac{f_{X_{j}, \boldsymbol{x}_{-j}} ( X_{tj}, \boldsymbol{x}_{-tj}) f_{\widetilde{X}_{j}^{\dagger}, \boldsymbol{x}_{-j}} ( \widetilde{X}_{tj}, \boldsymbol{x}_{-tj}) }{f_{X_{j}, \boldsymbol{x}_{-j}} ( \widetilde{X}_{tj}, \boldsymbol{x}_{-tj}) f_{\widetilde{X}_{j}^{\dagger}, \boldsymbol{x}_{-j}} ( X_{tj}, \boldsymbol{x}_{-tj}) }\right)}\\
& = \sum_{t=1}^{n}\log{\left(\frac{f_{X_{j}| \boldsymbol{x}_{-j}} ( X_{ij}| \boldsymbol{x}_{-tj}) f_{\widetilde{X}_{j}^{\dagger}| \boldsymbol{x}_{-j}} ( \widetilde{X}_{tj}| \boldsymbol{x}_{-tj}) }{f_{X_{j}| \boldsymbol{x}_{-j}} ( \widetilde{X}_{tj}| \boldsymbol{x}_{-tj}) f_{\widetilde{X}_{j}^{\dagger}| \boldsymbol{x}_{-j}} ( X_{tj}| \boldsymbol{x}_{-tj}) }\right)},
\end{split}
\end{equation}
where $f_{\boldsymbol{z}_{1}| \boldsymbol{z}_{2}} ( \boldsymbol{z}_{1}| \boldsymbol{z}_{2})$ denotes the conditional probability density function of $\boldsymbol{z}_{1}$ given $\boldsymbol{z}_{2}$.
\end{theorem}

The proof of Theorem \ref{theorem2.new} follows mainly those in \citep{barber2018robust, ren2022derandomized, wang2022false} and is presented in Section~\ref{SecA.2.new} later. 
Comparing \eqref{robust.eq.7} to \eqref{robust.eq.6}, we see that the KL divergences become invariant to $\widetilde{\boldsymbol{x}}_{-j}$ thanks to the additional assumption Condition~\ref{knockoff.generator.2}. The simplified form in \eqref{robust.eq.7} is important in proving the asymptotic FDR control as in \eqref{rate.1}. In addition, Condition~\ref{knockoff.generator.2} allows for deviation of the conditional distribution of $\widetilde{X}_{j}^{\dagger}|\boldsymbol{x}_{-j}$ from the true underlying conditional distribution of $X_j|\boldsymbol{x}_{-j}$, making the procedure more practically applicable, as verified in examples given in~\citep{barber2018robust}.


\subsection{Appendix of Section~\ref{Sec6}} \label{SecA1}

\subsubsection{ARX and ARXARCH models}
In this section, we present additional simulation examples: the autoregressive model with exogenous variables and the autoregressive conditional heteroskedasticity model with exogenous variables, as detailed in~\citep{engle1982autoregressive}.
All the symbols and notation are consistent with those in Section~\ref{Sec6}. 

\begin{model}[ARX model]\label{ARX1} For each integer $t$, we define 
\[Y_{t} = \sum_{j=1}^{2} (-0.5)^{j-1} \beta Y_{t-j} + \sum_{j=1}^{15}0.6\times H_{t, j}  + \varepsilon_{t}.\]
\end{model}

\begin{model}[ARX-ARCH model] \label{ARXGARCH1} For each $t$, we define 
\[Y_{t} = \sum_{j=1}^{2}(-0.5)^{j-1} \beta Y_{t - j} + \sum_{j=1}^{15} 0.6\times H_{t, j}  + \sigma_{t}\varepsilon_{t} \]
with 
$\sigma_{t}^{2} = 0.1 + 0.9 (\sigma_{t-1}\varepsilon_{t-1})^{2}$. 
\end{model}

The model error $\{\varepsilon_{t}\}$ is a sequence of i.i.d. standard Gaussian random variables and $\beta =0.7$. The time series covariates are given by $H_{t, j} =  \eta\times  H_{t-1, j} + \epsilon_{t, j}$ with $j\in\{1, \dots, 50\}$ and some $\eta\in \{0.2, 0.95\}$, where $(\epsilon_{t, 1}, \dots, \epsilon_{t, 50})$'s are i.i.d. Gaussian random vectors with zero mean and $\mathbb{E}(\epsilon_{t, k}\epsilon_{t, l}) = (0.2)^{|k-l|}$ for all $k, l$. We formulate the coveriate vector with respect to response $Y_t$ as $\boldsymbol{x}_{t} = (Y_{t-1}, \dots, Y_{t-20}, \boldsymbol{h}_{t}, \boldsymbol{h}_{t-1}, \boldsymbol{h}_{t-2}, \boldsymbol{h}_{t-3}, \boldsymbol{h}_{t-4})$, where   $\boldsymbol{h}_{t} = (H_{t, 1},\dots, H_{t, 50})$, giving rise to $p =270$. We set $p=270$ but vary the sample size $n$ across experiments with $n\in\{200, 300, 500\}$.

In Models~\ref{ARX1} and \ref{ARXGARCH1}, the mean functions both depend linearly on the covariates.  It is worth mentioning that because of the ARCH component, for Model~\ref{ARXGARCH1}, the relevant and null sets according to definition \ref{null.features} are $S_{\textnormal{arch}} = \{1, 2, 3, 21, \dots, 35, 71, \dots 85\}$ and $\mathcal{H}_{\textnormal{arch}} = \{4,\dots,20, 36, \dots, 70, 86,\dots, 270\}$, respectively. The sets $S_{0}$ and $\mathcal{H}_{0}$ defined previously are the sets of active and null covariates, respectively, in the mean regression function. Although $S_{0}$ and $\mathcal{H}_{0}$ differ from $ S_{\textnormal{arch}}$ and $\mathcal{H}_{\textnormal{arch}}$, respectively, in Model \ref{ARXGARCH1}, we examine the empirical power and FDR of the TSKI with respect to  $\mathcal S_{0}$ and $\mathcal{H}_{0}$ for two reasons: 1) this is an interesting problem in time series inference and 2) random forests and Lasso are both algorithms designed for fitting the mean regression and thus are not expected to detect variables that affect only the variance function.

For implementation, the target FDR level is set as $\tau^* = 0.2$, and the R packages \texttt{glmnet} and \texttt{randomForest} are used for calculating the Lasso estimates and the random forests MDA, respectively. We generate the approximate knockoff variables using the idea of the second-order approximation~\citep{CandesFanJansonLv2018, FanDemirkayaLiLv2020}. Specifically, 
for the ideal scenario with a zero-mean random vector $\boldsymbol{x}$ given, we sample its knockoff vector from the multivariate Gaussian distribution (also see \eqref{ko.sampling.1})
\begin{equation}
\widetilde{\boldsymbol{x}}|\boldsymbol{x} \sim N(\boldsymbol{x} - \textnormal{diag}(\vec{s}) \Sigma^{-1} \boldsymbol{x}, 2\textnormal{diag}(\vec{s}) - \textnormal{diag}(\vec{s})\Sigma^{-1} \textnormal{diag}(\vec{s})),
\end{equation}
where $\Sigma = \mathbb{E}(\boldsymbol{x}\boldsymbol{x}^{\top})$, $\vec{s}\in\mathbb{R}^p$ denotes the tuning parameters, and diag$(\vec{s})$ is a diagonal matrix with diagonal entries in $\vec{s}$. Larger components of $\vec{s}$ imply that the resulting knockoff variables deviate more from the original features, thereby providing higher power in distinguishing them. Further details about this knockoff variable sampling procedure can be found in~\citep{CandesFanJansonLv2018, FanDemirkayaLiLv2020}. In practical applications, we provide an estimate of the precision matrix $\widehat{\Sigma}^{-1}$ using the full sample and the method developed in \citep{fan2016innovated}, and select $\vec{s} = (\widehat{s}, \dots, \widehat{s})^{\top}$ with $\widehat{s}$ the inverse of the maximum eigenvalue of $\widehat{\Sigma}^{-1}$. Notably, this method matches the first two moments of the original covariates and their knockoffs counterparts and is thus termed the second-order approximation method. The TSKI Algorithm~\ref{Algorithm1} 
with subsampling parameter $q \in \{0, 1\}$, $\tau^* = 0.2$, and $\tau_{1}= \tau^* /(q+1)$ is considered in our simulation. The R  code for the simulation experiments is available in the online Supplementary Material.



\subsubsection{Empirical performance of TSKI}

\begin{table}[tp]		
\footnotesize
\begin{tabular}{clccc}
\hline
$n/p/\eta$   & Method & $q$ & FDR & Power  \\ \hline
\multirow{4}{*}{$200$/$270/0.2$} 
& TSKI-LCD &0  &0.237 & 0.992 \\
& TSKI-LCD &1  & 0.108 & 0.529 \\ 

& TSKI-MDA &0 & 0.273 &  0.391 \\ 
& TSKI-MDA &1 & 0.026 & 0.021 \\
\hline
\multirow{4}{*}{$300$/$270/0.2$} 
& TSKI-LCD &0  &0.189 & 0.999 \\
& TSKI-LCD &1  & 0.110 & 0.97 \\ 
& TSKI-MDA &0 & 0.222 &  0.520 \\ 
& TSKI-MDA &1 & 0.049 & 0.057 \\
\hline
\multirow{4}{*}{$500$/$270/0.2$} 
& TSKI-LCD &0  &0.188 & 1 \\
& TSKI-LCD &1  & 0.142 & 0.999 \\ 

& TSKI-MDA &0 & 0.208&  0.729 \\ 
& TSKI-MDA &1 & 0.068 & 0.164 \\
\hline
\multirow{4}{*}{$500$/$270/0.95$} 
& TSKI-LCD & 0  &0.299 & 0.972 \\        
& TSKI-LCD & 1  &0.172 & 0.979 \\
& TSKI-MDA &0  & 0.042  & 0.019 \\ 
& TSKI-MDA &1  & 0.000  & 0.000 \\ 
\hline
\end{tabular}
\begin{tabular}{clccc}
\hline
$n/p/\eta$   & Method & $q$ & FDR & Power  \\ \hline
\multirow{4}{*}{$200$/$270/0.2$} 
& TSKI-LCD &0  &0.203 & 0.985 \\
& TSKI-LCD &1  & 0.122 & 0.755 \\ 

& TSKI-MDA &0 & 0.220 &  0.292 \\ 
& TSKI-MDA &1 & 0.044 & 0.021 \\
\hline
\multirow{4}{*}{$300$/$270/0.2$} 
& TSKI-LCD &0  &0.233 & 0.999 \\
& TSKI-LCD &1  & 0.124 & 0.986 \\ 
& TSKI-MDA &0 & 0.185 &  0.387 \\ 
& TSKI-MDA &1 & 0.017 & 0.025 \\
\hline
\multirow{4}{*}{$500$/$270/0.2$} 
& TSKI-LCD &0  &0.181 & 0.999 \\
& TSKI-LCD &1  & 0.166  & 0.996 \\ 

& TSKI-MDA &0 & 0.142&  0.476 \\ 
& TSKI-MDA &1 & 0.037 & 0.076 \\
\hline
\multirow{4}{*}{$500$/$270/0.95$} 
& TSKI-LCD & 0  &0.312 & 0.961 \\        
& TSKI-LCD & 1  &0.195 & 0.969 \\
& TSKI-MDA &0  & 0.032  & 0.015 \\ 
& TSKI-MDA &1  & 0.000  & 0.000 \\ 
\hline
\end{tabular}

\caption{The simulation results on the empirical FDR and power for the TSKI with $\tau_{1} = \tau^* / (q+1)$ and $\tau^* = 0.2$ under Model \ref{ARX1} (left panel) and Model~\ref{ARXGARCH1} (right panel) in Section \ref{Sec6.1}.}\label{fig:arx}
\end{table}

\begin{table}[tp]	
\footnotesize
\centering
\begin{tabular}{cccc}
\multicolumn{3}{c}{Model~\ref{ARX1} (ARX)}\\
\hline
$n/p/\eta $   &  FDR & Power  \\ \hline
$200$/270/0.2& -- & -- \\
$300$/270/0.2 & 0.007 & 0.009 \\
$500$/270/0.2 & 0.029 & 0.989 \\
$500$/270/0.95 &0.099 & 0.999 \\
\hline

\end{tabular}
\begin{tabular}{cccc}
\multicolumn{3}{c}{Model~\ref{ARXGARCH1} (ARXARCH)}\\
\hline
$n/p/\eta$   &  FDR & Power  \\ \hline
$200$/270/0.2& -- & -- \\
$300$/270/0.2 & 0.025 & 0.04 \\
$500$/270/0.2 & 0.094 & 0.983 \\
$500$/270/0.95 & 0.233 & 0.987 \\
\hline

\end{tabular}

\caption{ The simulation results on the empirical FDR and power for the ordinary least squares +  Benjamini--Yekutieli (BY~\citep{benjamini2001control}) with the target FDR level at $0.2$. This approach does not apply to high-dimensional scenarios with more features than observations.}\label{fig:by}
\end{table}


\begin{table}[tp]	
\footnotesize
\centering
\begin{tabular}{cccc}
\multicolumn{3}{c}{Model~\ref{ARX1} (ARX)}\\
\hline
$n/p/\eta$   &   FDR & Power  \\ \hline
$200$/270/0.2& 0.001 & 0.999 \\
$300$/270/0.2 & 0.000 & 1.000 \\
$500$/270/0.2 & 0.131 & 1.000  \\
$500$/270/0.95 & 0.001 & 0.991 \\
\hline

\end{tabular}
\begin{tabular}{cccc}
\multicolumn{3}{c}{Model~\ref{ARXGARCH1} (ARX-ARCH)}\\
\hline
$n/p/\eta$   &   FDR & Power  \\ \hline
$200$/270/0.2 & 0.009 & 0.999 \\
$300$/270/0.2 & 0.001 & 1.000  \\
$500$/270/0.2 & 0.069 & 0.999  \\
$500$/270/0.95 & 0.008 & 1.000 \\
\hline

\end{tabular}

\caption{ The simulation results on the empirical FDR and power for the adaptive Lasso~\citep{medeiros2016l1, zou2006adaptive}. There is no target FDR level for the adaptive Lasso.}\label{fig:adalasso}
\end{table}

For all simulation experiments reported in Tables \ref{fig:setarx2} and \ref{fig:arx}, both TSKI-LCD and TSKI-MDA with $q=1$ control the FDR in finite samples at the target value of $\tau^*=0.2$, but at the cost of lower selection power compared to the case of $q=0$ (i.e., no subsampling). In contrast, TSKI with $q=0$  has FDR exceeding the target FDR level in some cases. 

When analyzing Models~\ref{ARX1}--\ref{ARXGARCH1} with linear mean regression functions, the LCD-based method demonstrates superior performance over the MDA-based method in terms of power, as evidenced in Table~\ref{fig:arx}. However, for Model~\ref{SETAR1} when $\iota=5$ (i.e., high nonlinearity), we observe from Table~\ref{fig:setarx2} that MDA outperforms LCD in power when the sample size is large (i.e., $n=500$ and $q=0$), an intuitive observation considering the nonparametric nature of the MDA measure. Indeed, the empirical performance of MDA decreases drastically in all settings when $q$ increases from 0 to 1, because of the smaller sample size when calculating the MDA measures.

In Table~\ref{fig:by}, LS-BY exhibits a slightly higher error rate than $\tau^* = 0.2$ for Model~\ref{ARXGARCH1} with $\eta = 0.95$. 
In addition, it is inapplicable to high-dimensional time series applications when $n<p$, such as our real data example in Section~\ref{real.data.1}, due to the lack of a reliable p-value calculation method. 
It is important to emphasize that obtaining valid p-values under high-dimensional linear or nonlinear time series models, such as Models~\ref{SETAR1}--\ref{ARXGARCH1}, presents a highly challenging and currently unresolved issue. Such a challenge is also reflected by the deteriorating performance of LS-BY when $p$ becomes comparable to $n$.   On the other hand, Table~\ref{fig:adalasso} highlights that adaLasso demonstrates the highest overall selection powers among all four methods. However, adaLasso fails to control error rates under Model~\ref{SETAR1}, which is expected because it is developed for the linear model and focuses on consistent model selection instead of FDR control.

To sum up, the additional simulation results in this section support our conclusion in Section~\ref{Sec6}. For implementation, we suggest that practitioners working on time series inference with limited sample sizes initiate their diagnosis using TSKI-LCD with parameter $q$ set to either $0$ or $1$, along with our knockoffs sampling procedure~\eqref{ko.sampling.1}.

\begin{remark}
Regarding the subsampling parameter $q$, we remark that by the construction of \eqref{tau.2} in Algorithm~\ref{Algorithm1}, TSKI may have decent asymptotic power only when the number of relevant features is no less than $\tau_{1}^{-1} = (q+1)/\tau^*$. To see the intuition, let us consider an ideal scenario when the number of relevant features is less than $\tau_{1}^{-1}$ and these relevant features' knockoff statistics are the largest (positive) among all knockoff statistics. Then, even if the other knockoff statistics are all zero, we have $T^k = \infty$ in \eqref{tau.2}, and, hence the knockoff filter screens out all features. 
It is worth mentioning that a similar requirement is assumed in Condition~\ref{power.new.2} for the asymptotic power. 

\end{remark}

\subsubsection{Additional simulation results for Model~\ref{SETAR1} with $q=2$}\label{q2addi}

\begin{table}[h]
\footnotesize
\centering
\begin{tabular}{c | cccc |cccc}
\hline
Method & $n/p/\eta/ \iota$   &  $q$ & FDR & Power  & $n/p/\eta/ \iota$   & $q$ & FDR & Power   \\   \hline
TSKI-LCD & \multirow{6}{*}{$200$/$270$/0.2/0} 
&0  & 0.157 &0.698 & \multirow{6}{*}{$300$/$270$/0.2/0} &0  & 0.164 & 0.870 \\    
TSKI-LCD & & 1  & 0.026 & 0.051 & &1  & 0.075 & 0.413 \\ 
TSKI-LCD & & 2  & 0.000  &  0.000 & & 2  & 0.007  & 0.012 \\ 
TSKI-MDA & & 0 & 0.173 &  0.456 & &0 & 0.157 &  0.718 \\ 
TSKI-MDA & & 1 &  0.026 &  0.028 & &1 & 0.041 &  0.102 \\ 
TSKI-MDA & & 2 &  0.000  &  0.000 & & 2 & 0.005  &  0.005 \\ 
\hline     
TSKI-LCD & \multirow{6}{*}{$200$/$270$/0.2/5} 
& 0  & 0.139 & 0.287 & \multirow{6}{*}{$300$/$270$/0.2/5} 
&0  &0.160 & 0.514 \\
TSKI-LCD & &1  & 0.023 & 0.019 & &1  & 0.032 & 0.048 \\ 
TSKI-LCD & & 2  & 0.000 & 0.000 & & 2  & 0.000  &  0.000\\ 
TSKI-MDA & & 0 & 0.138 &  0.215 & &0 & 0.196 &  0.506 \\ 
TSKI-MDA & & 1 & 0.012 & 0.011 &  &1 & 0.038 & 0.036 \\
TSKI-MDA & & 2 &  0.000 &  0.000 & & 2 &  0.000 &  0.000\\
\hline
\end{tabular}

\begin{tabular}{c | cccc |cccc}
\hline
Method & $n/p/\eta/ \iota$   &  $q$ & FDR & Power  & $n/p/\eta/ \iota$   & $q$ & FDR & Power   \\   \hline
TSKI-LCD & \multirow{6}{*}{$500$/$270$/0.2/0} 
&0  &  0.176 & 0.939   & \multirow{6}{*}{$1000$/$270$/0.2/0} &0  &  0.222  &  0.975 \\    
TSKI-LCD & & 1  &   0.099 & 0.872   &  & 1  & 0.119  &  0.946 \\ 
TSKI-LCD & & 2  &  0.047 &  0.216   &  & 2  & 0.126  &  0.879 \\ 
TSKI-MDA & & 0 & 0.181 &  0.922    &  & 0 & 0.178  & 0.971  \\ 
TSKI-MDA & & 1 &  0.092 & 0.550   & & 1 &  0.107 & 0.939   \\ 
TSKI-MDA & & 2 &  0.027  &  0.053   &  & 2 & 0.067 & 0.432   \\ 
\hline     
TSKI-LCD & \multirow{6}{*}{$500$/$270$/0.2/5} 
& 0  &   0.141 & 0.634   & \multirow{6}{*}{$1000$/$270$/0.2/5} 
&0  & 0.186 &  0.755  \\
TSKI-LCD & &1  &   0.086 & 0.267    & &1  &  0.117 &  0.613 \\ 
TSKI-LCD & &2   &  0.004 &  0.006  & & 2  &  0.022 &  0.054 \\ 
TSKI-MDA & & 0 &   0.166 &  0.679    & &0 &  0.199 &  0.849 \\ 
TSKI-MDA & & 1 &   0.084 & 0.216  &  &1 & 0.115  & 0.649   \\
TSKI-MDA & & 2 &  0.000 &  0.000  &  & 2 &  0.027 &  0.068 \\
\hline
\end{tabular}


\caption{The simulation results on the empirical FDR and power for the TSKI with $\tau_{1} = \tau^* / (q+1)$ and $\tau^* = 0.2$ under Model \ref{SETAR1} in Section \ref{Sec6.1}. The results in this table are the same as those in Table~\ref{fig:setarx2}, except for the experiments with $q=2$ or $n \in \{300, 1000\}$.}\label{fig:setarx3}
\end{table}

In this section, we present additional simulation for Model~\ref{SETAR1} with $q=2$ and $n \in \{300, 1000\}$. The results with $q=2$ and $n \in \{300, 1000\}$ in Table~\ref{fig:setarx3} complements those in Table~\ref{fig:setarx2}; the other results in these two tables are the same.


Let us begin with commenting on the results of $q=2$ in Table~\ref{fig:setarx3} here. For all simulation experiments reported in Table~\ref{fig:setarx3}, both TSKI-LCD and TSKI-MDA with $q=2$ control the FDR in finite samples below the target value of $\tau^*=0.2$, but at the cost of lower selection power compared to the case of $q\in \{0, 1\}$. When $q=2$ and the sample size is small,  TSKI becomes overly conservative with both low FDR and low power in most cases, with the MDA-based method suffering more severely from these issues. On the other hand, the additional results for $n \in \{300, 1000\}$ provide a clearer understanding of how the selection power increases with larger sample sizes. These results also suggest that the subsample size needs to be reasonably large for TSKI to have good power.

In summary, given the simulation results in Table~\ref{fig:setarx3}, we recommend the use of $q=1$ in practice as we are considering finite samples with limited sample sizes in our real data applications.

\subsection{Augmented Dickey--Fuller test for unit roots}\label{unit_root_sec}

\begin{figure}[t!]
\centering
\includegraphics[width=15cm]{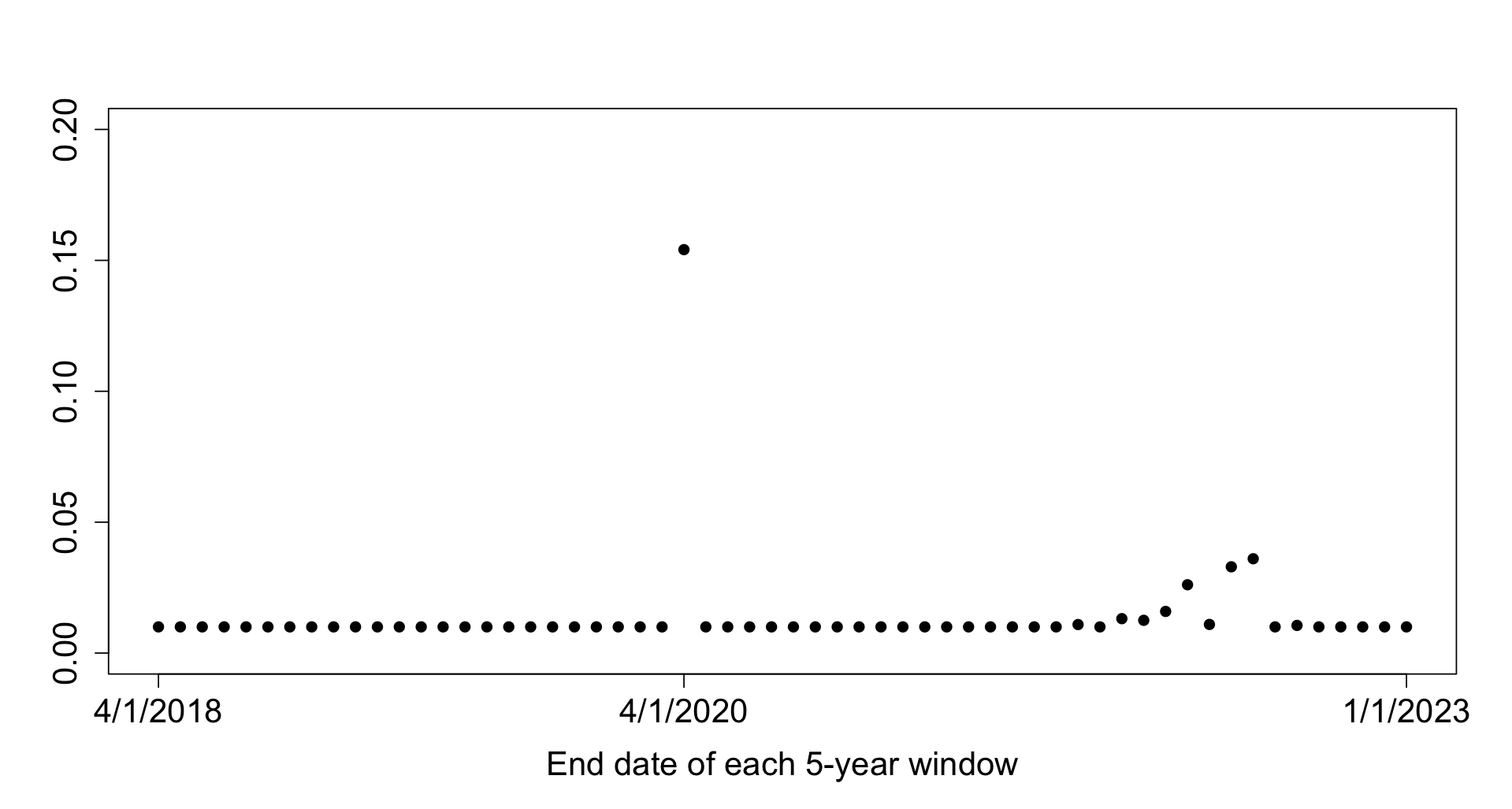} 
\includegraphics[width=15cm]{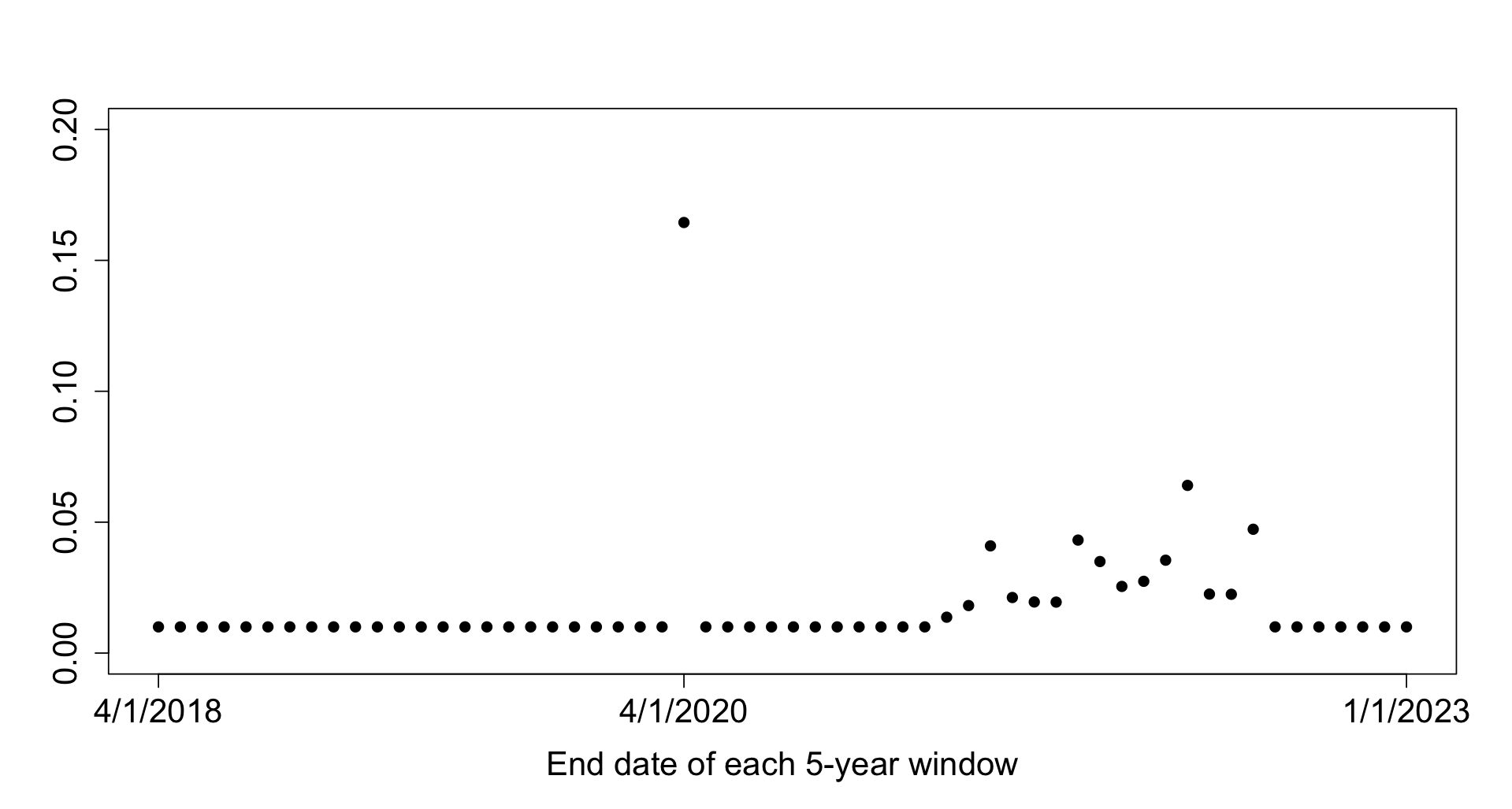} 
\captionsetup{width=1.0\linewidth}
\caption{The results of the augmented Dickey--Fuller (ADF) test, where the unit root AR models include $b$ lags. The y-axis displays the p-values of the tests. The top panel displays the results of the ADF test with one lag $(b = 1)$ for each rolling window. The bottom panel displays the results of the ADF test with no lags $(b = 0)$. }
\label{fig:adf_2}
\end{figure}

We run the augmented Dickey--Fuller (ADF) test implemented with the R package \pkg{aTSA} to test for unit roots. The null hypothesis of the ADF test is that the time series contains a unit root and is non-stationary, while the alternative hypothesis is that the time series is a stationary linear AR model. The unit root AR models considered by the ADF test may include $b$ lags, where $b \ge 0$ is a tuning parameter. The unit root model of the ADF test also encompasses a drift term and a trend term. For more details, we refer to the R package \pkg{aTSA}. It is noteworthy that the ADF test with no lags $(i.e., b = 0)$ is equivalent to the Dickey--Fuller test for the unit root.

The ADF test result for each rolling window is displayed in Figure~\ref{fig:adf_2}. From Figure~\ref{fig:adf_2}, it is observed that most periods do not exhibit clear numerical evidence of non-stationarity with unit roots (at p-value significance level $0.05$), except for 4/1/2020, when COVID-19 occurred. In addition to the tests for rolling windows, we run the ADF test for the entire inflation series from 5/1/2013 to 1/1/2023 and rejected the null hypothesis of non-stationarity at $\alpha = 0.01$ for $b\in \{0, 1\}$.

\renewcommand{\thesubsection}{B.\arabic{subsection}}

\section{Proofs of Theorems \ref{theorem1.new}--\ref{theorem4.new}, Corollaries \ref{collary.1}--\ref{collary.2}, and Proposition~\ref{glp1}} \label{SecC}

\subsection{Proof of Theorem~\ref{theorem1.new}} \label{SecA.1.new}

For simplicity, in this proof we use the notation $\boldsymbol{U}, \widetilde{\boldsymbol{U}}, \boldsymbol{V} $ to denote a generic subsample of $\{\boldsymbol{x}_{t}, \widetilde{\boldsymbol{x}}_{t}, Y_{t}\}_{t=1}^{n}$ or their independent and identically distributed (i.i.d.) counterparts $\{\boldsymbol{x}_{t}^\pi, \widetilde{\boldsymbol{x}}_{t}^\pi, Y_{t}^\pi\}_{t=1}^{n}$, where their exact meaning will be made explicit whenever confusion is possible. Let us define
\begin{equation}
\label{tski.eq.3}
\widehat{\textnormal{KL}}_{j}^{k} \equiv \log{\left(\frac{f_{\boldsymbol{U}_{j}, \widetilde{\boldsymbol{U}}_{j}, \boldsymbol{U}_{-j}, \widetilde{\boldsymbol{U}}_{-j}, \boldsymbol{V}} ( \boldsymbol{U}_{j}, \widetilde{\boldsymbol{U}}_{j}, \boldsymbol{U}_{-j}, \widetilde{\boldsymbol{U}}_{-j}, \boldsymbol{V})}{f_{\boldsymbol{U}_{j}, \widetilde{\boldsymbol{U}}_{j}, \boldsymbol{U}_{-j}, \widetilde{\boldsymbol{U}}_{-j}, \boldsymbol{V}} (\widetilde{\boldsymbol{U}}_{j}, \boldsymbol{U}_{j}, \boldsymbol{U}_{-j}, \widetilde{\boldsymbol{U}}_{-j}, \boldsymbol{V})}\right)},
\end{equation}
where $\boldsymbol{U} = (\boldsymbol{x}_{i}, i\in H_{k})^{\top}$, $\widetilde{\boldsymbol{U}} = (\widetilde{\boldsymbol{x}}_{i}, i\in H_{k})^{\top}$, and $\boldsymbol{V} = (Y_{i}, i\in H_{k})^{\top}$. 
We can use \eqref{robust.eq.19} in the proof of Theorem~\ref{theorem2.new} in Section \ref{SecA.2.new} to conclude that for each $k \in \{1, \dots, q+1\}$, 
\begin{equation*}
\mathbb{E}(\sum_{j\in\mathcal{H}_{0}}e_{j}^{k}\times \boldsymbol{1}_{\{\widehat{\textnormal{KL}}_{j}^{k}\le \varepsilon\}})\le p\times e^{\varepsilon}.    
\end{equation*}
Then it holds that 
$$\mathbb{E}(\sum_{j\in\mathcal{H}_{0}}e_{j}^{(\varepsilon)}) \le p \times e^{\varepsilon},$$
where 
$$e_{j}^{(\varepsilon)} = (q+1)^{-1}\sum_{k=1}^{q+1} e_{j}^{k} \times \boldsymbol{1}_{\{\widehat{\textnormal{KL}}_{j}^{k}\le \varepsilon\}}.$$

Denote by $\widehat{S}_{\varepsilon}$ be the set of selected features when applying the e-BH method to $e_{j}^{(\varepsilon)}$'s at the target FDR level $\tau^*$. Then using similar arguments to those for \eqref{robust.eq.20} in Section \ref{SecA.2.new}, we can show that 
$$\mathbb{E}\left(\frac{\#(\widehat{S}_{\varepsilon}\cap \mathcal{H}_{0})}{ (\# \widehat{S}_{\varepsilon}) \vee 1}\right)\le \tau^* \times e^{\varepsilon}.$$
Thus, it follows that 
\begin{equation}
\begin{split}\label{theorem1.eq.4}
& \mathbb{E}\left(\frac{\#(\widehat{S}\cap 
\mathcal{H}_{0})}{ (\# \widehat{S}) \vee 1}\right) \le \mathbb{E}\left(\frac{\#(\widehat{S}_{\varepsilon}\cap \mathcal{H}_{0})}{ (\# \widehat{S}_{\varepsilon}) \vee 1}\times \boldsymbol{1}_{\{\widehat{S} = \widehat{S}_{\varepsilon}\} } + \boldsymbol{1}_{\{\widehat{S} \not= \widehat{S}_{\varepsilon}\} }\right) \\
& \le \tau^* \times e^{\varepsilon}  + \sum_{k=1}^{q+1}\mathbb{P}(\max_{1\le j\le p}\widehat{\textnormal{KL}}_{j}^{k}> \varepsilon)
\end{split}
\end{equation}
in view of $\{\widehat{S} \not= \widehat{S}_{\varepsilon}\}\subset \{\max_{1\le k\le q+1}\max_{1\le j\le p}\widehat{\textnormal{KL}}_{j}^{k}> \varepsilon \}$. Since \eqref{theorem1.eq.4} holds for each $\varepsilon>0$, we can further obtain that 
\begin{equation}
\begin{split}\label{theorem1.eq.4b}
\mathbb{E}\left(\frac{\#(\widehat{S}\cap 
\mathcal{H}_{0})}{ (\# \widehat{S}) \vee 1}\right) \le &
\inf_{\varepsilon>0}\left\{ \tau^*  e^{\varepsilon}  + \sum_{k=1}^{q+1}\mathbb{P}(\max_{1\le j\le p}\widehat{\textnormal{KL}}_{j}^{k\pi}> \varepsilon)\right\} \\
&+ \sum_{k=1}^{q+1}\Big(\mathbb{P}(\max_{1\le j\le p}\widehat{\textnormal{KL}}_{j}^{k}> \varepsilon) - \mathbb{P}(\max_{1\le j\le p}\widehat{\textnormal{KL}}_{j}^{k\pi}> \varepsilon)\Big).
\end{split}
\end{equation}

It remains to bound the second term on the right-hand side of (\ref{theorem1.eq.4b}) above. Recall that $\widehat{\textnormal{KL}}_{j}^{k\pi}$ is defined analogously to \eqref{tski.eq.3} but based on i.i.d. sample $\{\boldsymbol{x}_{t}^\pi, \widetilde{\boldsymbol{x}}_{t}^\pi, Y_{t}^\pi\}_{t=1}^{n}$.  With $\boldsymbol{U} = (\boldsymbol{x}_{t}^{\pi}, t\in H_{k})^{\top}$, $\widetilde{\boldsymbol{U}} = (\widetilde{\boldsymbol{x}}_{t}^{\pi}, t\in H_{k})^{\top}$, and $\boldsymbol{V} = (Y_{t}^{\pi}, i\in H_{k})^{\top}$, we can deduce that
\begin{equation}
\begin{split}\label{tski.eq.1}
\widehat{\textnormal{KL}}_{j}^{k\pi} & = \sum_{t\in H_{k}}\log{\left(\frac{f_{X_{j}, \widetilde{X}_{j}, \boldsymbol{x}_{-j}, \widetilde{\boldsymbol{x}}_{-j}} ( X_{tj}^{\pi}, \widetilde{X}_{tj}^{\pi}, \boldsymbol{x}_{-tj}^{\pi}, \widetilde{\boldsymbol{x}}_{-tj}^{\pi})}{f_{X_{j}, \widetilde{X}_{j}, \boldsymbol{x}_{-j}, \widetilde{\boldsymbol{x}}_{-j}} (\widetilde{X}_{tj}^{\pi}, X_{tj}^{\pi}, \boldsymbol{x}_{-tj}^{\pi}, \widetilde{\boldsymbol{x}}_{-tj}^{\pi} )}\right)}\\
& = \log{\left(\frac{f_{\boldsymbol{U}_{j}, \widetilde{\boldsymbol{U}}_{j}, \boldsymbol{U}_{-j}, \widetilde{\boldsymbol{U}}_{-j}} ( \boldsymbol{U}_{j}, \widetilde{\boldsymbol{U}}_{j}, \boldsymbol{U}_{-j}, \widetilde{\boldsymbol{U}}_{-j})}{f_{\boldsymbol{U}_{j}, \widetilde{\boldsymbol{U}}_{j}, \boldsymbol{U}_{-j}, \widetilde{\boldsymbol{U}}_{-j}} (\widetilde{\boldsymbol{U}}_{j}, \boldsymbol{U}_{j}, \boldsymbol{U}_{-j}, \widetilde{\boldsymbol{U}}_{-j})}\right)} \\
& = \log{\left(\frac{f_{\boldsymbol{U}_{j}, \widetilde{\boldsymbol{U}}_{j}, \boldsymbol{U}_{-j}, \widetilde{\boldsymbol{U}}_{-j}, \boldsymbol{V}} ( \boldsymbol{U}_{j}, \widetilde{\boldsymbol{U}}_{j}, \boldsymbol{U}_{-j}, \widetilde{\boldsymbol{U}}_{-j}, \boldsymbol{V})}{f_{\boldsymbol{U}_{j}, \widetilde{\boldsymbol{U}}_{j}, \boldsymbol{U}_{-j}, \widetilde{\boldsymbol{U}}_{-j}, \boldsymbol{V}} (\widetilde{\boldsymbol{U}}_{j}, \boldsymbol{U}_{j}, \boldsymbol{U}_{-j}, \widetilde{\boldsymbol{U}}_{-j}, \boldsymbol{V})}\right)},
\end{split}
\end{equation}
where the third equality above follows from similar analysis to that of \eqref{theorem2.new.2} in Theorem~\ref{theorem1.new}. The conditional column independence required by \eqref{theorem2.new.2} holds for each $j\in\mathcal{H}_{0}$  because \eqref{tski.eq.1} involves i.i.d. samples.

We can now see that $\widehat{\textnormal{KL}}_{j}^{k}$ is well-defined thanks to Condition~\ref{knockoff.generator.3}. Moreover, by the fact that the supports of $(\boldsymbol{x}, \widetilde{\boldsymbol{x}})$ and $[\boldsymbol{x}, \widetilde{\boldsymbol{x}}]_{\textnormal{swap}(\{j\})}$ are the same (as guaranteed by Condition~\ref{knockoff.generator.3}) and the definition of $(\boldsymbol{x}_{i}^{\pi}, \widetilde{\boldsymbol{x}}_{i}^{\pi})$'s, $\widehat{\textnormal{KL}}_{j}^{k\pi}$ is well-defined. Hence, in light of \eqref{tski.eq.3} and \eqref{tski.eq.1}, there exists some measurable function $g:\mathbb{R}^{\#H_{k}\times (2p+1)}\longmapsto\mathbb{R}$ such that $\widehat{\textnormal{KL}}_{j}^{k} = g(\mathcal{X}_{k})$ and $\widehat{\textnormal{KL}}_{j}^{k\pi} = g(\mathcal{X}_{k}^{\pi})$ for each $\varepsilon\ge 0$, which entails that there exists some  $\mathcal{D}\in\mathcal{R}^{\# H_{k}\times (2p+1)}$ such that 
\begin{equation}
\begin{split}\label{theorem1.eq.2}
\{\mathcal{X}_{k}\in \mathcal{D}\} &= \{\max_{1\le j\le p}\widehat{\textnormal{KL}}_{j}^{k}> \varepsilon\}, \\
\{\mathcal{X}_{k}^{\pi}\in \mathcal{D}\} &= \{\max_{1\le j\le p}\widehat{\textnormal{KL}}_{j}^{k\pi}> \varepsilon\}.
\end{split}
\end{equation}
With the aid of \eqref{theorem1.eq.2}, it holds that 
\begin{equation}
\begin{split}\label{theorem1.eq.5}
& |\mathbb{P}(\max_{1\le j\le p}\widehat{\textnormal{KL}}_{j}^{k}> \varepsilon) - \mathbb{P}(\max_{1\le j\le p}\widehat{\textnormal{KL}}_{j}^{k\pi}> \varepsilon)| \\
& \le \sup_{\mathcal{D}\in\mathcal{R}^{\#H_{k} \times (2p+1)}}|\mathbb{P}(\mathcal{X}_{k}\in\mathcal{D})-\mathbb{P} (\mathcal{X}_{k}^{\pi} \in\mathcal{D})|.
\end{split}
\end{equation}
Therefore, from  \eqref{theorem1.eq.4b}--\eqref{tski.eq.1} and~\eqref{theorem1.eq.5} we 
can obtain the desired conclusion, which completes the proof of Theorem~\ref{theorem1.new}.

\subsection{Proof of Theorem~\ref{theorem4.new}} \label{SecA.3.new}

\noindent \textit{Proof of \eqref{theorem4.new.2}}. We aim to prove the second assertion \eqref{theorem4.new.2} of Theorem~\ref{theorem4.new}, and will defer the proof of \eqref{theorem4.eq.2} to the end. Let the knockoff thresholds $T^k$'s, knockoff statistics $W_{j}^k$'s, statistics $e_{j}^{k}$'s, and e-values $e_{j}$'s be given as in Algorithm~\ref{Algorithm1}. Let us first outline the proof idea for \eqref{theorem4.new.2} as follows. Using the inclusion-exclusion principle, we will show that $\cap_{k=1}^{q+1}\{j:e_{j}^{k}>0\}$ includes most features in $S^*$ with high probability. We then prove that each $e_{j}$ with $j$ in $\cap_{k=1}^{q+1}\{j:e_{j}^{k}>0\}$ is sufficiently large to be selected by the e-BH procedure; that is, $\cap_{k=1}^{q+1}\{j:e_{j}^{k}>0\}\subset \widehat{S}$. Combining all these results will complete the proof of \eqref{theorem4.new.2}. We will provide the full details of the proof next.

First, recall that $\widehat{S} = \{j: e_{j} \ge p(\tau^* \times \widehat{k})^{-1}\}$ with $\widehat{k} = \max\{k:e_{(k)} \ge p(\tau^* \times k)^{{-1}}\}$, where $e_{(j)}$'s are the ordered statistics of $e_{j}$'s such that $e_{(1)}\ge \dots \ge e_{(p)}$. Let $K>1$ be the constant specified in Theorem~\ref{theorem4.new}. Let us consider two events given by 
\begin{equation}
\label{power.new.23}
\cap_{k=1}^{{q+1}} \{\# \{j: W_{j}^k\ge T^k\} \le K (\# S^*)\}
\end{equation}
and
\begin{equation}
\label{power.new.24}
\cap_{k=1}^{q+1} \left\{\frac{\#(S^*\cap \{j: W_{j}^k\ge T^k\}) }{\# S^*} \ge 1- (1 + \phi)c_{0}k_{1n}^{-1}\right\},
\end{equation}
where $c_{0}$ is given in Condition~\ref{theorem4.new.4c} and that $\phi>0$ is some real constant such that  $\phi^2 - \phi - 1 = 0$. We will show that conditional on the two events in \eqref{power.new.23} and \eqref{power.new.24} above, it holds that
\begin{equation}\label{power.new.24a}
e_{(\#\mathcal S)} = \min_{j\in \mathcal S}e_j \geq p(\tau^*(\#\mathcal S))^{-1}, 
\end{equation}
where
\begin{equation}
\label{def.S}
\mathcal{S}\coloneqq \cap_{k=1}^{q+1} \{j: W_{j}^k\ge T^k\}.
\end{equation} 

Then it follows from \eqref{power.new.24a} and the definition of $\widehat{k}$ that $\hat k \geq \#\mathcal{S}$ and $\mathcal S\subset \widehat S$. Such results along with an application of the inclusion–exclusion principle entail that conditional on the intersection of events \eqref{power.new.23} and \eqref{power.new.24}, we have 
\begin{equation}
\label{power.new.15}
\frac{\#(S^*\cap \widehat S)}{\#S^*} \geq  \frac{\#(S^*\cap \mathcal S)}{\#S^*} \ge 1- (q+1)(1 + \phi)c_{0}k_{1n}^{-1}.
\end{equation}
Since it holds that 
\begin{align}
\begin{split}
\mathbb E & \Big[\frac{\#(S^*\cap \widehat S)}{\#S^*}\Big] \geq \Big(1- (q+1)(1 + \phi)c_{0}k_{1n}^{-1}\Big) \\
& \quad \times \mathbb P\Big(   \frac{\#(S^*\cap \widehat S)}{\#S^*}  \ge 1- (q+1)(1 + \phi)c_{0}k_{1n}^{-1}\Big)\\
&\ge \Big(1- (q+1)(1 + \phi)c_{0}k_{1n}^{-1}\Big)\mathbb P\Big(   \text{event in \eqref{power.new.23}}\cap\text{event in \eqref{power.new.24}}\Big),
\end{split}
\end{align}
to establish \eqref{theorem4.new.2} we need only to prove \eqref{power.new.24a} and construct the probability lower bounds for events in \eqref{power.new.23} and \eqref{power.new.24}.

To show \eqref{power.new.24a}, note that by the definition of $T^k$'s and the assumption that $T^k<\infty$, we have that conditional on event given in \eqref{power.new.23},
\begin{equation}
\begin{split}\label{power.new.19}
1+ \#\{j: W_{j}^k\le - T^k\} & \le \tau_{1}  (\# \{j: W_{j}^k\ge T^k\})\\
&\le \tau_{1} K (\# S^*)
\end{split}
\end{equation}
for each $k\in\{1, \dots, q+1\}$. In view of \eqref{power.new.19}, it holds that for each nonzero $e_{j}^k$, 
\begin{equation}
\begin{split}
\label{power.new.13}
e_{j}^k & = \frac{p}{\#\{s:W_{s}^{k}\le -T^{k}\}+1}\\
& \ge \frac{p}{\tau_{1} K (\# S^*)}
\\
&\ge  \frac{p}{\tau^* \times (1- (1+q)(1 + \phi)c_{0}k_{1n}^{-1}) \times (\# S^*)},
\end{split}
\end{equation}
where we have used the definitions of $e_{j}^k$'s, $\phi$, and $\tau_{1}$. Then from \eqref{power.new.13} and the definition $e_{j}= (q+1)^{-1}\sum_{k=1}^{q+1}e_{j}^{k}$, we can deduce that conditional on event  \eqref{power.new.23}, 
$$\min_{j\in \mathcal{S}} e_{j} \ge \frac{p}{\tau^* \times (1- (1+q)(1 + \phi)c_{0}k_{1n}^{-1}) \times (\# S^*)},$$
which establishes \eqref{power.new.24a}.

It remains to provide the probability upper bounds for the complementary events of \eqref{power.new.23} and \eqref{power.new.24}, which are given in \eqref{power.new.12} and \eqref{power.new.14}, respectively, below. By the assumptions (Conditions~\ref{power.new.1}--\ref{power.new.2} and that Condition \ref{theorem4.new.4c} is satisfied for the Lasso estimates applied to each subsample in $H_k$ in Algorithm~\ref{Algorithm1}), we can show that 
\begin{equation}
\begin{split}\label{power.new.14}
\mathbb{P}(\textnormal{complementary event of } \eqref{power.new.24}) \le  (q+1)\times (k_{2n}+k_{3n})
\end{split}
\end{equation}
for all large $n$. We postpone the detailed proof of \eqref{power.new.14} to a later part.

On the other hand, it follows from the assumption of $\# S^*>0$ that conditional on event $\{\# \{j: W_{j}^k\ge T^k\} > K (\# S^*)\}$, 
\begin{equation}
\begin{split}\label{power.new.11}
\frac{\# (\{j: W_{j}^k\ge T^k\} \cap ( S^*)^{c}) }{ \# \{j: W_{j}^k\ge T^k\} \vee 1 } & \ge \frac{\# \{j: W_{j}^k\ge T^k\} - \# S^* }{ \# \{j: W_{j}^k\ge T^k\} \vee 1} \\
& > \frac{\# f\{j: W_{j}^k\ge T^k\} -  \# \{j: W_{j}^k\ge T^k\} \times K^{-1} }{ \# \{j: W_{j}^k\ge T^k\} \vee 1} \\
& \ge \frac{K-1}{K}.
\end{split}
\end{equation}
Therefore, from \eqref{power.new.10} and  \eqref{power.new.11} some simple calculations give that 
\begin{equation*}
\begin{split}
& \mathbb{P}(\{\# \{j: W_{j}^k\ge T^k\} > K (\# S^*)\}) \times \frac{K-1}{K} + \mathbb{P}(\{\# \{j: W_{j}^k\ge T^k\} \le K (\# S^*)\})\times 0\\
& \le \mathbb{E}\left( \frac{\# (\{j: W_{j}^k\ge T^k\} \cap ( S^*)^{c}) }{ \# \{j: W_{j}^k\ge T^k\} \vee 1} \right)\\
& \le \tau_{1} +\theta_{\varepsilon},
\end{split}
\end{equation*}
which yields that 
\begin{equation}
\begin{split}\label{power.new.12}
\mathbb{P}(\textnormal{complementary event of } \eqref{power.new.23}) \le   (q+1)\times \frac{(\tau_{1}+\theta_{\varepsilon}) K}{K-1}.
\end{split}
\end{equation}
This establishes the desired conclusion in \eqref{theorem4.new.2} of Theorem~\ref{theorem4.new}.

\bigskip
\textit{Proof of \eqref{power.new.14}}. We need to prove for each $k\in \{1, \dots, p+1\}$ that conditional on event $\{\sum_{j=1}^{2p}|\widehat{\beta}_{j} - \beta_{j}^{*}|\le c_{0}(\# S^*)\lambda_{n}\} \cap \{\# \{j:W_{j}^k\ge T^k\} \ge c_{1}(\#S^*)\}$, it holds that 
\begin{equation}
\begin{split}\label{power.new.9a}
\frac{\#(S^*\cap \{j:W_{j}^k\ge T^k\}) }{\# S^*} \ge  1-(1+\phi)c_0 k_{1n}^{-1},
\end{split}
\end{equation}
where we recall that $c_{0}$ is from Condition~\ref{theorem4.new.4c} and that $\phi>0$ is a real constant such that $\phi^2 - \phi - 1 = 0$. Then \eqref{power.new.14} follows from Conditions~\ref{theorem4.new.4c}--\ref{power.new.1}. 

We now proceed with establishing \eqref{power.new.14}. Without loss of generality, we consider the case $k=1$, and omit the superscripts ``$k$'' on $e_{j}^k$'s, $W_{j}^k$'s, and $T^k$'s for simplicity.

Assume without loss of generality that 
$$|W_{1}|\ge \dots\ge |W_{p}|.$$ 
Let $j^*\in \{1, \dots, p\}$ be given such that $j^*\in \{s:|W_{s}| = T\}$. Such $j^*$ always exists because of the assumption that $T<\infty$. Then it follows that 
$$-T< W_{j^*+1} \le 0$$ 
by the definition of $T$ (because otherwise  $T$ would take a smaller value than $|W_{j*}|$) and the assumption that there are no ties in $\{|W_{j}|: |W_{j}|>0\}$. We will analyze two cases separately, where the first case considers $W_{j^*+1} = 0$ and the second case considers $-T<W_{j^*+1}<0$.

Let us consider the first case of $W_{j^*+1} = 0$. Denote by  $\widetilde{q} = \phi c_{0}k_{1n}^{-1}$ with $\phi>0$ and $\phi^2 - \phi - 1 = 0$. We will discuss the scenarios of $\#\{j: W_{j} < 0\} \le \widetilde{q} (\# S^*)$ and $\#\{j: W_{j} < 0\} > \widetilde{q} (\# S^*)$ separately, where the former case will be examined here and the latter one will be left to a later part. For the scenario of $\#\{j: W_{j} < 0\} \le \widetilde{q} (\# S^*)$, some simple calculations together with $W_{j*+1}=0$ give that 
\begin{equation}
\begin{split}\label{power.new.4}
\#(\{j:W_{j}\ge T\} \cap S^*) & = \# (\{j: |W_{j}| > 0 \} \cap S^*) - \# (\{j: W_{j} < 0 \} \cap S^*) \\
& \ge \# (\{j: |W_{j}| > 0 \} \cap S^*) - \widetilde{q}(\# S^*).
\end{split}
\end{equation}
We will deal with the term $\# (\{j: |W_{j}| > 0 \} \cap S^*)$ on the RHS of \eqref{power.new.4} below. On the event $\{\sum_{j=1}^{2p}|\widehat{\beta}_{j} - \beta_{j}^{*}|\le c_{0}(\# S^*)\lambda_{n}\}$, we can deduce that 
\begin{equation*}
\begin{split}
c_{0}\lambda_{n}(\# S^*) & \ge \sum_{j\in \widehat{S}_{1}\cap S^*} |\beta_{j}^*| \\ 
& \ge \# (\widehat{S}_{1}\cap S^*)\times (\min_{j\in S^*}|\beta_{j}^{*}|),
\end{split}
\end{equation*}
where $\widehat{S}_{1} =\{j: |\widehat{\beta}_{j}| = 0\}$. Such result and Condition~\ref{power.new.1} entail that 
$$c_{0} (\# S^*) k_{1n}^{-1} \ge \# (\widehat{S}_{1}\cap S^*). $$ 
Hence, it follows from the assumption that there are no ties in $\{|\widehat{\beta}_{j}|: |\widehat{\beta}_{j}|>0\}$ that
\begin{equation} 
\begin{split}\label{power.new.3}
\# (\{j: |W_{j}| > 0 \} \cap S^*) & = \# ((\widehat{S}_{1})^c\cap S^*)\\
& \ge (1 - c_{0}  k_{1n}^{-1}) \times (\# S^*) .
\end{split}
\end{equation}

Then combining \eqref{power.new.4}--\eqref{power.new.3}, we can obtain that conditional on event $\{\sum_{j=1}^{2p}|\widehat{\beta}_{j} - \beta_{j}^{*}|\le c_{0}(\# S^*)\lambda_{n}\}$,
\begin{equation}
\begin{split}\label{power.new.17}
\frac{\#(\{j:W_{j}\ge T\} \cap S^*) }{\# S^*} & \ge 1 - c_{0}  k_{1n}^{-1} - \widetilde{q} \\
& = 1-(1+\phi)c_{0}  k_{1n}^{-1},
\end{split}
\end{equation}
which establishes \eqref{power.new.9a}. Moreover, observe that the second scenario of $\#\{j: W_{j} < 0\} > \widetilde{q} (\# S^*)$ when $W_{j^*+1} = 0$ implies that 
$$\#\{j: W_{j} \le -T\} = \#\{j: W_{j} < 0\} > \widetilde{q} (\# S^*),$$ which reduces to the same form as in \eqref{power.new.5} below. Thus, the proof provided below can be applied here to conclude the proof for the first case $W_{j^*+1}  =0$.

We now consider the case of $-T < W_{j^*+1} < 0$. From the definition of $T$ and the assumption that there are no ties in $\{|W_{j}|: |W_{j}|>0\}$, it holds on  event $\{\# \{j:W_{j}\ge T\} \ge c_{1}(\#S^*)\}$ that 
\begin{equation}
\begin{split}
\label{power.new.5}
\# \{j : W_{j} \le -T\} + 2 & \ge  \tau_{1}\times \# \{j : W_{j} \ge T\} \\
& \ge \tau_{1} c_{1}(\# S^*).
\end{split}
\end{equation}
Meanwhile, on the event $\{\sum_{j=1}^{2p}|\widehat{\beta}_{j} - \beta_{j}^{*}|\le c_{0}(\# S^*)\lambda_{n}\}$, since $\beta_{j+p}^*=0$ for all $j>0$ and $|\widehat{\beta}_{j+p}| \ge T + |\widehat{\beta}_{j}|$ for all  $j\in \{s : W_{s} \le -T\}$, we have that 
\begin{equation}
\begin{split}\label{power.new.6}
c_{0}\lambda_{n}(\# S^*) & \ge \sum_{j : W_{j} \le -T} |\widehat{\beta}_{j+p} - \beta_{j+p}^{*}| \\ 
& = \sum_{j : W_{j} \le -T} |\widehat{\beta}_{j+p}| \\
& \ge \# \{j : W_{j} \le -T\} \times T.
\end{split}
\end{equation}
Then from \eqref{power.new.5}--\eqref{power.new.6} and Conditions~\ref{power.new.1}--\ref{power.new.2}, we can obtain that conditional on  event $\{\# \{j:W_{j}\ge T\} \ge c_{1}(\#S^*)\}\cap \{\# \{j:W_{j}\ge T\} \ge c_{1}(\#S^*)\}$, 
\begin{equation}
\begin{split}\label{power.new.7}
T \le \frac{c_{0}\lambda_{n}(\# S^*)}{c_{1}\tau_{1} (\# S^*)-2}\le k_{1n}\lambda_{n} \phi^{-1}
\end{split}
\end{equation}
for all large $n$. 

Further, conditional on event $\{\sum_{j=1}^{2p}|\widehat{\beta}_{j} - \beta_{j}^{*}|\le c_{0}(\# S^*)\lambda_{n}\} \cap \{\# \{j:W_{j}\ge T\} \ge c_{1}(\#S^*)\}$, we can deduce that for all large $n$,
\begin{equation}
\begin{split}\label{power.new.8}
c_{0}\lambda_{n}(\# S^*) & \ge \sum_{j\in S^* \cap (\{j:W_{j}\ge T\})^c} (|\widehat{\beta}_{j} - \beta_{j}^*| + |\widehat{\beta}_{j+p}|)\\
& \ge \sum_{j\in S^* \cap (\{j:W_{j}\ge T\})^c} (|\widehat{\beta}_{j} - \beta_{j}^*| + |\widehat{\beta}_{j}| - T)\\
& \ge \sum_{j\in S^* \cap (\{j:W_{j}\ge T\})^c} (| \beta_{j}^*|  - T)\\
&\ge \#(S^*\cap (\{j:W_{j}\ge T\})^{c}) \times k_{1n}(1 - \phi^{-1})\lambda_{n},
\end{split}
\end{equation}
where the second inequality above is from the fact that $|\widehat{\beta}_{j+p}| > |\widehat{\beta}_{j}| - T$ for $j$ in $\{j:W_{j}\ge T\}^c$, the third inequality above is due to the triangle inequality, and the last inequality above results from Condition~\ref{power.new.1} and \eqref{power.new.7}.

In light of \eqref{power.new.8}, it holds on event  $\{\sum_{j=1}^{2p}|\widehat{\beta}_{j} - \beta_{j}^{*}|\le c_{0}(\# S^*)\lambda_{n}\} \cap \{\# \{j:W_{j}\ge T\} \ge c_{1}(\#S^*)\}$ that for all large $n$,
\begin{equation}
\begin{split}\label{power.new.9}
\frac{\#(S^*\cap \{j:W_{j}\ge T\}) }{\# S^*} & = 1 - \frac{\#(S^*\cap (\{j:W_{j}\ge T\})^{c}) }{\# S^*} \\
& \ge 1- \frac{c_{0}}{k_{1n}(1 - \phi^{-1})} \\
& = 1- (1 + \phi)c_{0}k_{1n}^{-1},
\end{split}
\end{equation}
where the second equality above follows from the definition of $\phi$. This establishes \eqref{power.new.9a}. Thus, combining the above results concludes the proof for \eqref{power.new.14}.

\bigskip
\noindent \textit{Proof of \eqref{theorem4.eq.2}}. Finally, we show the second assertion \eqref{theorem4.eq.2} of Theorem~\ref{theorem4.new}. Let us observe that by the construction of Algorithm~\ref{Algorithm1} and the definition of $\mathcal{S}$ in \eqref{def.S}, it holds that 
$$\mathbb{P}(\{\widehat{S} = \emptyset \} \cup \{ \mathcal{S}\subset \widehat{S}\}) = 1.$$
Then by \eqref{power.new.24}--\eqref{power.new.15}, \eqref{power.new.14}, and the fact that $\phi\le 3$ (recall that $\phi$ is defined at the beginning of this proof), we can obtain the desired result in \eqref{theorem4.eq.2}. It is worth mentioning that we do not require $\tau_{1} \le \tau^*$ here. This completes the proof of Theorem~\ref{theorem4.new}.

\subsection{Proof of Theorem~\ref{theorem2.new}} \label{SecA.2.new}

Let us first make a useful claim that with $\widehat{\textnormal{KL}}_{j}$'s given in \eqref{theorem2.new.3}, it holds that for each $\varepsilon>0$, 
\begin{equation}\label{robust.eq.19}
\sum_{j\in \mathcal{H}_{0}}\mathbb{E}(e_{j} \times \boldsymbol{1}_{\{\widehat{\textnormal{KL}}_{j} \le \varepsilon\}}) \le p\times e^{\varepsilon}.
\end{equation}
Then we consider an application of the e-BH method~\citep{wang2022false} to  $e_{j}^{(\varepsilon)}\coloneqq e_{j} \times \boldsymbol{1}_{\{\widehat{\textnormal{KL}}_{j} \le \varepsilon\}}$ with the target FDR level $\tau^*$, yielding a set of selected features $\widehat{S}_{\varepsilon}\subset\{1, \dots, p\}$ defined as
$$\widehat{S}_{\varepsilon} = \{j: e_{j}^{(\varepsilon)} \ge p(\tau^* \times \widehat{k}_{\varepsilon})^{-1}\}$$ 
with $\widehat{k}_{\varepsilon} \equiv \max\{k:e_{(k)}^{(\varepsilon)} \ge p(\tau^* \times k)^{{-1}}\}$. Here, $e_{(j)}^{(\varepsilon)}$'s are the ordered statistics of $e_{j}^{(\varepsilon)}$'s such that $e_{(1)}^{(\varepsilon)}\ge \dots \ge e_{(p)}^{(\varepsilon)}$. It is easy to see that  $\#\widehat{S}_{\varepsilon} = \widehat{k}_{\varepsilon}$. 

In view of the definition of $\widehat{k}_{\varepsilon}$, we can deduce that 
\begin{equation}
\begin{split}\label{robust.eq.20}
\mathbb{E}\left(\frac{\#(\widehat{S}_{\varepsilon}\cap \mathcal{H}_{0})}{(\#\widehat{S}_{\varepsilon})\vee 1} \right) & = \mathbb{E}\left(\frac{\sum_{j\in\mathcal{H}_{0}}\boldsymbol{1}_{\{j\in \widehat{S}_{\varepsilon}\}} }{\widehat{k}_{\varepsilon}\vee 1} \right)\\
&\le  \mathbb{E}\left(\frac{\sum_{j\in\mathcal{H}_{0}}\boldsymbol{1}_{\{j\in \widehat{S}_{\varepsilon}\}} \times \tau^*\times e_{j}^{(\varepsilon)} }{p} \right)\\
& \le \tau^*p^{-1}\times \mathbb{E}\left(\sum_{j\in\mathcal{H}_{0}} e_{j}^{(\varepsilon)}  \right)\\
&\le  \tau^{*} e^{\varepsilon},
\end{split}
\end{equation}
where the last inequality above is from \eqref{robust.eq.19}. Then  combining \eqref{robust.eq.20} and  $\{\widehat{S} \not=\widehat{S}_{\varepsilon}\} \subset \cup_{j=1}^{p}\{\widehat{\textnormal{KL}}_{j} > \varepsilon\}$ leads to
\begin{equation*}
\begin{split}
\mathbb{E}\left(\frac{\#(\widehat{S}\cap \mathcal{H}_{0})}{(\#\widehat{S})\vee 1} \right) & \le \mathbb{E}\left(\boldsymbol{1}_{\{\widehat{S} =\widehat{S}_{\varepsilon}\}} \times \frac{\#(\widehat{S}_{\varepsilon}\cap \mathcal{H}_{0})}{(\#\widehat{S}_{\varepsilon})\vee 1} + \boldsymbol{1}_{\{\widehat{S} \not=\widehat{S}_{\varepsilon} \} } \right) \\
& \le \tau^* e^{\varepsilon}+ \mathbb{P}(\max_{1\le j\le p} \widehat{\textnormal{KL}}_{j} > \varepsilon)
\end{split}
\end{equation*}
for each $\varepsilon\ge 0$. This concludes the proof for the desired result \eqref{robust.eq.14}. We will  provide the proofs of \eqref{robust.eq.19} and \eqref{theorem2.new.2}--\eqref{robust.eq.7}, separately. 

\bigskip
\noindent \textit{Proof of \eqref{robust.eq.19}}. Let us define 
$$T_{j}  \equiv \min\left\{t \in \mathcal{W}_{+}^{\dagger}:  \frac{1 + \#\{ s : W_{s}^{\dagger}\le -t \}}{\#\{ s : W_{s}^{\dagger} \ge t\}\vee 1}\le \tau_{1} \right\},$$
where $W_{k}^{\dagger} = W_{k}$ if $k\not=j$ and $W_{j}^{\dagger} = |W_{j}|$,  $\mathcal{W}_{+}^{\dagger} = \{ |W_{s}^{\dagger}|: |W_{s}^{\dagger}| >0\}$, and $\min \emptyset$ is defined as infinity. We further define $\boldsymbol{X}_{j}^{(0)}$ and $\boldsymbol{X}_{j}^{(1)}$ such that $\boldsymbol{X}_{j}^{(0)} = \boldsymbol{X}_{j}$
and $\boldsymbol{X}_{j}^{(1)} = \widetilde{\boldsymbol{X}}_{j}$ if $W_{j}\ge  0$, and $\boldsymbol{X}_{j}^{(1)} = \boldsymbol{X}_{j}$
and $\boldsymbol{X}_{j}^{(0)} = \widetilde{\boldsymbol{X}}_{j}$ if $W_{j}<0$.

For each $\varepsilon\ge 0$, we can deduce that 
\begin{equation}
\begin{split}\label{robust.eq.15}
& \sum_{j\in \mathcal{H}_{0}} \mathbb{E} \left(\frac{\boldsymbol{1}_{\{W_{j} \ge T\} } \times \boldsymbol{1}_{\{\widehat{\textnormal{KL}}_{j} \le \varepsilon\}}}{ 1+\sum_{s=1}^{p} \boldsymbol{1}_{\{W_{s} \le - T\} } } \right)  \\
& = \sum_{j\in \mathcal{H}_{0}} \mathbb{E} \left(\frac{\boldsymbol{1}_{\{W_{j} \ge T_{j}\} } \times \boldsymbol{1}_{\{\widehat{\textnormal{KL}}_{j} \le \varepsilon\}}}{ 1+\sum_{s=1}^{p} \boldsymbol{1}_{\{W_{s} \le - T_{j}\} } } \right)\\
& = \sum_{j\in \mathcal{H}_{0}} \mathbb{E} \left(\frac{\boldsymbol{1}_{\{W_{j} \ge T_{j}\} } \times \boldsymbol{1}_{\{\widehat{\textnormal{KL}}_{j} \le \varepsilon\}}}{ 1+\sum_{s=1, s\not= j}^{p} \boldsymbol{1}_{\{W_{s} \le - T_{j}\} } } \right)\\
& = \sum_{j\in \mathcal{H}_{0}} \mathbb{E} \left(\frac{\boldsymbol{1}_{\{W_{j} > 0\} }\times \boldsymbol{1}_{\{|W_{j}| \ge T_{j}\} } \times \boldsymbol{1}_{\{\widehat{\textnormal{KL}}_{j} \le \varepsilon\}}}{ 1+\sum_{s=1, s\not= j}^{p} \boldsymbol{1}_{\{W_{s} \le - T_{j}\} } } \right)\\
& = \sum_{j\in \mathcal{H}_{0}} \mathbb{E} \left(\frac{ \mathbb{P}(W_{j}> 0, \widehat{\textnormal{KL}}_{j} \le \varepsilon \big|\ \boldsymbol{X}_{j}^{(0)}, \boldsymbol{X}_{j}^{(1)}, \boldsymbol{X}_{-j}, \widetilde{\boldsymbol{X}}_{-j}, \boldsymbol{Y} ) \times \boldsymbol{1}_{\{|W_{j}| \ge T_{j}\} }  }{ 1+\sum_{s=1, s\not= j}^{p} \boldsymbol{1}_{\{W_{s} \le - T_{j}\} } } \right),
\end{split}
\end{equation}
where the first three equalities above hold because when $\boldsymbol{1}_{\{W_{j} \geq T\} }=1$, we have $W_j>0$, $T=T_j$, and $\boldsymbol{1}_{\{W_{j} \leq -T_j \} }=0$,  and the last equality above holds since $|W_{j}|$, $T_{j}$, and $W_{1}, \dots, W_{j-1}$, $W_{j+1}, \dots, W_{p}$ are functions of $(\boldsymbol{X}_{j}^{(0)}, \boldsymbol{X}_{j}^{(1)}, \boldsymbol{X}_{-j}, \widetilde{\boldsymbol{X}}_{-j}, \boldsymbol{Y} )$ due to the sign-flip property \eqref{sign-flip.2}.

From the definitions of $\boldsymbol{X}_{j}^{(0)}$, $\boldsymbol{X}_{j}^{(1)}$, and $\widehat{\textnormal{KL}}_{j}$, we can obtain that 
\begin{equation}
\begin{split}\label{robust.eq.18}
& \mathbb{P}(W_{j}> 0, \widehat{\textnormal{KL}}_{j} \le \varepsilon \big|\ \boldsymbol{X}_{j}^{(0)}, \boldsymbol{X}_{j}^{(1)}, \boldsymbol{X}_{-j}, \widetilde{\boldsymbol{X}}_{-j}, \boldsymbol{Y} ) \\
& =  \mathbb{P}(W_{j}> 0, \widehat{\textnormal{KL}}_{j}^{(01)} \le \varepsilon \big|\ \boldsymbol{X}_{j}^{(0)}, \boldsymbol{X}_{j}^{(1)}, \boldsymbol{X}_{-j}, \widetilde{\boldsymbol{X}}_{-j}, \boldsymbol{Y} ) \\
& = \mathbb{P}(W_{j}> 0 \big|\ \boldsymbol{X}_{j}^{(0)}, \boldsymbol{X}_{j}^{(1)}, \boldsymbol{X}_{-j}, \widetilde{\boldsymbol{X}}_{-j}, \boldsymbol{Y} ) \times \boldsymbol{1}_{\{\widehat{\textnormal{KL}}_{j}^{(01)} \le \varepsilon\}},
\end{split}
\end{equation}
where 
$$\widehat{\textnormal{KL}}_{j}^{(01)} \equiv \log{\left(\frac{f_{\boldsymbol{X}_{j}, \widetilde{\boldsymbol{X}}_{j}, \boldsymbol{X}_{-j}, \widetilde{\boldsymbol{X}}_{-j}, \boldsymbol{Y}} (\boldsymbol{X}_{j}^{(0)}, \boldsymbol{X}_{j}^{(1)}, \boldsymbol{X}_{-j}, \widetilde{\boldsymbol{X}}_{-j}, \boldsymbol{Y}) }{f_{\boldsymbol{X}_{j}, \widetilde{\boldsymbol{X}}_{j}, \boldsymbol{X}_{-j}, \widetilde{\boldsymbol{X}}_{-j}, \boldsymbol{Y}} (\boldsymbol{X}_{j}^{(1)}, \boldsymbol{X}_{j}^{(0)}, \boldsymbol{X}_{-j}, \widetilde{\boldsymbol{X}}_{-j}, \boldsymbol{Y}) }\right)}. $$
Furthermore, we will show that it holds almost surely that 
\begin{equation}
\begin{split}
\label{robust.eq.2}
& \mathbb{P}(W_{j}> 0 \big|\ \boldsymbol{X}_{j}^{(0)}, \boldsymbol{X}_{j}^{(1)}, \boldsymbol{X}_{-j}, \widetilde{\boldsymbol{X}}_{-j}, \boldsymbol{Y} ) \\
& \le \frac{f_{\boldsymbol{X}_{j}, \widetilde{\boldsymbol{X}}_{j}, \boldsymbol{X}_{-j}, \widetilde{\boldsymbol{X}}_{-j}, \boldsymbol{Y}} (\boldsymbol{X}_{j}^{(0)}, \boldsymbol{X}_{j}^{(1)}, \boldsymbol{X}_{-j}, \widetilde{\boldsymbol{X}}_{-j}, \boldsymbol{Y}) }{f_{\boldsymbol{X}_{j}, \widetilde{\boldsymbol{X}}_{j}, \boldsymbol{X}_{-j}, \widetilde{\boldsymbol{X}}_{-j}, \boldsymbol{Y}} (\boldsymbol{X}_{j}^{(1)}, \boldsymbol{X}_{j}^{(0)}, \boldsymbol{X}_{-j}, \widetilde{\boldsymbol{X}}_{-j}, \boldsymbol{Y}) } \\
& \qquad \times \mathbb{P}(W_{j}< 0 \big| \ \boldsymbol{X}_{j}^{(0)}, \boldsymbol{X}_{j}^{(1)}, \boldsymbol{X}_{-j}, \widetilde{\boldsymbol{X}}_{-j}, \boldsymbol{Y})\\
&= e^{\widehat{\textnormal{KL}}_{j}^{(01)}} \times \mathbb{P}(W_{j}< 0 \big| \ \boldsymbol{X}_{j}^{(0)}, \boldsymbol{X}_{j}^{(1)}, \boldsymbol{X}_{-j}, \widetilde{\boldsymbol{X}}_{-j}, \boldsymbol{Y}),
\end{split}
\end{equation}
where Condition~\ref{knockoff.generator.3} is assumed to avoid division by zero on the right-hand side (RHS) of the second inequality above. The proof of \eqref{robust.eq.2} is deferred to right after the proof of \eqref{robust.eq.19}.

By \eqref{robust.eq.18}--\eqref{robust.eq.2}, it holds that
\begin{equation}
\begin{split}\label{robust.eq.23}
& \textnormal{RHS of \eqref{robust.eq.15}} \\
& \le \sum_{j\in \mathcal{H}_{0}} \mathbb{E} \left(\frac{ e^{\widehat{\textnormal{KL}}_{j}^{(01)} } 
\mathbb{P}(W_{j} < 0 \big|\ \boldsymbol{X}_{j}^{(0)}, \boldsymbol{X}_{j}^{(1)}, \boldsymbol{X}_{-j}, \widetilde{\boldsymbol{X}}_{-j}, \boldsymbol{Y} ) 
\, \boldsymbol{1}_{\{|W_{j}| \ge T_{j}\} } 
\, \boldsymbol{1}_{\{\widehat{\textnormal{KL}}_{j}^{(01)} \le \varepsilon\}} }{ 1+\sum_{s=1, s\not= j}^{p} \boldsymbol{1}_{\{W_{s} \le - T_{j}\} } } \right) \\
& \le e^{\varepsilon} 
\, \sum_{j=1}^{p} \mathbb{E} \left(\frac{   \boldsymbol{1}_{\{W_{j} \le -T_{j}\} }  }{ 1+\sum_{s=1, s\not= j}^{p} \boldsymbol{1}_{\{W_{s} \le - T_{j}\} } } \right)\\
& \le e^{\varepsilon} 
\, \mathbb{E} \left(\frac{   \sum_{j=1}^{p} \boldsymbol{1}_{\{W_{j} \le -T_{j}\} }  }{ 1+\sum_{s=1, s\not= j}^{p} \boldsymbol{1}_{\{W_{s} \le - T_{s}\} } } \right)\\
& \le e^{\varepsilon} 
\, \mathbb{E} \left(\frac{   \sum_{j=1}^{p} \boldsymbol{1}_{\{W_{j} \le -T_{j}\} }  }{ 1\vee\left(\sum_{s=1}^{p} \boldsymbol{1}_{\{W_{s} \le - T_{s}\} }\right) } \right)\\
&\le  e^{\varepsilon},
\end{split}
\end{equation}
where RHS is short for the right-hand side, and the third inequality above follows from Lemma 6 in \citep{barber2018robust}. Hence, by resorting to \eqref{robust.eq.15}, \eqref{robust.eq.23}, and the fact that 
$$\sum_{j\in \mathcal{H}_{0}}\mathbb{E}(e_{j} \times \boldsymbol{1}_{\{\widehat{\textnormal{KL}}_{j} \le \varepsilon\}}) = p\times \sum_{j\in \mathcal{H}_{0}} \mathbb{E} \left(\frac{\boldsymbol{1}_{\{W_{j} \ge T\} } \times \boldsymbol{1}_{\{\widehat{\textnormal{KL}}_{j} \le \varepsilon\}}}{ 1+\sum_{s=1}^{p} \boldsymbol{1}_{\{W_{s} \le - T\} } } \right),$$
we can establish \eqref{robust.eq.19}.

\bigskip
\noindent \textit{Proof of \eqref{robust.eq.2}}. 
Denote by 
$$F_{> 0}(\boldsymbol{X}_{j}^{(0)}, \boldsymbol{X}_{j}^{(1)}, \boldsymbol{X}_{-j}, \widetilde{\boldsymbol{X}}_{-j}, \boldsymbol{Y}) \  \text{ and } \ F_{<0}(\boldsymbol{X}_{j}^{(0)}, \boldsymbol{X}_{j}^{(1)}, \boldsymbol{X}_{-j}, \widetilde{\boldsymbol{X}}_{-j}, \boldsymbol{Y})$$ the versions of $$\mathbb{P}(W_{j}>0 | \boldsymbol{X}_{j}^{(0)}, \boldsymbol{X}_{j}^{(1)}, \boldsymbol{X}_{-j}, \widetilde{\boldsymbol{X}}_{-j}, \boldsymbol{Y} ) \  \text{ and } \ \mathbb{P}(W_{j}<0 | \boldsymbol{X}_{j}^{(0)}, \boldsymbol{X}_{j}^{(1)}, \boldsymbol{X}_{-j}, \widetilde{\boldsymbol{X}}_{-j}, \boldsymbol{Y} ), $$
respectively. We will show that functions $F_{> 0}:\mathbb{R}^{n(2p+1)}\longmapsto\mathbb{R}$ and $F_{< 0}:\mathbb{R}^{n(2p+1)}\longmapsto\mathbb{R}$ satisfiy that 
\begin{equation}
\begin{split}
\label{robust.eq.13}
F_{> 0}(\vec{z}) &= \frac{\boldsymbol{1}_{\{w _{j}(\vec{z}) > 0 \} }\times f_{\boldsymbol{X}_{j}, \widetilde{\boldsymbol{X}}_{j}, \boldsymbol{X}_{-j}, \widetilde{\boldsymbol{X}}_{-j}, \boldsymbol{Y} } (\vec{z}) 
}{
f_{\boldsymbol{X}_{j}^{(0)}, \boldsymbol{X}_{j}^{(1)}, \boldsymbol{X}_{-j}, \widetilde{\boldsymbol{X}}_{-j}, \boldsymbol{Y} }
(\vec{z})
},\\
F_{<0}(\vec{z}) &= \frac{ 
\boldsymbol{1}_{\{w _{j}(\vec{z}_{\textnormal{swap}}) <0 \} } \times f_{\boldsymbol{X}_{j}, \widetilde{\boldsymbol{X}}_{j}, \boldsymbol{X}_{-j}, \widetilde{\boldsymbol{X}}_{-j}, \boldsymbol{Y} } (\vec{z}_{\textnormal{swap}}) }{
f_{\boldsymbol{X}_{j}^{(0)}, \boldsymbol{X}_{j}^{(1)}, \boldsymbol{X}_{-j}, \widetilde{\boldsymbol{X}}_{-j}, \boldsymbol{Y} }
(\vec{z})
},
\end{split}
\end{equation}
respectively, where $\vec{z} = (\vec{z}_{1}, \vec{z}_{2}, \vec{z}_{3}, \vec{z}_{4}, \vec{z}_{5})$,  $\vec{z}_{\textnormal{swap}} = (\vec{z}_{2}, \vec{z}_{1}, \vec{z}_{3}, \vec{z}_{4}, \vec{z}_{5})$ with $\vec{z}_{1}\in\mathbb{R}^{n}$, $\vec{z}_{2}\in\mathbb{R}^{n}$, $\vec{z}_{3}\in\mathbb{R}^{n(p-1)}$, $\vec{z}_{4}\in\mathbb{R}^{n(p-1)}$, $\vec{z}_{5}\in\mathbb{R}^{n}$, and  $w_{j}:\mathbb{R}^{n(2p+1)}\longmapsto\mathbb{R}$ denotes the knockoff statistic function of $W_{j}$.  From the definitions of $\boldsymbol{X}_{j}^{(0)}, \boldsymbol{X}_{j}^{(1)}$, and $w_{j}(\cdot)$ along with the sign-flip property \eqref{sign-flip.2}, we have that almost surely, 
\begin{equation}
\begin{split}\label{robust.eq.17}
w_{j}(\boldsymbol{X}_{j}^{(0)}, \boldsymbol{X}_{j}^{(1)}, \boldsymbol{X}_{-j}, \widetilde{\boldsymbol{X}}_{-j}, \boldsymbol{Y}) & \ge 0,\\
w_{j}(\boldsymbol{X}_{j}^{(1)}, \boldsymbol{X}_{j}^{(0)}, \boldsymbol{X}_{-j}, \widetilde{\boldsymbol{X}}_{-j}, \boldsymbol{Y}) & < 0.
\end{split}
\end{equation}

Then an application of  \eqref{robust.eq.13}--\eqref{robust.eq.17} and the fact that the probability density function is nonnegative yields that 
\begin{equation*}
\begin{split} 
& F_{> 0}(\boldsymbol{X}_{j}^{(0)}, \boldsymbol{X}_{j}^{(1)}, \boldsymbol{X}_{-j}, \widetilde{\boldsymbol{X}}_{-j}, \boldsymbol{Y})  \le \frac{
f_{\boldsymbol{X}_{j}, \widetilde{\boldsymbol{X}}_{j}, \boldsymbol{X}_{-j}, \widetilde{\boldsymbol{X}}_{-j}, \boldsymbol{Y} } (\boldsymbol{X}_{j}^{(0)}, \boldsymbol{X}_{j}^{(1)}, \boldsymbol{X}_{-j}, \widetilde{\boldsymbol{X}}_{-j}, \boldsymbol{Y})
}{
f_{\boldsymbol{X}_{j}^{(0)}, \boldsymbol{X}_{j}^{(1)}, \boldsymbol{X}_{-j}, \widetilde{\boldsymbol{X}}_{-j}, \boldsymbol{Y}}(\boldsymbol{X}_{j}^{(0)}, \boldsymbol{X}_{j}^{(1)}, \boldsymbol{X}_{-j}, \widetilde{\boldsymbol{X}}_{-j}, \boldsymbol{Y})
}, \\
& F_{<0}(\boldsymbol{X}_{j}^{(0)}, \boldsymbol{X}_{j}^{(1)}, \boldsymbol{X}_{-j}, \widetilde{\boldsymbol{X}}_{-j}, \boldsymbol{Y})  = \frac{
f_{\boldsymbol{X}_{j}, \widetilde{\boldsymbol{X}}_{j}, \boldsymbol{X}_{-j}, \widetilde{\boldsymbol{X}}_{-j}, \boldsymbol{Y} } (\boldsymbol{X}_{j}^{(1)}, \boldsymbol{X}_{j}^{(0)}, \boldsymbol{X}_{-j}, \widetilde{\boldsymbol{X}}_{-j}, \boldsymbol{Y})
}{
f_{\boldsymbol{X}_{j}^{(0)}, \boldsymbol{X}_{j}^{(1)}, \boldsymbol{X}_{-j}, \widetilde{\boldsymbol{X}}_{-j}, \boldsymbol{Y}}(\boldsymbol{X}_{j}^{(0)}, \boldsymbol{X}_{j}^{(1)}, \boldsymbol{X}_{-j}, \widetilde{\boldsymbol{X}}_{-j}, \boldsymbol{Y})
},
\end{split}
\end{equation*}
which entail that
\begin{equation}
\begin{split} \label{robust.eq.4}
&\mathbb{P}(W_{j}> 0 | \boldsymbol{X}_{j}^{(0)}, \boldsymbol{X}_{j}^{(1)}, \boldsymbol{X}_{-j}, \widetilde{\boldsymbol{X}}_{-j}, \boldsymbol{Y} ) \\
& = F_{> 0}(\boldsymbol{X}_{j}^{(0)}, \boldsymbol{X}_{j}^{(1)}, \boldsymbol{X}_{-j}, \widetilde{\boldsymbol{X}}_{-j}, \boldsymbol{Y}) \\
& \le \frac{
f_{\boldsymbol{X}_{j}, \widetilde{\boldsymbol{X}}_{j}, \boldsymbol{X}_{-j}, \widetilde{\boldsymbol{X}}_{-j}, \boldsymbol{Y} } (\boldsymbol{X}_{j}^{(0)}, \boldsymbol{X}_{j}^{(1)}, \boldsymbol{X}_{-j}, \widetilde{\boldsymbol{X}}_{-j}, \boldsymbol{Y})
}{
f_{\boldsymbol{X}_{j}^{(0)}, \boldsymbol{X}_{j}^{(1)}, \boldsymbol{X}_{-j}, \widetilde{\boldsymbol{X}}_{-j}, \boldsymbol{Y}}(\boldsymbol{X}_{j}^{(0)}, \boldsymbol{X}_{j}^{(1)}, \boldsymbol{X}_{-j}, \widetilde{\boldsymbol{X}}_{-j}, \boldsymbol{Y})
},\\
&\mathbb{P}(W_{j}<0 | \boldsymbol{X}_{j}^{(0)}, \boldsymbol{X}_{j}^{(1)}, \boldsymbol{X}_{-j}, \widetilde{\boldsymbol{X}}_{-j}, \boldsymbol{Y} ) \\
& = F_{<0}(\boldsymbol{X}_{j}^{(0)}, \boldsymbol{X}_{j}^{(1)}, \boldsymbol{X}_{-j}, \widetilde{\boldsymbol{X}}_{-j}, \boldsymbol{Y}) \\
& = \frac{
f_{\boldsymbol{X}_{j}, \widetilde{\boldsymbol{X}}_{j}, \boldsymbol{X}_{-j}, \widetilde{\boldsymbol{X}}_{-j}, \boldsymbol{Y} } (\boldsymbol{X}_{j}^{(1)}, \boldsymbol{X}_{j}^{(0)}, \boldsymbol{X}_{-j}, \widetilde{\boldsymbol{X}}_{-j}, \boldsymbol{Y})
}{
f_{\boldsymbol{X}_{j}^{(0)}, \boldsymbol{X}_{j}^{(1)}, \boldsymbol{X}_{-j}, \widetilde{\boldsymbol{X}}_{-j}, \boldsymbol{Y}}(\boldsymbol{X}_{j}^{(0)}, \boldsymbol{X}_{j}^{(1)}, \boldsymbol{X}_{-j}, \widetilde{\boldsymbol{X}}_{-j}, \boldsymbol{Y})
}.
\end{split}
\end{equation}
Hence, a combination of \eqref{robust.eq.4} and  Condition~\ref{knockoff.generator.3} (which ensures that the denominator is nonzero) establishes the result in \eqref{robust.eq.2}.

It remains to prove \eqref{robust.eq.13}. To this end, observe that for any Borel sets $A_{1}\in\mathcal{R}^{n}$, $A_{2}\in\mathcal{R}^{n}$, $A_{3}\in\mathcal{R}^{n(p-1)}$, $A_{4}\in\mathcal{R}^{n(p-1)}$, and $A_{5}\in\mathcal{R}^{n}$, it holds that 
\begin{equation}
\begin{split} \label{robust.eq.12}
& \int_{\vec{z}\in A_{1}\times \dots\times A_{5}} F_{> 0}(\vec{z}) f_{\boldsymbol{X}_{j}^{(0)}, \boldsymbol{X}_{j}^{(1)}, \boldsymbol{X}_{-j}, \widetilde{\boldsymbol{X}}_{-j}, \boldsymbol{Y} }
(\vec{z})d\vec{z}\\
& = \mathbb{P}(W_{j}> 0,  \boldsymbol{X}_{j} \in A_{1}, \widetilde{\boldsymbol{X}}_{j} \in A_{2}, \boldsymbol{X}_{-j} \in A_{3}, \widetilde{\boldsymbol{X}}_{-j} \in A_{4}, \boldsymbol{Y}\in A_{5} )\\
& = \mathbb{P}(W_{j}> 0,  \boldsymbol{X}_{j}^{(0)} \in A_{1}, \boldsymbol{X}_{j}^{(1)} \in A_{2}, \boldsymbol{X}_{-j} \in A_{3}, \widetilde{\boldsymbol{X}}_{-j} \in A_{4}, \boldsymbol{Y}\in A_{5} )\\
& =
\int_{\{\boldsymbol{X}_{j}^{(0)} \in A_{1}, \boldsymbol{X}_{j}^{(1)} \in A_{2}, \boldsymbol{X}_{-j} \in A_{3}, \widetilde{\boldsymbol{X}}_{-j} \in A_{4}, \boldsymbol{Y}\in A_{5}\}}
\mathbb{P}(W_{j}> 0 \big|\ \boldsymbol{X}_{j}^{(0)}, \boldsymbol{X}_{j}^{(1)}, \boldsymbol{X}_{-j}, \widetilde{\boldsymbol{X}}_{-j}, \boldsymbol{Y} ) d\mathbb{P},
\end{split}
\end{equation}
where the second equality above holds by the definitions of $\boldsymbol{X}_{j}^{(0)}$ and $\boldsymbol{X}_{j}^{(1)}$. Similarly, for any Borel sets $A_{1}\in\mathcal{R}^{n}$, $A_{2}\in\mathcal{R}^{n}$, $A_{3}\in\mathcal{R}^{n(p-1)}$, $A_{4}\in\mathcal{R}^{n(p-1)}$, and $A_{5}\in\mathcal{R}^{n}$, we have 
\begin{equation}
\begin{split}\label{robust.eq.25}
& \int_{\vec{z}\in A_{1}\times \dots\times A_{5}} F_{< 0}(\vec{z}) f_{\boldsymbol{X}_{j}^{(0)}, \boldsymbol{X}_{j}^{(1)}, \boldsymbol{X}_{-j}, \widetilde{\boldsymbol{X}}_{-j}, \boldsymbol{Y} }
(\vec{z})d\vec{z}\\
& = \mathbb{P}(W_{j}< 0,  \widetilde{\boldsymbol{X}}_{j} \in A_{1}, \boldsymbol{X}_{j} \in A_{2}, \boldsymbol{X}_{-j} \in A_{3}, \widetilde{\boldsymbol{X}}_{-j} \in A_{4}, \boldsymbol{Y}\in A_{5} )\\
& = \mathbb{P}(W_{j}< 0,  \boldsymbol{X}_{j}^{(0)} \in A_{1}, \boldsymbol{X}_{j}^{(1)} \in A_{2}, \boldsymbol{X}_{-j} \in A_{3}, \widetilde{\boldsymbol{X}}_{-j} \in A_{4}, \boldsymbol{Y}\in A_{5} )\\
& =
\int_{\{\boldsymbol{X}_{j}^{(0)} \in A_{1}, \boldsymbol{X}_{j}^{(1)} \in A_{2}, \boldsymbol{X}_{-j} \in A_{3}, \widetilde{\boldsymbol{X}}_{-j} \in A_{4}, \boldsymbol{Y}\in A_{5}\}}
\mathbb{P}(W_{j}< 0 \big|\ \boldsymbol{X}_{j}^{(0)}, \boldsymbol{X}_{j}^{(1)}, \boldsymbol{X}_{-j}, \widetilde{\boldsymbol{X}}_{-j}, \boldsymbol{Y} ) d\mathbb{P},
\end{split}
\end{equation}
where the second equality above  holds by the definitions of $\boldsymbol{X}_{j}^{(0)}$ and $\boldsymbol{X}_{j}^{(1)}$. With the aid of  \eqref{robust.eq.12}--\eqref{robust.eq.25}, we can resort to the $\pi-\lambda$ Theorem~\citep{Durrett2019} and the definition of the conditional expectation to obtain  \eqref{robust.eq.13}. 
Thus, we have established \eqref{robust.eq.2}.

\bigskip
\noindent \textit{Proof of \eqref{theorem2.new.2}}. 
We now aim to show \eqref{theorem2.new.2} under Condition~\ref{knockoff.generator.4}  and the assumption that $\boldsymbol{X}_{j}$ is independent of $\boldsymbol{Y}$ conditional on $\boldsymbol{X}_{-j}$ for each $j\in\mathcal{H}_{0}$. First, in light of Condition~\ref{knockoff.generator.4} and Definition~\ref{knockoff.generator.1} of the knockoff generator, we see that $\boldsymbol{Y}$ is independent of $\widetilde{\boldsymbol{X}}$ conditional on $\boldsymbol{X}$. Next, we can deduce that
\begin{equation}
\begin{split} \label{robust.eq.1}
& f_{\boldsymbol{X}_{j}, \widetilde{\boldsymbol{X}}_{j}, \boldsymbol{X}_{-j}, \widetilde{\boldsymbol{X}}_{-j}, \boldsymbol{Y}} (\vec{z}_{1}, \vec{z}_{2}, \vec{z}_{3}, \vec{z}_{4}, \vec{z}_{5}) \\
& = f_{\boldsymbol{X}}(\vec{z}_{1}, \vec{z}_{3}) \times f_{\widetilde{\boldsymbol{X}}_{j}, \widetilde{\boldsymbol{X}}_{-j}, \boldsymbol{Y}|\boldsymbol{X}} (\vec{z}_{2}, \vec{z}_{4}, \vec{z}_{5}|\vec{z}_{1}, \vec{z}_{3})\\
& = f_{\boldsymbol{X}}(\vec{z}_{1}, \vec{z}_{3}) \times f_{\boldsymbol{Y}|\boldsymbol{X}} (\vec{z}_{5}|\vec{z}_{1}, \vec{z}_{3})
\times 
f_{\widetilde{\boldsymbol{X}}|\boldsymbol{X}} (\vec{z}_{2}, \vec{z}_{4}|\vec{z}_{1}, \vec{z}_{3})\\
& = f_{\boldsymbol{X}, \widetilde{\boldsymbol{X}}}(\vec{z}_{1}, \vec{z}_{3}, \vec{z}_{2}, \vec{z}_{4}) \times f_{\boldsymbol{Y}|\boldsymbol{X}} (\vec{z}_{5}|\vec{z}_{1}, \vec{z}_{3})\\
& = f_{\boldsymbol{X}, \widetilde{\boldsymbol{X}}}(\vec{z}_{1}, \vec{z}_{3}, \vec{z}_{2}, \vec{z}_{4}) \times f_{\boldsymbol{Y}|\boldsymbol{X}_{-j}} (\vec{z}_{5}| \vec{z}_{3}),
\end{split}
\end{equation}
where the second equality above follows from the conditional independence property of the knockoffs, and the last equality above holds because of the column-wise conditional independence assumption on the null features.

From \eqref{robust.eq.1}, it holds that  
\begin{equation*}
\begin{split} 
& 
f_{\boldsymbol{X}_{j}, \widetilde{\boldsymbol{X}}_{j}, \boldsymbol{X}_{-j}, \widetilde{\boldsymbol{X}}_{-j}, \boldsymbol{Y} } (\boldsymbol{X}_{j}, \widetilde{\boldsymbol{X}}_{j}, \boldsymbol{X}_{-j}, \widetilde{\boldsymbol{X}}_{-j}, \boldsymbol{Y})
\\
& = 
f_{\boldsymbol{X}_{j}, \widetilde{\boldsymbol{X}}_{j}, \boldsymbol{X}_{-j}, \widetilde{\boldsymbol{X}}_{-j}} (\boldsymbol{X}_{j}, \widetilde{\boldsymbol{X}}_{j}, \boldsymbol{X}_{-j}, \widetilde{\boldsymbol{X}}_{-j}) \times f_{\boldsymbol{Y}|\boldsymbol{X}_{-j}}( \boldsymbol{Y}|\boldsymbol{X}_{-j})
\end{split}
\end{equation*}
and
\begin{equation*}
\begin{split} 
& 
f_{\boldsymbol{X}_{j}, \widetilde{\boldsymbol{X}}_{j}, \boldsymbol{X}_{-j}, \widetilde{\boldsymbol{X}}_{-j}, \boldsymbol{Y} } (\widetilde{\boldsymbol{X}}_{j}, \boldsymbol{X}_{j}, \boldsymbol{X}_{-j}, \widetilde{\boldsymbol{X}}_{-j}, \boldsymbol{Y})
\\
& = 
f_{\boldsymbol{X}_{j}, \widetilde{\boldsymbol{X}}_{j}, \boldsymbol{X}_{-j}, \widetilde{\boldsymbol{X}}_{-j}} (\widetilde{\boldsymbol{X}}_{j}, \boldsymbol{X}_{j}, \boldsymbol{X}_{-j}, \widetilde{\boldsymbol{X}}_{-j}) \times f_{\boldsymbol{Y}|\boldsymbol{X}_{-j}}( \boldsymbol{Y}|\boldsymbol{X}_{-j}),
\end{split}
\end{equation*}
which establish \eqref{theorem2.new.2}.

\bigskip
\noindent \textit{Proofs of \eqref{robust.eq.6} and \eqref{robust.eq.7}}.  
The proof of \eqref{robust.eq.6} is straightforward using the additional assumption of i.i.d. observations and hence, is omitted here for simplicity. We now focus on proving \eqref{robust.eq.7}. Fixing a feature index $j$, let us consider a random vector $(\widetilde{X}_{j}^{\dagger}, \boldsymbol{x}_{-j}, \widetilde{\boldsymbol{z}})$ such that $\widetilde{X}_{j}^{\dagger}$ is generated by the $j$th coordinatewise knockoff generator $\kappa_j(\boldsymbol{x}_{-j},)$ given $\boldsymbol{x}_{-j}$ and that $\widetilde{\boldsymbol{z}}=(\widetilde{Z}_{j}, \widetilde{\boldsymbol{z}}_{-j})$ is a knockoff vector of $(\widetilde{X}_{j}^{\dagger}, \boldsymbol{x}_{-j})$ generated from the knockoff filter $\kappa((\widetilde{X}_{j}^{\dagger}, \boldsymbol{x}_{-j}),)$, where $\kappa_j$ and $\kappa$ are as given in Condition~\ref{knockoff.generator.2}. In view of  Condition~\ref{knockoff.generator.2}, we see that $(\widetilde{X}_{j}^{\dagger}, \widetilde{Z}_{j}, \boldsymbol{x}_{-j}, \widetilde{\boldsymbol{z}}_{-j})$ and $(\widetilde{Z}_{j}, \widetilde{X}_{j}^{\dagger}, \boldsymbol{x}_{-j}, \widetilde{\boldsymbol{z}}_{-j})$ have the same distribution and the corresponding density functions exist, which entail that for each $(z_{1}, z_{2}, \vec{z}_{3}, \vec{z}_{4})\in\mathbb{R}^{2p}$,
\begin{equation}
\label{robust.eq.8}
f_{\widetilde{X}_{j}^{\dagger}, \widetilde{Z}_{j}, \boldsymbol{x}_{-j}, \widetilde{\boldsymbol{z}}_{-j}}(z_{1}, z_{2}, \vec{z}_{3}, \vec{z}_{4}) = f_{\widetilde{X}_{j}^{\dagger}, \widetilde{Z}_{j}, \boldsymbol{x}_{-j}, \widetilde{\boldsymbol{z}}_{-j}}(z_{2}, z_{1}, \vec{z}_{3}, \vec{z}_{4}).
\end{equation}

Next, it follows from the definition of $(\widetilde{X}_{j}^{\dagger}, \widetilde{Z}_{j}, \boldsymbol{x}_{-j}, \widetilde{\boldsymbol{z}}_{-j})$ that 
\begin{equation}
\begin{split}
f_{\widetilde{X}_{j}^{\dagger}, \widetilde{Z}_{j}, \boldsymbol{x}_{-j}, \widetilde{\boldsymbol{z}}_{-j}}(z_{1}, z_{2}, \vec{z}_{3}, \vec{z}_{4}) &= 
f_{\boldsymbol{x}_{-j}}(\vec{z}_{3}) f_{\widetilde{X}_{j}^{\dagger} | \boldsymbol{x}_{-j}} (z_{1}|\vec{z}_{3}) 
f_{\widetilde{Z}_{j}, \widetilde{\boldsymbol{z}}_{-j} | \widetilde{X}_{j}^{\dagger}, \boldsymbol{x}_{-j}}(z_{2}, \vec{z}_{4} | z_{1}, \vec{z}_{3}) \\
& = f_{\boldsymbol{x}_{-j}}(\vec{z}_{3}) f_{\widetilde{X}_{j}^{\dagger} | \boldsymbol{x}_{-j}} (z_{1}|\vec{z}_{3}) 
f_{\widetilde{X}_{j}, \widetilde{\boldsymbol{x}}_{-j} | X_{j}, \boldsymbol{x}_{-j}}(z_{2}, \vec{z}_{4} | z_{1}, \vec{z}_{3}),
\end{split}
\end{equation}
where $f_{\widetilde{Z}_{j}, \widetilde{\boldsymbol{z}}_{-j}  | \widetilde{X}_{j}^{\dagger}, \boldsymbol{x}_{-j}}(z_{2}, \vec{z}_{4} | z_{1}, \vec{z}_{3}) = f_{\widetilde{X}_{j}, \widetilde{\boldsymbol{x}}_{-j} | X_{j}, \boldsymbol{x}_{-j}}(z_{2}, \vec{z}_{4} | z_{1}, \vec{z}_{3})$ because a knockoff generator outputs random vectors with the same distribution if the input values are the same due to Definition~\ref{knockoff.generator.1}. Similarly, it holds that 
\begin{equation}
\begin{split}\label{robust.eq.11}
f_{\widetilde{X}_{j}^{\dagger}, \widetilde{Z}_{j}, \boldsymbol{x}_{-j}, \widetilde{\boldsymbol{z}}_{-j}}(z_{2}, z_{1}, \vec{z}_{3}, \vec{z}_{4}) &= 
f_{\boldsymbol{x}_{-j}}(\vec{z}_{3}) f_{\widetilde{X}_{j}^{\dagger} | \boldsymbol{x}_{-j}} (z_{2}|\vec{z}_{3}) 
f_{\widetilde{Z}_{j}, \widetilde{\boldsymbol{z}}_{-j} | \widetilde{X}_{j}^{\dagger}, \boldsymbol{x}_{-j}}(z_{1}, \vec{z}_{4} | z_{2}, \vec{z}_{3}) \\
& = f_{\boldsymbol{x}_{-j}}(\vec{z}_{3}) f_{\widetilde{X}_{j}^{\dagger} | \boldsymbol{x}_{-j}} (z_{2}|\vec{z}_{3}) 
f_{\widetilde{X}_{j}, \widetilde{\boldsymbol{x}}_{-j} | X_{j}, \boldsymbol{x}_{-j}}(z_{1}, \vec{z}_{4} | z_{2}, \vec{z}_{3}).
\end{split}
\end{equation}

From \eqref{robust.eq.8}--\eqref{robust.eq.11}, we can deduce that 
\begin{equation*}
\begin{split}
f_{\widetilde{X}_{j}^{\dagger} | \boldsymbol{x}_{-j}} (z_{1}|\vec{z}_{3}) f_{\widetilde{X}_{j}, \widetilde{\boldsymbol{x}}_{-j} | X_{j}, \boldsymbol{x}_{-j}}(z_{2}, \vec{z}_{4} | z_{1}, \vec{z}_{3}) = f_{\widetilde{X}_{j}^{\dagger} | \boldsymbol{x}_{-j}} (z_{2}|\vec{z}_{3}) f_{\widetilde{X}_{j}, \widetilde{\boldsymbol{x}}_{-j} | X_{j}, \boldsymbol{x}_{-j}}(z_{1}, \vec{z}_{4} | z_{2}, \vec{z}_{3}),
\end{split}
\end{equation*}
which results in 
\begin{equation}
\begin{split}\label{robust.eq.9}
\frac{f_{ X_{j}, \widetilde{X}_{j}, \boldsymbol{x}_{-j}, \widetilde{\boldsymbol{x}}_{-j}}(z_{1}, z_{2}, \vec{z}_{3}, \vec{z}_{4})}{f_{ X_{j}, \widetilde{X}_{j},\boldsymbol{x}_{-j},  \widetilde{\boldsymbol{x}}_{-j}}(z_{2}, z_{1}, \vec{z}_{3}, \vec{z}_{4})} = \frac{f_{\widetilde{X}_{j}^{\dagger}, \boldsymbol{x}_{-j}} (z_{2}, \vec{z}_{3}) \times f_{X_{j}, \boldsymbol{x}_{-j}}(z_{1}, \vec{z}_{3})}{f_{\widetilde{X}_{j}^{\dagger}, \boldsymbol{x}_{-j}} (z_{1},\vec{z}_{3}) \times f_{ X_{j}, \boldsymbol{x}_{-j}}(  z_{2}, \vec{z}_{3})}.
\end{split}
\end{equation}
Setting $(z_{1}, z_{2}, \vec{z}_{3}, \vec{z}_{4}) = (X_{ij}, \widetilde{X}_{ij}, \boldsymbol{x}_{-ij}, \widetilde{\boldsymbol{x}}_{-ij})$ in \eqref{robust.eq.9} above, we can obtain that 
\begin{equation*}
\begin{split}
& \sum_{t=1}^{n}\log{\left(\frac{f_{X_{j}, \widetilde{X}_{j}, \boldsymbol{x}_{-j}, \widetilde{\boldsymbol{x}}_{-j}} ( X_{tj}, \widetilde{X}_{tj}, \boldsymbol{x}_{-tj}, \widetilde{\boldsymbol{x}}_{-ij})}{f_{X_{j}, \widetilde{X}_{j}, \boldsymbol{x}_{-j}, \widetilde{\boldsymbol{x}}_{-j}} (\widetilde{X}_{tj}, X_{tj}, \boldsymbol{x}_{-tj}, \widetilde{\boldsymbol{x}}_{-tj})}\right)} \\
& = \sum_{t=1}^{n}\log{\left(\frac{f_{X_{j}, \boldsymbol{x}_{-j}} ( X_{ij}, \boldsymbol{x}_{-tj}) f_{\widetilde{X}_{j}^{\dagger}, \boldsymbol{x}_{-j}} ( \widetilde{X}_{tj}, \boldsymbol{x}_{-tj}) }{f_{X_{j}, \boldsymbol{x}_{-j}} ( \widetilde{X}_{ij}, \boldsymbol{x}_{-tj}) f_{\widetilde{X}_{j}^{\dagger}, \boldsymbol{x}_{-j}} ( X_{tj}, \boldsymbol{x}_{-tj}) }\right)},
\end{split}
\end{equation*}
which establishes \eqref{robust.eq.7}. This concludes the proof of Theorem~\ref{theorem2.new}.

\subsection{Proof of Corollary~\ref{collary.1}} \label{collary.1.proof}

The conclusion of Corollary~\ref{collary.1} follows from the proof of \eqref{robust.eq.7} in Theorem~\ref{theorem2.new} given in Section~\ref{SecA.2.new}.

\subsection{Proof of Corollary~\ref{collary.2}} \label{collary.2.proof}
Under Condition~\ref{knockoff.generator.4} and the assumptions that $\{\boldsymbol{x}_{t}\}_{t\ge 1}$ satisfies Condition~\ref{ge4} with $h$-step and constants $C_{0}>0$ and $0 \le \rho<1$ and $Y_{t}$ is $\boldsymbol{x}_{t+1}$-measurable, we will prove in Section~\ref{temp.1.2} that for each $1\le k\le q+1$,
\begin{equation}\label{cor12.1}
\sup_{\mathcal{D}\in\mathcal{R}^{\#H_{k} \times (2p+1)}}|\mathbb{P}(\mathcal{X}_{k}\in\mathcal{D})-\mathbb{P} (\mathcal{X}_{k}^{\pi} \in\mathcal{D})|\\
\le \#H_{k}\times \rho^{q} \times C_{0},
\end{equation}
where we recall that $\mathcal{X}_{k} = \{\boldsymbol{x}_{t}, \widetilde{\boldsymbol{x}}_{t}, Y_{t}\}_{t\in H_{k}}$ and  $\mathcal{X}_{k}^{\pi} = \{\boldsymbol{x}_{t}^{\pi}, \widetilde{\boldsymbol{x}}_{t}^{\pi}, Y_{t}^{\pi}\}_{t\in  H_{k}}$ for each $k\in \{1, \dots,  q+1\}$. Combining \eqref{cor12.1} and $\sum_{k=1}^{q+1}\# H_{k} \le n$ leads to the desired result in \eqref{asym.1}.

Next, we deal with the second assertion of Corollary~\ref{collary.2}. When Condition~\ref{knockoff.generator.4} holds and $\{Y_{t}, \boldsymbol{x}_{t}\}_{t=1}^{n}$ is also an i.i.d. sample, $\{Y_{t}, \boldsymbol{x}_{t}, \widetilde{\boldsymbol{x}}_{t}\}_{t=1}^{n}$ is an i.i.d. sample. Therefore, it follows from the fact that $(Y_{t}^{\pi}, \boldsymbol{x}_{i}^{\pi}, \widetilde{\boldsymbol{x}}_{t}^{\pi})$'s are i.i.d. with $(Y_{t}^{\pi}, \boldsymbol{x}_{t}^{\pi}, \widetilde{\boldsymbol{x}}_{t}^{\pi})$ having the same distribution as $(Y_{1}, \boldsymbol{x}_{1}, \widetilde{\boldsymbol{x}}_{1})$ that \eqref{cor12.1} holds with $\rho =0$, 
which concludes the proof of Corollary~\ref{collary.2}.

\subsection{Proof of Proposition~\ref{glp1}} \label{Bglp7}

For the reader's convenience, we provide some basic knowledge about time-homogeneous Markov chains here. Two sufficient conditions for a process $\{\boldsymbol{Q}_{t}\}$ to admit a transition kernel are 1) for each Borel set $A$ and each $t$,
$$\mathbb{P}(\boldsymbol{Q}_{t}\in A| \boldsymbol{Q}_{t-1}) = \mathbb{P}(\boldsymbol{Q}_{t}\in A| \boldsymbol{Q}_{t-j}, j < 1),$$
the so-called  the ``Markov property;'' and  2) the conditional distribution of $\boldsymbol{Q}_{t}$ given $\boldsymbol{Q}_{t-1}$ are the same for each $t$. Processes satisfying these two conditions are known as time-homogeneous Markov chains. A useful sufficient condition for verifying that a process is a time-homogeneous Markov chain is to check whether the process can be written as
$\boldsymbol{Q}_{t} = F(\boldsymbol{Q}_{t-1}, \boldsymbol{\varepsilon}_{t})$ for some measurable $F(\cdot, \cdot)$ and  identically distributed innovative random vectors $\{\boldsymbol{\varepsilon}_{t}\}$ such that $\boldsymbol{\varepsilon}_{t}$ is independent of $\boldsymbol{Q}_{t-j}$ with $j \ge 1$. It can be shown that $\{\boldsymbol{x}_{t}\}$ in Example~\ref{ARX_example} is a time-homogeneous Markov chain, and we omit the details on proving such claim for simplicity. 

Next, let us consider Example~\ref{glp} below, which is more general than Example~\ref{ARX_example}. In particular, $\{\boldsymbol{x}_{t}\}$ in Example~\ref{ARX_example} is a special case of $\{\boldsymbol{z}_{t}\}$ in Example~\ref{glp}.

\begin{exmp}[Gaussian linear processes]\label{glp}
Let  $\{\boldsymbol{z}_{t} \coloneqq(Y_{t1}, \dots, Y_{tp })^{\top}\}$ be such that for $l = 1,\dots, p$, $Y_{tl} = \sum_{i=0}^{\infty} (\vec{w}_{i}(l))^{\top} \boldsymbol{\delta}_{t-i}$, where  $\vec{w}_{i}(l)$ is an $\iota$-dimensional coefficient vector such that for each $h\ge 0$,
\begin{equation}\label{glp10}
\max_{1\le l \le p }\sum_{i \ge h} \norm{\vec{w}_{i}(l)}_{1} \le C_{1}e^{-s_{1}h}
\end{equation}
with some positive $C_{1}$ and $s_{1}$, and $\boldsymbol{\delta}_{t}$'s are i.i.d. $\iota$-dimensional Gaussian random vectors with zero mean and covariance matrix $\Sigma$. In addition, assume that $\lambda_{\max}(\Sigma) < L_{3}$  and $\lambda_{\min}(\mathbb{E} (\boldsymbol{z}_{1} \boldsymbol{z}_{1}^{\top} )) >l_{1}$ for some positive $L_{3}$ and $l_{1}$, where $\lambda_{\max}(\cdot)$ and $\lambda_{\min}(\cdot)$ denote the largest and smallest eigenvalues of a given matrix, respectively.
\end{exmp}

We use Propositions 3.1.1--3.1.2 of Brockwell and Davis~\citep{brockwell1991time} to obtain the stationarity of Example~\ref{glp}; the details on this are omitted. Thus,  $\{\boldsymbol{x}_{t}\}$ in Example~\ref{ARX_example} is a stationary time-homogeneous Markov chain; equivalently, the stationary distribution and transition kernel of $\{\boldsymbol{x}_{t}\}$ exist. 

\begin{remark}
Notice that time-homogeneous Markov chains are not always stationary; particularly, a random walk process can be a time-homogeneous Markov chain. Also, note that Example~\ref{glp} may not be a time-homogeneous Markov chain. With regularity conditions assumed, for a stationary $Q_{t} = \sum_{j=0}^{\infty} \beta_{j}\varepsilon_{t-j}$ to be a time-homogeneous Markov chain, it is usually required that $Q_{t}$ can be written as $Q_{t} = \sum_{j=1}^{k}\gamma_{j}Q_{t-j}+\varepsilon_{t}$ for some positive integer $k$. Without further assumptions, the linear processes in Example~\ref{glp} may not admit such representation.
\end{remark}

Let $\{\boldsymbol{z}_{t}^{(h)}\}$ in Example~\ref{glp} with dimensionality $p_{h}$ and stationary distribution $\pi_{h}(\cdot)$ be given. Note that we do not assume a transition kernel for Example~\ref{glp} and that all parameters (except for constant $C_{1}$, $s_{1}$, $l_{1}$, and $L_{3}$) in Example~\ref{glp} may change for each $h$, but we drop the superscript or subscript $h$ for simplicity of presentation whenever there is no confusion. Since Example~\ref{ARX_example} admits a transition kernel and it is a special case of Example~\ref{glp}, to prove Proposition~\ref{glp1} it suffices to show that for all large $h$, there exist some constants $0 \le \rho<1$, $0  <C_{0}< \infty$, and measurable functions $V_{h}:  \mathbb{R}^{p_{h}}\longrightarrow[0,\infty)$ such that for each integer $t$,
\begin{equation}\label{need.1}
\begin{split}
&\sup_{\mathcal{D}\in \mathcal{R}^{p_{h}}} \left| \mathbb{P} (\boldsymbol{z}_{t+h}^{(h)}\in \mathcal{D})  - \mathbb{P} (\boldsymbol{z}_{t+h}^{(h)} \in \mathcal{D}\ | \ \boldsymbol{z}_{t}^{(h)}) \right| \le V_{h}(\boldsymbol{z}_{t}^{(h)})\rho^{h}C_{3} 
\end{split}
\end{equation}
almost surely for some constant $C_{3} > 0$, and 
\begin{equation*} C_{0} \ge \sup_{h > 0} \int _{\mathbb{R}^{p_{h}}} V_{h}(x)\pi_{h}(dx).
\end{equation*}

To facilitate the technical presentation, we first introduce some necessary notations. For each $h$, denote by 
\begin{equation}\begin{split}
U_{1t}^{(h)} & \coloneqq \left( \sum_{i=h}^{\infty} (\vec{w}_{i}(1))^{\top} \boldsymbol{\delta}_{t-i}, \dots, \sum_{i=h}^{\infty} (\vec{w}_{i}(p_{h}))^{\top} \boldsymbol{\delta}_{t-i} \right)^{\top},\\
U_{2t}^{(h)} & \coloneqq \left(\sum_{i=0}^{h-1} (\vec{w}_{i}(1))^{\top} \boldsymbol{\delta}_{t-i}, \dots, \sum_{i=0}^{h-1} (\vec{w}_{i}(p_{h}))^{\top} \boldsymbol{\delta}_{t-i} \right)^{\top},
\end{split}\end{equation}
and let $V_{1t}^{(h)}$ and $V_{2t}^{(h)}$ be independent copies of $U_{1t}^{(h)}$ and $U_{2t}^{(h)}$, respectively, where the superscript or subscript $h$ represents the truncation length. Observe that $U_{1t}^{(h)} + U_{2t}^{(h)}$ is an instance of $\boldsymbol{z}_{t}$ in Example~\ref{glp}. Due to the Gaussian innovations, the stationary distribution $\pi_{h}$ is the distribution of $V_{11}^{(0)}$, which is the same as that of $V_{1t}^{(h)} + V_{2t}^{(h)}$ for each $t$ and $h$.

Let us repeat the needed statement \eqref{need.1} with the newly defined notation. For all large $h$, there exist some constants $0 \le \rho<1$, $0  <C_{0}< \infty$, and measurable functions $V_{h}:  \mathbb{R}^{p_{h}}\longrightarrow[0,\infty)$ such that for each integer $t$,
\begin{equation}\label{glp2}
\begin{split}
&\sup_{\mathcal{D}\in \mathcal{R}^{p_{h}}} \left| \mathbb{P} (V_{1t}^{(h)} + V_{2t}^{(h)} \in \mathcal{D})  - \mathbb{P} (U_{1(t+h)}^{(h)}  + U_{2(t+h)}^{(h)} \in \mathcal{D}\ | \ U_{1t}^{(h)} + U_{2t}^{(h)}) \right| \\
&\quad\le V_{h}(U_{1t}^{(h)} + U_{2t}^{(h)})\rho^{h}C_{3} 
\end{split}
\end{equation}
almost surely for some constant $C_{3} > 0$, and 
\begin{equation}\label{glp3} C_{0} \ge \sup_{h > 0} \int _{\mathbb{R}^{p_{h}}} V_{h}(x)\pi_{h}(dx).
\end{equation}
If (\ref{glp2}) holds for some $t$, it holds for each integer $t$ because the process is stationary. Notice that the technical analysis here does not depend on the Markov property. For the remaining proof of Proposition~\ref{glp1}, we tend to omit the term almost surely when the equality or inequality holds clearly almost surely.

Let us begin with establishing (\ref{glp2}). In view of assumption (\ref{hds1}), let $s_{3} > 0$ and $0 < \delta_{0}< 1$ be given such that $0 < s_{3} < s_{1}$ and $s_{2} < \delta_{0}s_{3}$. For each positive integer $h$, we have 
\begin{equation}
\label{glp6}
p_{h} \exp{(-\delta_{0}s_{3} h)} \le C_{2}\exp{(( s_{2} - \delta_{0}s_{3})h) }.
\end{equation}
We claim that for all large $h$ and each $t$, it holds that for each $\mathcal{D}\in \mathcal{R}^{p_{h}}$,
\begin{equation}\begin{split} 
\label{glp7}
& \left|\mathbb{P} \left(V_{1t}^{(h)} + V_{2t}^{(h)} \in \mathcal{D}\right)  - \mathbb{P} \left(U_{1(t + h)}^{(h)} + U_{2(t + h)}^{(h)} \in \mathcal{D} \ \Big\vert \ U_{1t}^{(h)} + U_{2t}^{(h)}\right)\right| \\
& \le  \mathbb{P}\left(\norm{V_{11}^{(h)}}_{\infty} \ge e^{-s_{3}h} \right) + \mathbb{P}(\norm{U_{1(t+h)}^{(h)}}_{\infty}  \ge e^{(-s_{3}h)} \ | \ U_{1t}^{(h)} + U_{2t}^{(h)})\\
& \quad+ 2\mathbb{P}\left(\norm{V_{21}^{(h)}}_{\infty} \ge e^{(1-\delta_{0})s_{3}h} - 2e^{-s_{3}h} \right) 
+ \frac{8p_{h}}{\underbar{c}}e^{-\delta_{0} s_{3} h} ,
\end{split}\end{equation}
where $\underbar{c}>0$ is some constant and $\norm{\vec{z}}_{\infty}\coloneqq \max_{1 \le i\le k }|z_{i}|$ for $\vec{z} = (z_{1}, \dots ,z_{k})^{\top} \in \mathbb{R}^k$. The proof of claim (\ref{glp7}) above is presented in Section \ref{proof-(A.26)}.

We next construct some upper bounds for the first and third terms on the RHS of (\ref{glp7}). It follows from $\norm{\cdot}_{1}^2\ge \norm{\cdot}_{2}^2$ and (\ref{glp10}) that for each $q$ and $t$,
\begin{equation*}
\begin{split}
\textnormal{Var}(\sum_{i\ge h} \vec{w}_{i}^{\top}(q) \boldsymbol{\delta}_{t-i} ) & \le \sum_{i\ge h} \norm{\vec{w}_{i}(q)}_{2}^{2} \lambda_{\max}(\Sigma_{h})\\
& \le \left( \sum_{i\ge h} \norm{\vec{w}_{i}(q)}_{1} \right)^{2} \lambda_{\max}(\Sigma_{h})\\
& \le \left( C_{1}\exp{(-s_{1}h)} \right)^{2} \lambda_{\max}(\Sigma_{h}),
\end{split}\end{equation*}
where $\Sigma_{h}$ denotes the covariance matrix of the underlying Gaussian random vectors $\boldsymbol{\delta}_{t}$'s (the superscript $h$ is dropped) associated with $\{\boldsymbol{z}_{t}^{(h)}\}$ in Example~\ref{glp}. Combining this, the fact that $\boldsymbol{\delta}_{t}$'s are Gaussian random vectors, and Markov's inequality, it holds that for each  $h\ge 0$,
\begin{equation}\begin{split}
\label{glp4}
\mathbb{P} & \left(\norm{V_{11}^{(h)}}_{\infty}\ge \exp{(-s_{3}h)}\right) 
\\
&\quad\le p_{h} \mathbb{P}\left(C_{1}\exp{(-s_{1}h)}  \sqrt{\lambda_{\max}(\Sigma_{h})} |Z| \ge  \exp{(-s_{3}h)}\right)\\
& \quad\le p_{h} \mathbb{E}(e^{|Z|})\exp{\left[- \left(C_{1} \sqrt{\lambda_{\max}(\Sigma_{h})} \right)^{-1} \exp{\left((s_{1} - s_{3})h \right)} \right] },
\end{split}\end{equation}
where $Z$ denotes a Gaussian random variable with zero mean and unit variance. Similarly, we can show that for all large $h$, it holds that 
\begin{equation}
\begin{split}
\label{glp5}
&\mathbb{P}\left(\norm{V_{21}^{(h)}}_{\infty}\ge \exp{((1-\delta_{0})s_{3}h)} - 2\exp{(-s_{3}h)} \right) \\
& \quad\le p_{h}\mathbb{P}\left(C_{1} \sqrt{\lambda_{\max}(\Sigma_{h})} |Z| \ge \exp{((1-\delta_{0})s_{3}h)} - 1\right)\\
& \quad\le p_{h} \mathbb{E}(e^{|Z|}) \exp{\left[- \left(C_{1} \sqrt{\lambda_{\max}(\Sigma_{h})} \right)^{-1} (\exp{((1-\delta_{0})s_{3}h)}-1) \right] }.
\end{split}
\end{equation}

We are now ready to construct the $V_{h}$ function. Let $g$ be a measurable function such that  $g(U_{1t}^{(h)} + U_{2t}^{(h)})$ is a version of $\mathbb{P}(\norm{U_{1(t+h)}^{(h)}}_{\infty}  \ge e^{(-s_{3}h)} \ | \ U_{1t}^{(h)} + U_{2t}^{(h)})$. It follows from the assumption that $\lambda_{\max}(\Sigma_{h})$ is bounded by a constant,  $\mathbb{E}(e^{|Z|})< \infty$, \eqref{glp6}, \eqref{glp4}, and \eqref{glp5} that there exist some constants $C_{3} > 0$ and $0 \le \rho < 1$ such that for each positive $h$, $C_{3}\rho^{h}$ is larger than the summation of the first, third, and fourth terms on the RHS of (\ref{glp7}). For each $h$, let us define function $V_{h}$ as 
\[V_{h}(x) \coloneqq
\begin{cases}
2 \quad \textnormal{ if } g(x) \le C_{3}\rho^{h}, \\
2\rho^{-h}C_{3}^{-1} \quad \textnormal{ otherwise}.
\end{cases}\]
Then combining (\ref{glp7}) and the definitions of $\rho$, $C_{3}$, and $V_{h}$ leads to (\ref{glp2}).

Finally, we deal with \eqref{glp3}. By Markov's inequality, the definition of $g$, and $\mathbb{P}  (\norm{V_{11}^{(h)}}_{\infty}\ge \exp{(-s_{3}h)})$ $= \mathbb{P}  (\norm{U_{1(t+h)}^{(h)}}_{\infty}\ge \exp{(-s_{3}h)})$, we can deduce that 
\begin{equation}
\begin{split}
\label{glp222}
\int V_{h}(x)d\pi_{h}(x)&= \mathbb{E}(V_{h} (U_{1t}^{(h)} + U_{2t}^{(h)})) \\
& \le 2 + 2(C_{3}\rho^{h})^{-1}\mathbb{P}(g(U_{1t}^{(h)} + U_{2t}^{(h)}) > C_{3}\rho^{h} )\\
& \le 2 + 2(C_{3}\rho^{h})^{-2}\mathbb{E}(g(U_{1t}^{(h)} + U_{2t}^{(h)}))\\
& = 2 + 2(C_{3}\rho^{h})^{-2}\mathbb{P}  \left(\norm{V_{11}^{(h)}}_{\infty}\ge \exp{(-s_{3}h)}\right).
\end{split}
\end{equation}
For the first equality, recall that $U_{1t}^{(h)} + U_{2t}^{(h)}$ has the stationary distribution. Therefore, by \eqref{glp222}, \eqref{glp4}, and \eqref{glp6}, for all large $h$, it holds that $\int V_{h}(x)d\pi_{h}(x)$ is bounded by a constant, which leads to \eqref{glp3}. This completes the proof of Proposition~\ref{glp1}.

\renewcommand{\thesubsection}{C.\arabic{subsection}}

\section{Some key lemmas and additional technical details} \label{SecB}

In this section, we will provide additional technical details and some key lemmas. 
In particular, we rely on measure theory for valid arguments for the manipulation of integration when conditional distributions are involved. 

\subsection{Proof of Claim \eqref{cor12.1}}\label{temp.1.2}
Let us first make a simple observation. For any $q_{1}$-dimensional random vectors $X_{1},  Y_{1}$ and $q_{2}$-dimensional random vectors $X_{2}, Y_{2}$ such that $X_{2} = F(X_{1})$ and $Y_{2}= F(Y_{1})$ for some measurable $F:\mathbb{R}^{q_{1}}\longmapsto \mathbb{R}^{q_{2}}$, it holds that
\begin{equation}
\begin{split}	
\label{cancel1}
\sup_{\mathcal{D} \in\mathcal{R}^{q_{2}}}|\mu_{X_{2}} (\mathcal{D})  -  \mu_{Y_{2}} (\mathcal{D})| & = \sup_{\mathcal{D} \in\mathcal{R}^{q_{2} }}|\mu_{X_{1}} (F^{-1}(\mathcal{D}))  -  \mu_{Y_{1}} (F^{-1}(\mathcal{D}))| \\
& \le \sup_{\mathcal{A} \in\mathcal{R}^{q_{1}}}|\mu_{X_{1}} (\mathcal{A})  -  \mu_{Y_{1}} (\mathcal{A})|,
\end{split}\end{equation}
where $F^{-1}(\cdot)$ denotes the inverse mapping of $F(\cdot)$.
With the aid of \eqref{cancel1}, we now deal with the case of $k=1$ below. Let $\boldsymbol{M}$ be given as in (\ref{M0}) with $l=2$, $h=q-1$, and $\boldsymbol{z}_{i} = \boldsymbol{x}_{i}$.  Similarly, let $\boldsymbol{M}^{\pi}$ be given as in (\ref{M0}) with $l=2$, $h=q-1$, and i.i.d. random vectors $(\boldsymbol{z}_{1}^{\pi_{i}}, \boldsymbol{z}_{2}^{\pi_{i}})$'s such that $(\boldsymbol{z}_{1}^{\pi_{1}}, \boldsymbol{z}_{2}^{\pi_{1}})$ and $(\boldsymbol{x}_{1}, \boldsymbol{x}_{2})$ have the same distribution. Then it follows from Lemma~\ref{airv} in Section \ref{app_airv} that
\begin{equation}
\begin{split}
\label{cancel2}
& \sup_{ \mathcal{D}\in\mathcal{R}^{\# H_{1}\times(2p) }}  |\mathbb{P}( (\boldsymbol{x}_{i+1}, \boldsymbol{x}_{i}, i \in H_{1}) \in \mathcal{D})- \mathbb{P}( (\boldsymbol{z}_{2}^{\pi_{i}}, \boldsymbol{z}_{1}^{\pi_{i}}, i \in H_{1}) \in \mathcal{D}) | \\
& \quad \le \# H_{1}\times \rho^{q} \times C_{0}.
\end{split}
\end{equation}

By the assumption that $Y_{i}$ is $\boldsymbol{x}_{i+1}$-measurable, we have that
$(Y_{i}, \boldsymbol{x}_{i})  = F(\boldsymbol{x}_{i+1}, \boldsymbol{x}_{i})$
for some measurable $F:\mathbb{R}^{2p} \longmapsto\mathbb{R}^{1+p}$. Then it follows from the assumption that each $(Y_{i}^{\pi}, \boldsymbol{x}_{i}^{\pi})$ and $(Y_{1}, \boldsymbol{x}_{1})$ have the same distribution and the assumption that each $(\boldsymbol{z}_{1}^{\pi_{i}}, \boldsymbol{z}_{2}^{\pi_{i}})$ and $(\boldsymbol{x}_{1}, \boldsymbol{x}_{2})$ have the same distribution that
\begin{equation}
\label{cancel3}
\{F(\boldsymbol{z}_{2}^{\pi_{i}}, \boldsymbol{z}_{1}^{\pi_{i}})\}_{i=1}^{n} \ \textnormal{ and } \  \{(Y_{i}^{\pi}, \boldsymbol{x}_{i}^{\pi}) \}_{i=1}^{n}
\end{equation} 
have the same distribution.
Hence, from  \eqref{cancel1}--\eqref{cancel3}, we can deduce that
\begin{equation}
\begin{split}
& \sup_{ \mathcal{D}\in\mathcal{R}^{\# H_{1}\times(1+p) }} |\mathbb{P}( (Y_{i}, \boldsymbol{x}_{i}, i \in H_{1}) \in \mathcal{D})- \mathbb{P}( (Y_{i}^{\pi}, \boldsymbol{x}_{i}^{\pi}, i \in H_{1}) \in \mathcal{D}) |
\\
& = \sup_{ \mathcal{D}\in\mathcal{R}^{\# H_{1}\times(1+p) }} |\mathbb{P}( (\boldsymbol{x}_{i+1}, \boldsymbol{x}_{i}, i \in H_{1}) \in F^{-1}(\mathcal{D}))- \mathbb{P}( (\boldsymbol{z}_{2}^{\pi_{i}}, \boldsymbol{z}_{1}^{\pi_{i}}, i \in H_{1}) \in F^{-1}(\mathcal{D})) |\\
&\le \# H_{1}\times\rho^{q}\times C_{0} .
\end{split}
\end{equation}

Finally, we can apply Lemma~\ref{knockoff1} in Section \ref{SecA.9} to control the distributional variation results from the inclusion of knockoffs. Using Definition~\ref{knockoff.generator.1} and Conditions~\ref{knockoff.generator.3}--\ref{knockoff.generator.4}, we can show that 1) $(Y_{i}, \boldsymbol{x}_{i}, \widetilde{\boldsymbol{x}}_{i})$'s are identically distributed; 2) $(\widetilde{\boldsymbol{x}}_{i}, i\in H_{1})$ are independent conditional on $(\boldsymbol{x}_{i}, i\in H_{1})$, and that $\widetilde{\boldsymbol{x}}_{i}$ is independent of $(\boldsymbol{x}_{q}, q\in H_{1}\backslash\{i\})$ conditional on $\boldsymbol{x}_{i}$ for each $i\in H_{1}$; 3) $(Y_{i}, i\in H_{1})$ is independent of $(\widetilde{\boldsymbol{x}}_{i}, i\in H_{1})$ conditional on $(\boldsymbol{x}_{i}, i\in H_{1})$; and 4) the above results also hold for $(Y_{i}^{\pi}, \boldsymbol{x}_{i}^{\pi}, \widetilde{\boldsymbol{x}}_{i}^{\pi})$'s. By these results, an application of the first assertion of Lemma~\ref{knockoff1} concludes the proof of \eqref{cor12.1} for the case with $k=1$. The other cases with $2\le k \le q+1$ can be dealt with similarly. This concludes the proof of Claim \eqref{cor12.1}.



\subsection{Proof of Claim (\ref{glp7})} \label{proof-(A.26)}

We denote by $f_{h}(x)$ the density function of $V_{21}^{(h)}$; that is,
\[f_{h}(x) \coloneqq \frac{ 1 }{ \sqrt{(2\pi)^{p_{h}} |\Sigma_{h}^{V}|} } \exp{\left(-\frac{1}{2} x^{\top}(\Sigma_{h}^{V})^{-1} x\right) },\]
where $\Sigma_{h}^{V}$ is the corresponding covariance matrix and $|A|$ stands for the determinant of a given matrix $A$. By the assumptions of Gaussian linear processes (this is where we need $\lambda_{\min}(\mathbb{E} (\boldsymbol{z}_{1}^{(h)} (\boldsymbol{z}_{1}^{(h)})^{\top} )) >l_{1}$), there exist some constant $\underbar{c}>0$ and positive integer $\bar{h}$ such that		
\begin{equation}\label{glp14}
\min_{k\ge \bar{h}}\lambda_{\min}(\mathbb{E}(V_{21}^{(k)} V_{21}^{(k) \top}) ) > \underbar{c} > 0.\end{equation}
In view of (\ref{glp14}), for all $h\ge \bar{h}$ we have  $\lambda_{\min}(\Sigma_{h}^{V}) > \underbar{c}$. 

To support the technical analysis, we will make use of the following facts.
\begin{enumerate}
\item[1)] For all $x < 1.79$, it holds that 
\begin{equation}
\label{glp9}
\exp{(x)} \le  1 + x + x^{2}.
\end{equation}
To get the specific value of $1.79$, we use the first and second order derivatives of $\exp{(x)}$ and $1+x+x^{2}$ to conclude that there exists some positive number $x_{0}$ such that when $x \le x_{0}$, (\ref{glp9}) holds, and when $x > x_{0}$, it holds that $\exp{(x)} > 1 + x + x^{2}$. Then a direct calculation shows that $\exp{(1.79)} < 5.994 < 1 + 1.79 + 1.79^{2} $, which gives $x_{0} \ge 1.79$.
\item[2)] By (\ref{glp14}), for all large $h$, we have that for each $\Delta, x \in \mathbb{R}^{p_{h}}$ with $\norm{\Delta}_{2} \le \norm{x}_{2}$, 
\begin{equation}
\label{glp21}			
|x^{\top} (\Sigma_{h}^{V})^{-1}\Delta + \frac{1}{2}\Delta^{\top} (\Sigma_{h}^{V})^{-1}\Delta |\le 2\norm{x}_{2}\norm{\Delta}_{2}\underbar{c}^{-1}.
\end{equation}
If furthermore   $2\norm{x}_{2}\norm{\Delta}_{2}\underbar{c}^{-1} < 1.79$, then it follows from \eqref{glp9}--\eqref{glp21} that 
\begin{equation}
\begin{split}\label{glp8}
|f_{h}(x + \Delta) - f_{h}(x)| & \le f_{h}(x)\Big| \exp{\left(-x^{\top} (\Sigma_{h}^{V})^{-1}\Delta - \frac{1}{2}\Delta^{\top}(\Sigma_{h}^{V})^{-1}\Delta \right)} - 1\Big|\\
&\le f_{h}(x)\big(2\norm{x}_{2}\norm{\Delta}_{2}\underbar{c}^{-1} + (2\norm{x}_{2}\norm{\Delta}_{2}\underbar{c}^{-1})^{2}\big).
\end{split}
\end{equation}

\item[3)] We show that for each $\mathcal{D}\in\mathcal{R}^{p_{h}}$, $\mu_{V_{21}^{(h)}} (\mathcal{D} - V_{1t}^{(h)})$ is a version of $\mathbb{P} (V_{1t}^{(h)} + V_{2t}^{(h)} \in \mathcal{D}\ \vert\ V_{1t}^{(h)})$ and in particular, for each $t$, $h > 0$, and $\mathcal{D}\in \mathcal{R}^{p_{h}}$, 
\[\mathbb{P} (V_{1t}^{(h)} + V_{2t}^{(h)} \in \mathcal{D}\ \vert\ V_{1t}^{(h)}) = \mu_{V_{21}^{(h)}} (\mathcal{D} - V_{1t}^{(h)}). \]

To this end, let us define a measurable function $g(x) \coloneqq \int_{\mathbb{R}^{p_{h}}}\mu_{ V_{21}^{(h)}}( dx_{2}) \boldsymbol{1}_{ x_{2} \in \mathcal{D}-x }$ with $\mathcal{D} - x \coloneqq \{z - x : z \in \mathcal{D}\}$ and write $\mu_{V_{21}^{(h)}} (\mathcal{D} - V_{1t}^{(h)}) = g(V_{1t}^{(h)})$ to see that $\mu_{V_{21}^{(h)}} (\mathcal{D} - V_{1t}^{(h)})$ is $\sigma(V_{1t}^{(h)})$-measurable. Observe that if we can show that for each $\mathcal{A} \in\mathcal{R}^{p_{h}}$,
\begin{equation}\label{glp223}\int_{\mathcal{A}}  \mu_{V_{1t}^{(h)}}(dx_{1}) \mu_{ V_{21}^{(h)}}(  \mathcal{D}-x_{1}) = \int_{ \{V_{1t}^{(h)} \in \mathcal{A}\}  } \mathbb{P} (V_{1t}^{(h)} + V_{2t}^{(h)} \in \mathcal{D} \ | \ V_{1t}^{(h)}) d\mathbb{P},\end{equation}
then we can apply the change of variables formula to the left-hand side of (\ref{glp223}) and use the definition of conditional expectation to obtain the desired result. It remains to prove (\ref{glp223}).

\indent\qquad Since $V_{1t}^{(h)}$ and $V_{2t}^{(h)}$ are independent for each $t$ and $h > 0$,
it holds that for each $\mathcal{D}, \mathcal{A} \in\mathcal{R}^{p_{h}}$,
\begin{equation}\label{CM.add.1}
\begin{split}
& \mathbb{P} (\{V_{1t}^{(h)} + V_{2t}^{(h)} \in \mathcal{D}\}\cap\{V_{1t}^{(h)}\in\mathcal{A}\})  \\
&\quad= \int_{\mathbb{R}^{2p_{h}}} \mu_{V_{1t}^{(h)}, V_{2t}^{(h)}}(dx_{1} \times dx_{2}) \boldsymbol{1}_{x_{1} + x_{2} \in \mathcal{D}} \boldsymbol{1}_{x_{1} \in \mathcal{A}}\\
& \quad= \int_{\mathbb{R}^{2p_{h}}} \mu_{V_{1t}^{(h)}}(dx_{1})\mu_{ V_{2t}^{(h)}}( dx_{2}) \boldsymbol{1}_{x_{1} + x_{2} \in \mathcal{D}} \boldsymbol{1}_{x_{1} \in \mathcal{A}}\\
& \quad= \int_{\mathbb{R}^{p_{h}}} \boldsymbol{1}_{(x_{1} \in \mathcal{A})}\mu_{V_{1t}^{(h)}}(dx_{1}) \int_{\mathbb{R}^{p_{h}}}\mu_{ V_{2t}^{(h)}}( dx_{2}) \boldsymbol{1}_{ x_{2} \in \mathcal{D}-x_{1}} \\
& \quad= \int_{\mathbb{R}^{p_{h}}} \boldsymbol{1}_{(x_{1} \in \mathcal{A})} \mu_{V_{1t}^{(h)}}(dx_{1}) \mu_{ V_{2t}^{(h)}}(  \mathcal{D}-x_{1})\\
& \quad= \int_{\mathbb{R}^{p_{h}}} \boldsymbol{1}_{(x_{1} \in \mathcal{A})} \mu_{V_{1t}^{(h)}}(dx_{1}) \mu_{ V_{21}^{(h)}}(  \mathcal{D}-x_{1})\\
& \quad= \int_{\mathcal{A}}  \mu_{V_{1t}^{(h)}}(dx_{1}) \mu_{ V_{21}^{(h)}}(  \mathcal{D}-x_{1}),
\end{split}
\end{equation}
where the second equality above is due to independence. Moreover, since $\mathbb{P} (\{V_{1t}^{(h)} + V_{2t}^{(h)} \in \mathcal{D}\}\cap\{V_{1t}^{(h)}\in\mathcal{A}\} \ | \ V_{1t}^{(h)}) = \boldsymbol{1}_{V_{1t}^{(h)} \in \mathcal{A}} \mathbb{P} (V_{1t}^{(h)} + V_{2t}^{(h)} \in \mathcal{D} \ | \ V_{1t}^{(h)})$, by the law of total expectation we have that 
\begin{equation}\label{CM.add.2}
\begin{split}
& \mathbb{P} (\{V_{1t}^{(h)} + V_{2t}^{(h)}  \in \mathcal{D}\}\cap\{V_{1t}^{(h)}\in\mathcal{A}\})   \\
&= \mathbb{E}( \mathbb{P} (\{V_{1t}^{(h)} + V_{2t}^{(h)} \in \mathcal{D}\}\cap\{V_{1t}^{(h)}\in\mathcal{A}\} \ | \ V_{1t}^{(h)} ) )
\\
& = \int_{ \{V_{1t}^{(h)} \in \mathcal{A}\} }  \mathbb{P} (V_{1t}^{(h)} + V_{2t}^{(h)} \in \mathcal{D} \ | \ V_{1t}^{(h)}) d\mathbb{P}.
\end{split}
\end{equation}
Hence, combining  \eqref{CM.add.1}--\eqref{CM.add.2} leads to (\ref{glp223}). 

\item[4)] Observe that $U_{2(t+h)}^{(h)}$ is independent of $U_{1t}^{(h)} + U_{2t}^{(h)}$ and $U_{1(t+h)}^{(h)}$. Thus, for each $\mathcal{D}\in\mathcal{R}^{p_{h}}$, we have that 
\begin{equation*}
\begin{split}
& \mathbb{P} (U_{1(t+h)}^{(h)} + U_{2(t+h)}^{(h)} \in \mathcal{D}\ \vert\ U_{1t}^{(h)} + U_{2t}^{(h)}, U_{1(t+h)}^{(h)} ) \\
& =  \mathbb{P} (U_{1(t+h)}^{(h)} + U_{2(t+h)}^{(h)} \in \mathcal{D}\ \vert\  U_{1(t+h)}^{(h)} ).
\end{split}
\end{equation*}
Using such representation, similar arguments as in 3) above, and the fact that $\mu_{U_{2(t+h)}^{(h)}}$ is identical to $\mu_{V_{21}^{(h)}}$,  we can show that 
\[\mathbb{P} (U_{1(t+h)}^{(h)} + U_{2(t+h)}^{(h)} \in \mathcal{D}\ \vert\ U_{1t}^{(h)} + U_{2t}^{(h)}, U_{1(t+h)}^{(h)} ) = \mu_{V_{21}^{(h)}} (\mathcal{D} - U_{1(t+h)}^{(h)}) . \]

\item[5)] Denote by $Q \coloneqq \left\{\norm{V_{1t}^{(h)}}_{\infty}  \ge e^{(-s_{3}h)}\right\}$ and $G \coloneqq \left\{\norm{U_{1(t+h)}^{(h)}}_{\infty}  \ge e^{(-s_{3}h)}\right\}$. Then it follows from the definition of $\mu_{V_{21}^{(h)}}$ that for each $\mathcal{D}\in \mathcal{R}^{p_{h}}$,
\begin{equation}\label{glp19}
\begin{split}
&\boldsymbol{1}_{Q^{c}\cap G^{c}}  \left(\mu_{V_{21}^{(h)}} (\mathcal{D} - V_{1t}^{(h)}) - \mu_{V_{21}^{(h)}} (\mathcal{D} - U_{1(t+h)}^{(h)})\right) \\
&\le \sup_{\norm{\Delta_{i}}_{\infty} < e^{-s_{3}h}}\Big|  \int_{\mathcal{D} - \Delta_{2}} \big(f_{h}(x+\Delta_{2} - \Delta_{1}) -   f_{h}(x)\big)dx \Big|.
\end{split}
\end{equation}
\end{enumerate}

We are now ready to establish the desired upper bound. For each integer $h>0$, $t$, and  $\mathcal{D}\in \mathcal{R}^{p_{h}}$, it holds that 
\begin{equation}\begin{split}\label{glp11}
& \left|\mathbb{P} \left(V_{1t}^{(h)} + V_{2t}^{(h)} \in \mathcal{D}\right)  - \mathbb{P} \left(U_{1(t + h)}^{(h)} + U_{2(t + h)}^{(h)} \in \mathcal{D} \ \Big\vert \ U_{1t}^{(h)} + U_{2t}^{(h)}\right)\right| \\
& = \Bigg| \mathbb{E}\left[   \mathbb{P} (V_{1t}^{(h)} + V_{2t}^{(h)} \in \mathcal{D}\ \vert\ V_{1t}^{(h)}) \right]  \\
& \quad - \mathbb{E}\Bigg[   \mathbb{P} (U_{1(t + h)}^{(h)} + U_{2(t + h)}^{(h)} \in \mathcal{D} \ \vert\ U_{1t}^{(h)} + U_{2t}^{(h)}, U_{1(t+h)}^{(h)}) \ \Bigg\vert \ U_{1t}^{(h)} + U_{2t}^{(h)} \Bigg] \Bigg|.
\end{split}\end{equation}
By 3) and 4) above, for each $\mathcal{D}\in \mathcal{R}^{p_{h}}$, we have 
\begin{equation}\begin{split}\label{glp12}
& \textnormal{RHS of } (\ref{glp11})  = \Big|   \mathbb{E} \left[ \mu_{V_{21}^{(h)}} (\mathcal{D} - V_{1t}^{(h)}) \right]    -    \mathbb{E} \left[\mu_{V_{21}^{(h)}} (\mathcal{D} - U_{1(t+h)}^{(h)}) \ \Big\vert \ U_{1t}^{(h)} + U_{2t}^{(h)}\right]  \Big|.
\end{split}\end{equation}
Since $V_{1t}^{(h)}$ is an independent copy, it follows that 
\begin{equation}\begin{split}\label{glp16}
\textnormal{RHS of (\ref{glp12})}& = \Big|   \mathbb{E} \left[ \mu_{V_{21}^{(h)}} (\mathcal{D} - V_{1t}^{(h)}) - \mu_{V_{21}^{(h)}} (\mathcal{D} - U_{1(t+h)}^{(h)}) \ \Big\vert \ U_{1t}^{(h)} + U_{2t}^{(h)}\right]  \Big|.
\end{split}\end{equation}

Next, we separate the expectation according to events $Q$ and $G$ as  
\begin{equation}\begin{split}\label{glp17}
&\textnormal{RHS of (\ref{glp16})} \\
& = \Big|   \mathbb{E} \left[ \left( \boldsymbol{1}_{Q \cup G} + \boldsymbol{1}_{Q^{c}\cap G^{c}}  \right)\left(\mu_{V_{21}^{(h)}} (\mathcal{D} - V_{1t}^{(h)}) - \mu_{V_{21}^{(h)}} (\mathcal{D} - U_{1(t+h)}^{(h)})\right) \ \Big\vert \ U_{1t}^{(h)} + U_{2t}^{(h)} \right]  \Big|.
\end{split}\end{equation}
Then by (\ref{glp19}) and some simple calculations, we can show that 
\begin{equation}\begin{split}\label{glp18}
&\textnormal{RHS of (\ref{glp17})} \\
& \le \mathbb{P}( G\cup Q \ \vert \ U_{1t}^{(h)} + U_{2t}^{(h)}) +  \sup_{\norm{\Delta_{i}}_{\infty} < e^{-s_{3}h}}\Big|  \int_{\mathcal{D} - \Delta_{2}} \big(f_{h}(x+\Delta_{2} - \Delta_{1}) -   f_{h}(x)\big)dx \Big|.
\end{split}\end{equation}
For the first term on the RHS of (\ref{glp18}), it follows from the definitions of $Q$ and $G$ and the stationarity of $V_{1t}^{(h)}$ that 
\begin{equation}
\begin{split}\label{glp217}
& \mathbb{P}( G\cup Q \ \vert \ U_{1t}^{(h)} + U_{2t}^{(h)}) \le 
\mathbb{P}\left(\norm{V_{11}^{(h)}}_{\infty}\ge e^{(-s_{3}h)}\right) \\
&\quad+ \mathbb{P}\Big(\norm{U_{1(t+h)}^{(h)}}_{\infty}  \ge e^{(-s_{3}h)} \ | \ U_{1t}^{(h)} + U_{2t}^{(h)}\Big).
\end{split}
\end{equation}
For the second term on the RHS of (\ref{glp18}), it holds that 
\begin{equation}\begin{split}\label{glp215}
& \sup_{\norm{\Delta_{i}}_{\infty} < e^{-s_{3}h}}\Big|  \int_{\mathcal{D} - \Delta_{2}} \big(f_{h}(x+\Delta_{2} - \Delta_{1}) -   f_{h}(x)\big)dx \Big|\\ &  \le   2\mathbb{P}\left(\norm{V_{21}^{(h)}}_{\infty}\ge \exp{((1-\delta_{0})s_{3}h)} - 2\exp{(-s_{3}h)} \right) \\ 
& \quad +  \sup_{\norm{\Delta_{i}}_{\infty} < e^{-s_{3}h}}\Big|  \int_{x \in \mathcal{D} - \Delta_{2}, \,  \norm{x}_{\infty} < e^{(1 - \delta_{0})s_{3}h}} \big( f_{h}(x+\Delta_{2} - \Delta_{1}) -   f_{h}(x)\big)dx \Big|.
\end{split}\end{equation}

We proceed with dealing with the last term on the RHS of (\ref{glp215}). Let $x_{1}, x_{2}$ denote two vectors in $ \mathbb{R}^{p_{h}}$. Then, it follows from \eqref{glp6} that
\begin{equation}
\label{glp20}
\lim\sup_{h\uparrow\infty} \sup_{ \substack{\norm{x_{2}}_{2} \le 2\sqrt{p_{h}}e^{-s_{3}h} \\ \norm{x_{1}}_{2} \le \sqrt{p_{h}} e^{(1 - \delta_{0})s_{3}h } } } 
\norm{x_{1}}_{2}\norm{x_{2}}_{2}\underbar{c}^{-1} =0.
\end{equation}
Let us define $\Delta \coloneqq \Delta_{2} - \Delta_{1}$. In light of the fact that $\norm{z}_{2} \le \sqrt{p_{h}}\norm{z}_{\infty}$ for all $z\in \mathbb{R}^{p_{h}}$, (\ref{glp8}), (\ref{glp20}), and the fact that $\int_{x\in \mathbb{R}^{p_{h}} }f_{h}(x) dx = 1$, it holds that for all large $h$,
\begin{equation}\label{glp13}
\begin{split}
& \sup_{\norm{\Delta_{i}}_{\infty} < e^{-s_{3}h}}\Big|  \int_{\substack{x \in \mathcal{D} - \Delta_{2}, \,   \norm{x}_{\infty} < e^{(1 - \delta_{0})s_{3}h}} } \big(f_{h}(x+\Delta_{2} - \Delta_{1}) -   f_{h}(x)\big)dx \Big| \\
&\quad \le \sup_{\norm{\Delta}_{\infty} < 2 e^{-s_{3}h}, \norm{\Delta_{2}}_{\infty} <  e^{-s_{3}h}}\Big|  \int_{\substack{x \in \mathcal{D} - \Delta_{2}, \,   \norm{x}_{\infty} < e^{(1 - \delta_{0})s_{3}h}} } \big(f_{h}(x+\Delta ) -   f_{h}(x)\big)dx \Big| \\
& \quad \le \sup_{ \substack{\norm{\Delta}_{2} \le 2\sqrt{p_{h}}e^{-s_{3}h} \\ \norm{x}_{2} \le \sqrt{p_{h}} e^{(1 - \delta_{0})s_{3}h }}  }(2\norm{x}_{2}\norm{\Delta}_{2}\underbar{c}^{-1} + (2\norm{x}_{2}\norm{\Delta}_{2}\underbar{c}^{-1})^{2})\\
& \quad \le \sup_{ \substack{\norm{\Delta}_{2} \le 2\sqrt{p_{h}}e^{-s_{3}h} \\ \norm{x}_{2} \le \sqrt{p_{h}} e^{(1 - \delta_{0})s_{3}h }}  }4\norm{x}_{2}\norm{\Delta}_{2}\underbar{c}^{-1} \\
& \quad \le 8 \underbar{c}^{-1}p_{h}\exp{(-\delta_{0}s_{3}h)}.
\end{split}
\end{equation}
Therefore, combining (\ref{glp11})--(\ref{glp215}), (\ref{glp13}), and the stationarity of the process yields the desired conclusion. This completes the proof of Claim (\ref{glp7}).

\subsection{Lemma~\ref{airv} and its proof} \label{app_airv}

The theoretical foundation of our subsampling method is provided by  Lemma~\ref{airv}, which concerns the asymptotic independence of the $\beta$-mixing random vectors in each subsample. Since \eqref{M0} below for  Lemma~\ref{airv} involves stacking up stationary elements column-wise (there are $l$ elements in a row of matrices in \eqref{M0} below), it is unclear whether we can directly apply Lemma 4.1 of \citep{yu1994rates} to our setting. Thus, we provide our self-contained proof for Lemma~\ref{airv}. Our technical analysis of Lemma~\ref{airv} seems to be the first formal proof for results on the asymptotically independent blocks due to the $\beta$-mixing and subsampling. 


Consider a $p$-dimensional vector-valued stationary process $\{\boldsymbol{z}_{t}\}$. Let $n$, $\bar{n}$, $h$, and $l$ be positive integers such that 
\begin{equation}\label{nbar1}
\bar{n} = \sup\{s \in \mathbb{N}: s(l+h) - h\ \le n\} > 0.
\end{equation} 
We construct two $\bar{n} \times (l p)$ design matrices as 
\begin{equation}\begin{split}\label{M0}
& \boldsymbol{M}  \coloneqq\left(\begin{array}{c}
\boldsymbol{z}_{(1- 1)\times (l+h) + 1}^{\top}, \dots, \boldsymbol{z}_{  (2-1)\times (l + h) - h}^{\top} \\
\dots \\
\boldsymbol{z}_{(\bar{n} - 1 )\times (l+h)+1}^{\top}, \dots, \boldsymbol{z}_{   \bar{n}\times (l+h) - h }^{\top} 
\end{array}\right)\\  & \text{and } \  
\boldsymbol{M}^{\pi}  \coloneqq\left(\begin{array}{c}
(\boldsymbol{z}_{1}^{\pi_{1}})^{\top}, \dots, (\boldsymbol{z}_{l}^{\pi_{1}})^{\top} \\
\dots \\
(\boldsymbol{z}_{1}^{\pi_{\bar{n}}})^{\top}, \dots,  (\boldsymbol{z}_{l}^{\pi_{\bar{n}}})^{\top}
\end{array}\right),
\end{split}
\end{equation}
where $\{(\boldsymbol{z}_{1}^{\pi_{t}}, \dots, \boldsymbol{z}_{l}^{\pi_{t}})\}_{t}$ is an i.i.d. sequence such that $(\boldsymbol{z}_{1}^{\pi_{1}}, \dots, \boldsymbol{z}_{l}^{\pi_{1}})$ has the same distribution as $(\boldsymbol{z}_{1}, \dots, \boldsymbol{z}_{l})$.  Here, matrix $\boldsymbol{M}$ is obtained by removing $h$ random vectors in the process after each consecutive $l$ random vectors, and then stacking up the remaining random vectors. Lemma~\ref{airv} below characterizes the distributional distance between  $\boldsymbol{M}$ with dependent rows and $\boldsymbol{M}^{\pi}$ with i.i.d. rows. 

\begin{lemma}\label{airv}
Assume that the $p$-dimensional process $\{\boldsymbol{z}_{t}\}$ satisfies Condition~\ref{ge4} with $(h+1)$-step and constants $0 \le \rho < 1$ and $C_{0} > 0$. Then it holds that 
\begin{equation}\label{asym1}
\sup_{\mathcal{D}\in \mathcal{R}^{l p \bar{n} } }\Big| \mathbb{P} (\boldsymbol{M}\in \mathcal{D})  -  \mathbb{P} (\boldsymbol{M}^{\pi}\in \mathcal{D})\Big| \le n\rho^{h+1}C_{0},
\end{equation}
where random matrices $\boldsymbol{M}$ and $\boldsymbol{M}^{\pi}$ are defined in (\ref{M0}). 
\end{lemma}

\noindent\textit{Proof}. If we view the rows of $\boldsymbol{M}$ as random mappings, a simple version of this problem is to establish an upper bound of the total variation distance between the distributions of $U_{1}, \dots, U_{\bar{n}}$ and their i.i.d. counterparts denoted as  $U_{1}^{\pi}, \dots, U_{\bar{n}}^{\pi}$. The main technique used here is to separate the total variation distance into $\textnormal{TV}_{1}, \dots, \textnormal{TV}_{\bar{n}}$ introduced below and control them separately as 
\begin{equation}
\begin{split}
\label{airv9}
U_{1}, U_{2}, U_{3}, \dots, U_{\bar{n}} & \quad\underbrace{\longleftrightarrow}_{\textnormal{TV}_{1}}\quad
U_{1}^{\pi}, U_{2}, U_{3}, \dots, U_{\bar{n}} \quad\underbrace{\longleftrightarrow}_{\textnormal{TV}_{2}}\quad
U_{1}^{\pi}, U_{2}^{\pi}, U_{3}, \dots, U_{\bar{n}}\\
&\quad\underbrace{\longleftrightarrow}_{\textnormal{TV}_{3}} \quad\dots\quad \underbrace{\longleftrightarrow}_{\textnormal{TV}_{\bar{n}}} \quad U_{1}^{\pi}, U_{2}^{\pi}, \dots, U_{\bar{n}}^{\pi}.
\end{split}
\end{equation}
By the technique in (\ref{airv9}) above, for each step, we can focus on the total variation distance between two processes with only one distinct part. For example, for the $j$th and $(j+1)$th processes, the distinct part is $U_{j}$ and $U_{j}^{\pi}$. We will present the formal proof next.

To facilitate the technical presentation, let us first introduce some notations. Denote the distributions as 
\begin{equation}
\begin{split}
\label{notation1}
& \mu_{1} \coloneqq \mu_{(\boldsymbol{z}_{(1-1)(l+h) + 1 \ : \ 1(l+h) -h}, \dots, \boldsymbol{z}_{(i-1)(l+h) + 1 \ : \ i(l+h) -h}, \dots, \boldsymbol{z}_{(\bar{n}-1)(l+h) + 1 \ : \ \bar{n}(l+h) -h}  )},\\
& \mu_{\bar{n}+1} \coloneqq \mu_{(\boldsymbol{z}^{\pi_{1}}, \dots, \boldsymbol{z}^{\pi_{\bar{n}}})},\\
& \mu_{j} \coloneqq \mu_{ (\boldsymbol{z}^{\pi_{1}}, \dots, \boldsymbol{z}_{(j-1)(l+h) + 1 \ : \ j(l+h) -h}, \dots, \boldsymbol{z}_{(\bar{n}-1)(l+h) + 1 \ : \ \bar{n}(l+h) -h} ) }
\end{split}
\end{equation}
for $j = 2, \dots, \bar{n}$. Observe that we have $\mu_{\bar{n}} = \mu_{\bar{n}+1}$ since the process is stationary. With the notation introduced above, the desired conclusion is an upper bound for $\frac{1}{2}\norm{\mu_{1} - \mu_{\bar{n}+1}}_{TV}$.

For completeness, we state some important properties of the transition kernel $p: \mathbb{R}^{p}\times \mathcal{R}^{p} \longrightarrow [0, 1]$ of a stationary Markov chain with stationary distribution $\pi$. 1) For each integer $t$ and $\mathcal{D} \in \mathcal{R}^{p}$,  $p(\boldsymbol{z}_{t}, \mathcal{D})$ is a version of $\mathbb{P}(\boldsymbol{z}_{t+1} \in \mathcal{D}| \boldsymbol{z}_{t}) $. 
2) For each measurable function $f$ and $\mathcal{D}\in\mathcal{R}^{p}$, $\int_{\mathcal{D}} p(\vec{x}, d\vec{y}) f(\vec{y})$ is a measurable function of $\vec{x}$, and hence for each $\mathcal{D}_{k}\in\mathcal{R}^{p}$,
\[\int_{\mathcal{D}_{1}} \pi(d\vec{x}_{1}) \int_{\mathcal{D}_{2}} p (\vec{x}_{1}, d\vec{x}_{2}) \dots \int_{\mathcal{D}_{k}}p(\vec{x}_{k-1}, d\vec{x}_{k})f(\vec{x}_{k}) \]
is a well-defined integral. 3) For each measurable function $f$ and $\mathcal{D}\in\mathcal{R}^{p}$,
\begin{equation}
\label{airv8}
\int_{\mathbb{R}^{p}}\pi(d\vec{x})\int_{\mathcal{D}}p(d\vec{x}, d\vec{y})f(\vec{y}) = \int_{\mathcal{D}}\pi(d\vec{x})f(\vec{x}).
\end{equation} 4) We have an expression of $\mu_{j}$ as given in (\ref{airv1}) below, where we indicate each part of the distribution of $\mu_{j}$ according to
$$\underbrace{\boldsymbol{z}^{\pi_{1}}}_{\textnormal{1st part}}, \dots, \underbrace{\boldsymbol{z}_{(j-1)(l+h) + 1 \ : \ j(l+h) -h}}_{\textnormal{$j$th part}} , \dots, \underbrace{\boldsymbol{z}_{(\bar{n}-1)(l+h) + 1 \ : \ \bar{n}(l+h) -h}}_{\textnormal{$\bar{n}$th part}}. $$
Such representation follows from the first two properties, and the details on deriving it can be found in Section 5.2 of~\citep{Durrett2019}. 
For each $\mathcal{D}\in \mathcal{R}^{lp\bar{n}}$, $\mu_{j}(\mathcal{D})$ admits the following representation
\begin{equation}\begin{split}
\label{airv1}
& \mu_{j}(\mathcal{D}) = \int_{\mathcal{D}} \underbrace{\pi(d\vec{x}_{1})\times \dots\times p(\vec{x}_{l-1}, d\vec{x}_{l})}_{\textnormal{$1$st part}}\times \dots\\
& \hspace{1em}\times \underbrace{\pi(d\vec{x}_{(j-1)(l+h) +1 }) \times \dots\times p(\vec{x}_{j(l+h)-h-1}, d\vec{x}_{j(l+h)-h})}_{\textnormal{$j$th part}}\times \dots\\
& \hspace{1em} \times \underbrace{p^{h+1}(\vec{x}_{j(l+h) -h }, d\vec{x}_{j(l+h) +1 }) \times \dots\times p(\vec{x}_{(j+1)(l+h)-h-1}, d\vec{x}_{(j+1)(l+h)-h})}_{\textnormal{$(j+1)$th part}}\times \dots\\
& \hspace{1em} \times \underbrace{p^{h+1}(\vec{x}_{(\bar{n}-1)(l+h) -h }, d\vec{x}_{(\bar{n}-1)(l+h) +1 }) \times \dots\times p(\vec{x}_{\bar{n}(l+h)-h-1}, d\vec{x}_{\bar{n}(l+h)-h})}_{\bar{n}\textnormal{th part}},
\end{split}
\end{equation}
where $\underbrace{\vec{x}_{1}, \dots, \vec{x}_{l}}_{\textnormal{$1$st part}}, \dots, \underbrace{\vec{x}_{(\bar{n}-1)(l+h) +1 }, \dots, \vec{x}_{\bar{n}(l+h)-h}}_{\bar{n}\textnormal{th part}}$ stand for the corresponding running variables with  $\vec{x}_{k}\in\mathbb{R}^{p}$ for each $k$. 

Let us make use of a critical observation that 
\begin{equation}
\label{airv2}
\frac{1}{2}\norm{\mu_{1} - \mu_{\bar{n} + 1}}_{TV}\le \sum_{i=1}^{\bar{n} }\sup_{\mathcal{D}\in\mathcal{R}^{lp\bar{n}}}|\mu_{j}(\mathcal{D}) - \mu_{j+1}(\mathcal{D})|.
\end{equation}
We will bound each term in the above summation separately. Let us fix $1\le j \le \bar{n}$. We notice that $\mu_{j}$ and $\mu_{j+1}$ are almost identical except for the $(j+1)$th part in (\ref{airv1}). By a careful comparison, we see that for $\mu_{j}$, the $(j+1)$th part starts with $p^{h+1}(\vec{x}_{j(l+h) -h }, d\vec{x}_{j(l+h) +1 })$, whereas the $(j+1)$th part of $\mu_{j+1}$ starts with $\pi(d\vec{x}_{j(l+h) +1 })$. To see the difficulty for bounding each term on the right-hand side (RHS) of (\ref{airv2}) using such observation, note that
$\int_{\mathcal{D}} |\mu_{1} (dx) - \mu_{2}(dx)|$ is not a well-defined integral for a Borel set $\mathcal{D}$ and two measures $\mu_{1}$ and $\mu_{2}$ since there are two $dx$'s inside the integration. To have a valid argument for this bound, we use the Radon--Nikodym theorem~\citep{Durrett2019} to replace the underlying measures with measurable functions (the Radon--Nikodym derivatives). The arguments follow mainly those for the proof of Lemma~\ref{sup1} in Section \ref{SecA.6}.

By Condition~\ref{ge4}, for each $\vec{x}\in\mathbb{R}^{p}$ it holds that $p(\vec{x}, \cdot)$ is dominated by the Lebesgue measure. Since $p$ is the transition kernel of the stationary Markov chain, this entails that 1) $p^{h+1}(\vec{x}, \cdot)$ is dominated by the Lebesgue measure for each $\vec{x}\in\mathbb{R}^{p}$ and 2) $\pi(\cdot)$ is also dominated by the Lebesgue measure. By 1) and the Radon--Nikodym Theorem, there exists a nonnegative measurable function on $\mathbb{R}^{2p}$, which is denoted as $r$, such that for each $\vec{x}\in\mathbb{R}^{p}$ and $\mathcal{D}\in \mathcal{R}^{p}$,
\[p^{h+1}(\vec{x}, \mathcal{D} ) = \int_{\mathcal{D}}  r(\vec{x}, \vec{y} )\ d\vec{y}.\]
This measurable function is simply the Radon--Nikodym derivative~\citep{Durrett2019}, and $r(\vec{x}, \vec{y} )$ is also called the probability density functions of $\boldsymbol{z}_{t+h+1}$ conditional on $\boldsymbol{z}_{t}$. In particular, for each $\mathcal{D}_{1}, \mathcal{D}_{2}\in\mathcal{R}^{p}$, we have that 
\[\mathbb{P}( (\boldsymbol{z}_{t}, \boldsymbol{z}_{t+h+1} ) \in \mathcal{D}_{1}\times\mathcal{D}_{2} ) = \int_{\vec{x} \in \mathcal{D}_{1}} \pi(d\vec{x}) \int_{\vec{y} \in \mathcal{D}_{2}} r(\vec{x}, \vec{y}) d\vec{y}.\]

For more details on the conditional probability density functions, see Example 4.1.6 of \citep{Durrett2019}. Furthermore, by 2) we denote by $r_{\pi}(\vec{x})$ the Radon--Nikodym derivative such that for each $\mathcal{D}\in\mathcal{R}^{p}$,
\[\pi(\mathcal{D}) = \int_{\mathcal{D}} r_{\pi}(\vec{x}) d\vec{x}.\]
Thus, we can obtain that 
\begin{equation}\label{airv3}
\begin{split}
& \mu_{j}(\mathcal{D}) = \int_{\mathcal{D}} \dots \\
&\quad \times \underbrace{r(\vec{x}_{j(l+h)-h}, \vec{x}_{j(l+h)+1} ) d\vec{x}_{j(l+h) +1}  \times \dots\times p(\vec{x}_{(j+1)(l+h)-h-1}, d\vec{x}_{(j+1)(l+h)-h})}_{\textnormal{$(j+1)$th part}}\\
&\quad \times \dots\times \underbrace{p^{h+1}(\vec{x}_{(\bar{n}-1)(l+h) -h }, d\vec{x}_{(\bar{n}-1)(l+h) +1 }) \times \dots\times p(\vec{x}_{\bar{n}(l+h)-h-1}, d\vec{x}_{\bar{n}(l+h)-h})}_{\bar{n}\textnormal{th part}}.
\end{split}
\end{equation}
A similar expression also holds for $\mu_{j+1}$ with $r_{\pi}$. We will bound $|\mu_{j}(\mathcal{D}) -  \mu_{j+1}(\mathcal{D})|$ next.

It follows from (\ref{airv3}) and the fact of $p\le 1$ that for each $\mathcal{D}\in\mathcal{R}^{lp\bar{n}}$,
\begin{equation}\begin{split}
\label{airv4}
|&\mu_{j}(\mathcal{D}) -  \mu_{j+1}(\mathcal{D})|  \\
& =\left| \int_{\mathcal{D}} \underbrace{\cdots}_{1,\dots, j} \times \left(r(\vec{x}_{j(l+h)-h}, \vec{x}_{j(l+h)+1} ) - r_{\pi}(\vec{x}_{j(l+ h) + 1} ) \right) d\vec{x}_{j(l+h) +1 }\times \cdots\right|\\
&\le \int_{\mathcal{D}} \underbrace{\cdots}_{1,\dots, j} \times \left| r(\vec{x}_{j(l+h)-h}, \vec{x}_{j(l+h)+1} ) - r_{\pi}(\vec{x}_{j(l+ h) + 1} ) \right| d\vec{x}_{j(l+h) +1 } \times \cdots\\
& \le \int_{\mathbb{R}^{lpj + p}}\underbrace{\cdots}_{1\textnormal{st},\dots, j\textnormal{th parts}} \times \left| r(\vec{x}_{j(l+h)-h}, \vec{x}_{j(l+h)+1} ) - r_{\pi}(\vec{x}_{j(l+ h) + 1} ) \right| d\vec{x}_{j(l+h) +1 }.
\end{split}
\end{equation}
To bound the RHS of  (\ref{airv4}), we separate the modulus into positive and negative parts and get rid of the modulus operation. Let $\mathcal{D}_{+}$ and $\mathcal{D}_{-}$ be two disjoint Borel sets such that 
\begin{equation}
\begin{split}
\label{airv5}
&\int_{\mathbb{R}^{lpj + p}}\underbrace{\cdots}_{1,\dots, j} \times \left| r(\vec{x}_{j(l+h)-h}, \vec{x}_{j(l+h)+1} ) - r_{\pi}(\vec{x}_{j(l+ h) + 1} ) \right| d\vec{x}_{j(l+h) +1 } 
\\ &  = \int_{\mathcal{D}_{+} }\underbrace{\cdots}_{1,\dots, j} \times \left( r(\vec{x}_{j(l+h)-h}, \vec{x}_{j(l+h)+1} ) - r_{\pi}(\vec{x}_{j(l+ h) + 1} ) \right) d\vec{x}_{j(l+h) +1 } \\
& \quad+ \int_{\mathcal{D}_{-} }\underbrace{\cdots}_{1\textnormal{st},\dots, j\textnormal{th parts}} \times \left( r_{\pi}(\vec{x}_{j(l+ h) + 1} ) - r(\vec{x}_{j(l+h)-h}, \vec{x}_{j(l+h)+1} )  \right) d\vec{x}_{j(l+h) +1 }.
\end{split}
\end{equation}

To proceed, we exploit arguments involving ``cross-sections.'' For any $\mathcal{D} \in \mathbb{R}^{k_{1}+ k_{2}}$ and $\vec{x}\in \mathbb{R}^{k_{1}}$ with $k_{1}$ and $k_{2}$ some positive integers, let us define the cross-section at $\vec{x}$ as $\mathcal{D}_{\vec{x}} \coloneqq \{\vec{y} : (\vec{x}, \vec{y}) \in \mathcal{D}\}$. See Section~\ref{SecA.3} for more detail on cross-sections. Then it holds that 
\begin{equation}\begin{split}
\label{airv6}
& \int_{\mathcal{D}_{+} }\underbrace{\cdots}_{1,\dots, j} \times \left(r(\vec{x}_{j(l+h)-h}, \vec{x}_{j(l+h)+1} ) - r_{\pi}(\vec{x}_{j(l+ h) + 1} ) \right) d\vec{x}_{j(l+h) +1 } \\
& = \int_{\mathcal{D}_{+} }\underbrace{\cdots}_{1,\dots, j} \times  p^{h+1}(\vec{x}_{j(l+h) -h }, d\vec{x}_{j(l+h) + 1 } ) -  \int_{\mathcal{D}_{+} }\underbrace{\cdots}_{1,\dots, j} \times  \pi(d\vec{x}_{j(l+h) + 1 }) \\
& = \int_{\mathbb{R}^{lpj}}\underbrace{\cdots}_{1,\dots, j} \times  p^{h+1}(\vec{x}_{j(l+h) -h }, (\mathcal{D}_{+})_{z}) -  \int_{\mathbb{R}^{lpj} }\underbrace{\cdots}_{1,\dots, j} \times  \pi((\mathcal{D}_{+})_{z}) \\	
& = \int_{\mathbb{R}^{lpj} }\underbrace{\cdots}_{1\textnormal{st},\dots, j\textnormal{th parts}} \times \left( p^{h+1}(\vec{x}_{j(l+h) -h }, (\mathcal{D}_{+})_{z} ) - \pi((\mathcal{D}_{+})_{z}) \right),
\end{split}
\end{equation}
where $z$ represents $(\vec{x}_{1}^{\top}, \dots, \vec{x}_{j(l+h)-h}^{\top})^{\top}$ in the integration. Here, we have used the definition of the Radon--Nikodym derivative to get the first equality in (\ref{airv6}). The second equality in (\ref{airv6}) is justified by the fact that $\pi$ is a distribution and hence a transition kernel, and an application of Lemma~\ref{le4} in Section~\ref{SecA.4}. To apply Lemma~\ref{le4} in (\ref{airv6}), we can regard $\boldsymbol{z}_{j(l+h) + 1}$ as $\boldsymbol{X}_{3}$, $\boldsymbol{z}_{j(l+h) - h}$ as $\boldsymbol{X}_{2}$, and the remaining variables as $\boldsymbol{X}_{1}$ in Lemma~\ref{le4}, and notice that (\ref{le41}) is satisfied due to the Markov property. Similar arguments can be applied to  $\mathcal{D}_{-}$ too. 

In view of (\ref{airv5}) and (\ref{airv6}), it follows from the fact of $\mathcal{D}_{+}\cap \mathcal{D}_{-} = \emptyset$,  (\ref{airv8}), and  Condition~\ref{ge4} that 
\begin{equation}
\begin{split}
\label{airv7}
\textnormal{RHS of } (\ref{airv4}) 
&= \int_{\mathbb{R}^{lpj} }\underbrace{\cdots}_{1,\dots, j} \times \Big[\left( p^{h+1}(\vec{x}_{j(l+h) -h }, (\mathcal{D}_{+})_{z} ) - \pi((\mathcal{D}_{+})_{z}) \right) \\
&\quad- \left( p^{h+1}(\vec{x}_{j(l+h) -h }, (\mathcal{D}_{-})_{z} ) - \pi((\mathcal{D}_{-})_{z}) \right)\Big]\\
& \le \int_{\mathbb{R}^{lpj} }\underbrace{\cdots}_{1\textnormal{st},\dots, j\textnormal{th parts}} \times \norm{p^{h+1}(\vec{x}_{j(l+h) -h }, \cdot ) - \pi(\cdot) }_{TV}\\
& = \int_{\mathbb{R}^{p}} \pi(d\vec{x}_{j(l+h)-h}) \norm{p^{h+1}(\vec{x}_{j(l+h) -h }, \cdot ) - \pi(\cdot) }_{TV}\\
& \le \int_{\mathbb{R}^{p}}V(\vec{x}) \pi(d\vec{x}) \rho^{h+1}C,
\end{split}
\end{equation}
where $C>0$ is given in Condition~\ref{ge4}.  By Condition~\ref{ge4}, we can further show that 
\[\textnormal{RHS of } (\ref{airv7})\le  C_{0}\rho^{h + 1}, \] 
where $C_{0} > 0$ is a constant such that  $\int_{\mathbb{R}^{p}}V(\vec{x})\pi(d\vec{x})C\le C_{0}$. Therefore, combining (\ref{airv2}), (\ref{airv7}), and the fact of $\bar{n} \le n$, we can see that the upper bound is given by $n\rho^{h+1}C_{0}$, which concludes the proof of Lemma~\ref{airv}.

\subsection{Lemma \ref{le3} and its proof} \label{SecA.3}

To facilitate the technical presentation, let us first introduce some necessary notation. For any $\mathcal{D} \subset \mathbb{R}^{k_{1}+ k_{2}}$ and $x\in \mathbb{R}^{k_{1}}$ with $k_{1}$ and $k_{2}$ some positive integers, we define the cross-section at $x$ as $\mathcal{D}_{x} \coloneqq \{y : (x, y) \in \mathcal{D}\}$. A standard operation on $\mathcal{D}_{x}$ is described as follows. If $\mathcal{D}_{1}\subset\mathcal{D}_{2}$, then we have that for each $x$, $(\mathcal{D}_{1})_{x} \subset (\mathcal{D}_{2})_{x}$ and $(\mathcal{D}_{2}\backslash \mathcal{D}_{1})_{x} = (\mathcal{D}_{2})_{x}\backslash(\mathcal{D}_{1})_{x}$. In addition, for any set $\mathcal{D}\subset \mathbb{R}^{k}$ and $x\in\mathbb{R}^{k}$, we denote by $\mathcal{D} - x$ the set $\{y - x: y \in \mathcal{D}\}$. The expectation $\int f(x)\mu(dx)$ for a measurable function $f$ with respect to some random vector $\boldsymbol{X}$ with distribution $\mu$ is written as $\int_{\mathbb{R}^{k}} fd\mu$ whenever the running variables are obvious.  We also use the product notation in the integration to specify the running variables such as $\int f(x_{2}) \mu( dx_{1} \times dx_{2})$.

\begin{lemma}
\label{le3}
Let $\boldsymbol{X}$ and $\boldsymbol{Y}$ be $k_{1}$-dimensional and $k_{2}$-dimensional random vectors, respectively. Assume that $h : \mathbb{R}^{k_{1}} \times \mathcal{R}^{k_{2}} \longrightarrow [0, 1]
$ is a transition kernel such that for each $\mathcal{D}\in\mathcal{R}^{k_{2}}$, $h(\boldsymbol{X}, \mathcal{D})$ is a version of $\mathbb{P}( \boldsymbol{Y} \in\mathcal{D}\ \vert\ \boldsymbol{X})$. Then it holds that 
\begin{enumerate}
\item[1)] For each $\mathcal{D} \in \mathcal{R}^{k_{1} + k_{2}}$ and $x \in \mathbb{R}^{k_{1}}$, $\mathcal{D}_{x}\in\mathcal{R}^{k_{2}}$.
\item[2)] For each $\mathcal{D} \in\mathcal{R}^{k_{1} + k_{2}}$, $h(\cdot, \mathcal{D}_{\cdot})$ is $\mathcal{R}^{k_{1}}$-measurable.
\item[3)] For each $\mathcal{D} \in \mathcal{R}^{k_{1}+k_{2}}$, $\mathbb{P}( (\boldsymbol{X}^{\top}, \boldsymbol{Y}^{\top}) \in \mathcal{D}) = \int_{\mathbb{R}^{k_{1}}} \mu_{\boldsymbol{X}}(dx) h(x, \mathcal{D}_{x})$.
\end{enumerate}
\end{lemma}

\noindent \textit{Proof}. We begin with showing part 1). Let $L$ be the collection of sets in $\mathcal{R}^{k_{1} + k_{2}}$ satisfying the required conditions; that is, for each $\mathcal{D} \in L$, it holds that for each $x \in\mathbb{R}^{k_{1}}$, $\mathcal{D}_{x}\in\mathcal{R}^{k_{2}}$. Then it is easy to verify that $L$ contains all the rectangles of form $\mathcal{A}\times \mathcal{B}$, where $\mathcal{A}\in \mathcal{R}^{k_{1}}$ and $\mathcal{B}\in\mathcal{R}^{k_{2}}$. By the basic set operations, it holds that for each $x\in\mathbb{R}^{k_{1}}$ and $E, E_{i} \in \mathcal{R}^{k_{1}+k_{2}}$,
\begin{enumerate}
\item[a)] $(E_{x})^{c} = (\{ y :  (x, y) \in E\})^{c} = \{y : (x, y) \in E^{c}\} = (E^{c})_{x}$;
\item[b)] $\cup_{i} (E_{i})_{x} = \cap_{i} ((E_{i})_{x})^{c} = \cap_{i}(E_{i}^{c})_{x} = (\cap_{i}E_{i}^{c})_{x} = (\cup_{i}E_{i})_{x}$.
\end{enumerate}
Thus, for $E, E_{i} \in L$, we have that $E^{c}, \cup_{i}E_{i} \in L$. This shows that $L$ is a $\sigma$-algebra. Since $L$ contains all the rectangles, we obtain the conclusion in part 1) of Lemma~\ref{le3}.

We next proceed to establish part 2). Since $\mathcal{D}_{x}$ is a measurable set, $h(x, \mathcal{D}_{x})$ is a well-defined function of $x$. 
Let $L$ be the collection of sets such that for each $\mathcal{D}\in L$, $h(\cdot, \mathcal{D}_{\cdot})$ is $\mathcal{R}^{k_{1}}$-measurable. Since for each $\mathcal{A}\in\mathcal{R}^{k_{1}}$ and $\mathcal{B}\in\mathcal{R}^{k_{2}}$, it holds that \[ h(x, (\mathcal{A} \times \mathcal{B})_{x}) = h(x, \mathcal{B})\boldsymbol{1}_{\{x\in\mathcal{A}\}}, \] which is a measurable function of $x$, we can see that $L$ contains all such rectangles. Moreover, if $\mathcal{D}_{1}, \mathcal{D}_{2} \in L$ with $\mathcal{D}_{1}\subset \mathcal{D}_{2}$, then it follows that for each $x$, $(\mathcal{D}_{1})_{x} \subset (\mathcal{D}_{2})_{x}$ and $(\mathcal{D}_{2}\backslash\mathcal{D}_{1})_{x} = (\mathcal{D}_{2})_{x}\backslash(\mathcal{D}_{1})_{x}$, and hence 
\[ h(x, (\mathcal{D}_{2}\backslash\mathcal{D}_{1})_{x}) = h(x, (\mathcal{D}_{2})_{x}\backslash(\mathcal{D}_{1})_{x}) = h(x, (\mathcal{D}_{2})_{x}) - h(x, (\mathcal{D}_{1})_{x}).\]
Observe that the RHS of the equality above is measurable since the subtraction of measurable functions is still measurable. Next, if $\mathcal{D}_{i} \in L $ and $\mathcal{D}_{i}\subset \mathcal{D}_{i+1}$, by the continuity of measure, we have that for each $x\in\mathbb{R}^{k_{1}}$, $\lim_{n\uparrow \infty}h(x, (\cup_{i=1}^{n}\mathcal{D}_{i})_{x}) = h(x, (\cup_{i=1}^{\infty}\mathcal{D}_{i})_{x})$. Thus, $h(x, (\cup_{i=1}^{\infty}\mathcal{D}_{i})_{x})$ is a measurable function of $x$,  and we have  $\cup_{i=1}^{\infty}\mathcal{D}_{i}\in L$. This shows that $L$ is a $\lambda$-system containing the set of all the rectangles. Hence, by Lemma~\ref{pilambda} in Section \ref{SecA.2}, we see that $L$ contains the $\sigma$-algebra generated by the set, which concludes the proof for part 2) of Lemma~\ref{le3}.

Finally, let us show part 3). Note that the RHS of the assertion is well-defined due to part 2) of Lemma~\ref{le3}. By the definition of the conditional expectation, the change of variables formula, and the fact that for each $x\in \mathbb{R}^{k_{1}}$, $\mathcal{A}\in\mathcal{R}^{k_{1}}$, $\mathcal{B} \in \mathcal{R}^{k_{2}}$,
\[\boldsymbol{1}_{\{x\in\mathcal{A}\}} h(x, \mathcal{B}) = h(x, (\mathcal{A}\times \mathcal{B})_{x}), \]
we can deduce that 
\begin{equation}\begin{split}\label{le31}
\mathbb{P}((\boldsymbol{X}^{\top}, \boldsymbol{Y}^{\top}) \in \mathcal{A} \times \mathcal{B}) & = \int_{\Omega} \boldsymbol{1}_{\{\boldsymbol{X}\in\mathcal{A}\}} \ \mathbb{P} (\boldsymbol{Y}\in\mathcal{B} \ \vert \ \boldsymbol{X}) d\mathbb{P}\\
& = \int_{\Omega} \boldsymbol{1}_{\{\boldsymbol{X} \in \mathcal{A}\}} \ h(\boldsymbol{X}, \mathcal{B}) d\mathbb{P}\\
& = \int_{\mathbb{R}^{k_{1}}} \mu_{\boldsymbol{X}}(dx)\ \boldsymbol{1}_{\{x\in\mathcal{A}\}} \ h(x, \mathcal{B})\\
& =\int_{\mathbb{R}^{k_{1}}} \mu_{\boldsymbol{X}}(dx)\ h(x, (\mathcal{A}\times\mathcal{B} )_{x}),
\end{split}	
\end{equation}
where $\Omega$ represents the underlying probability space.

Denote by $L$ the collection of sets in $\mathcal{R}^{k_{1} + k_{2}}$ satisfying the required condition; that is, for each $\mathcal{D} \in L$, it holds that $\mathbb{P}((\boldsymbol{X}^{\top}, \boldsymbol{Y}^{\top}) \in \mathcal{D})=\int_{\mathbb{R}^{k_{1}}} \mu_{\boldsymbol{X}}(dx)\ h(x, (\mathcal{D} )_{x})$. In view of (\ref{le31}), $L$ contains all the rectangles in $\mathbb{R}^{k_{1} + k_{2}}$. In addition, we will make use of the following two facts.
\begin{enumerate}
\item[a)] If $\mathcal{D}_{1}, \mathcal{D}_{2}\in L$ and $\mathcal{D}_{1}\subset \mathcal{D}_{2}$, then an application of similar arguments to those in the proof for part 2) of Lemma~\ref{le3} leads to 
\begin{equation*}\begin{split}
\mathbb{P}( (\boldsymbol{X}^{\top} , \boldsymbol{Y}^{\top}) \in \mathcal{D}_{2} \backslash\mathcal{D}_{1} ) & = \mathbb{P}( (\boldsymbol{X}^{\top} , \boldsymbol{Y}^{\top}) \in \mathcal{D}_{2}  ) - \mathbb{P}( (\boldsymbol{X}^{\top} , \boldsymbol{Y}^{\top}) \in\mathcal{D}_{1} )\\
& = \int_{\mathbb{R}^{k_{1}}}\mu_{\boldsymbol{X}}(dx) \ \Big( h(x, (\mathcal{D}_{2})_{x}) - h(x, (\mathcal{D}_{1})_{x}) \Big)\\
& = \int_{\mathbb{R}^{k_{1}}}\mu_{\boldsymbol{X}}(dx) \  h(x, (\mathcal{D}_{2})_{x} \backslash (\mathcal{D}_{1})_{x})\\
& = \int_{\mathbb{R}^{k_{1}}}\mu_{\boldsymbol{X}}(dx) \  h(x, (\mathcal{D}_{2} \backslash \mathcal{D}_{1})_{x} ),
\end{split}		
\end{equation*}
which shows that $\mathcal{D}_{2} \backslash \mathcal{D}_{1} \in L$.
\item[b)] Assume that $\mathcal{D}_{i} \in L $ for each $n$ and $\mathcal{D}_{i} \subset \mathcal{D}_{i+1}$. Then it follows from the continuity of measure, the definition of $L$, and the monotone convergence theorem that 
\begin{equation*}
\begin{split}
\mathbb{P}( (\boldsymbol{X}^{\top} , \boldsymbol{Y}^{\top}) \in \cup_{i=1}^{\infty}\mathcal{D}_{i}  ) & = \lim_{n\uparrow \infty} \mathbb{P}( (\boldsymbol{X}^{\top} , \boldsymbol{Y}^{\top}) \in \mathcal{D}_{n}  ) \\
&  = \lim_{n\uparrow \infty} \int_{\mathbb{R}^{k_{1}}} \mu_{\boldsymbol{X} }(dx) \ h(x, (\mathcal{D}_{n})_{x} )\\
& =  \int_{\mathbb{R}^{k_{1}}} \mu_{\boldsymbol{X} }(dx) \ \lim_{n\uparrow \infty} h(x, (\mathcal{D}_{n})_{x} )\\
& = \int_{\mathbb{R}^{k_{1}}} \mu_{\boldsymbol{X} }(dx) \  h(x, (\cup_{i=1}^{\infty} \mathcal{D}_{i})_{x}),
\end{split}
\end{equation*}
which shows that  $\cup_{i=1}^{\infty} D_{i}\in L$.
\end{enumerate}
Therefore, using the aforementioned facts, an application of Lemma~\ref{pilambda} leads to the conclusion in part 3) of Lemma~\ref{le3}. This completes the proof of Lemma~\ref{le3}.

\subsection{Lemma \ref{le4} and its proof} \label{SecA.4}

\begin{lemma}
\label{le4}
Let $\boldsymbol{X}_{1}$, $\boldsymbol{X}_{2}$, and $\boldsymbol{X}_{3}$ be $k_{1}$-dimensional, $k_{2}$-dimensional, and $k_{3}$-dimensional random vectors, respectively, such that for each $\mathcal{D}\in \mathcal{R}^{k_{3}}$,
\begin{equation}
\label{le41}\mathbb{P}(\boldsymbol{X}_{3} \in \mathcal{D} \ \vert \ \boldsymbol{X}_{2})  = \mathbb{P}(\boldsymbol{X}_{3} \in \mathcal{D} \ \vert \ \boldsymbol{X}_{2}, \boldsymbol{X}_{1}).
\end{equation}
We define a transition kernel $h: \mathbb{R}^{k_{2}} \times \mathcal{R}^{k_{3}} \longrightarrow[0, 1]
$ such that for each $\mathcal{D}\in\mathcal{R}^{k_{3}}$, $h(\boldsymbol{X}_{2}, \mathcal{D})$ is a version of $\mathbb{P}(\boldsymbol{X}_{3} \in \mathcal{D} \ \vert \ \boldsymbol{X}_{2})$. Then it holds that 
\begin{enumerate}
\item[1)] For each $\mathcal{D} \in \mathcal{R}^{k_{3}}$, $h(\boldsymbol{X}_{2}, \mathcal{D})$ is a version of $\mathbb{P}(\boldsymbol{X}_{3}\in\mathcal{D} \ \vert \ \boldsymbol{X}_{2}, \boldsymbol{X}_{1} )$.
\item[2)] For each $\mathcal{D}\in\mathcal{R}^{k_{1} + k_{2} + k_{3}}$,
$h(\boldsymbol{X}_{2}, \mathcal{D}_{(\boldsymbol{X}_{1}, \boldsymbol{X}_{2} )} ) \textnormal{ is a version of } \mathbb{P}( (\boldsymbol{X}_{1} , \boldsymbol{X}_{2}, \boldsymbol{X}_{3}) \in\mathcal{D} \ \vert \ \boldsymbol{X}_{2},\\ \boldsymbol{X}_{1})$.
\end{enumerate}
\end{lemma}

\noindent \textit{Proof}. We first show part 1). Since $h(\boldsymbol{X}_{2}, \mathcal{D})$ is a version of $\mathbb{P}(\boldsymbol{X}_{3} \in \mathcal{D} \ \vert \ \boldsymbol{X}_{2})$, $h(\boldsymbol{X}_{2}, \mathcal{D})$ is $\sigma(\boldsymbol{X}_{2})$-measurable, and hence $\sigma(\boldsymbol{X}_{1}, \boldsymbol{X}_{2})$-measurable. In conjunction with (\ref{le41}) and the definition of conditional expectation, we see that $h(\boldsymbol{X}_{2}, \mathcal{D})$ a version of $\mathbb{P}(\boldsymbol{X}_{3}\in\mathcal{D} \ \vert \ \boldsymbol{X}_{2}, \boldsymbol{X}_{1} )$.  This yields the conclusion in part 1) of Lemma~\ref{le4}. We then establish part 2). Let us first verify that $h(\boldsymbol{X}_{2}, \mathcal{D}_{\boldsymbol{X}_{1}, \boldsymbol{X}_{2}})$ is $\sigma(\boldsymbol{X}_{1}, \boldsymbol{X}_{2})$-measurable for each $\mathcal{D}\in\mathcal{R}^{k_{1} + k_{2} + k_{3}}$. We start with an observation that for each $\mathcal{D}_{1} \in \mathcal{R}^{k_{1}}$, $\mathcal{D}_{2}\in\mathcal{R}^{k_{2}}$, $\mathcal{D}_{3} \in \mathcal{R}^{k_{3}}$, it holds that 
\begin{equation*}
h(\boldsymbol{X}_{2}, (\mathcal{D}_{1}\times \mathcal{D}_{2} \times \mathcal{D}_{3})_{\boldsymbol{X}_{1}, \boldsymbol{X}_{2}} ) = h(\boldsymbol{X}_{2} ,\mathcal{D}_{3}) \boldsymbol{1}_{\{\boldsymbol{X}_{1} \in \mathcal{D}_{1}\}} \boldsymbol{1}_{\{\boldsymbol{X}_{2} \in \mathcal{D}_{2}\}},
\end{equation*}
which is $\sigma(\boldsymbol{X}_{1}, \boldsymbol{X}_{2})$-measurable. This shows that for each Borel rectangle $\mathcal{D}$, $h(\boldsymbol{X}_{2}, \mathcal{D}_{\boldsymbol{X}_{1}, \boldsymbol{X}_{2}})$ is $\sigma(\boldsymbol{X}_{1}, \boldsymbol{X}_{2})$-measurable. In conjunction with similar arguments to those in the proof of Lemma~\ref{le3} in Section \ref{SecA.3}, the desired result follows.

Let  $\mathcal{D}\in\mathcal{R}^{k_{1} + k_{2} + k_{3}}$ be given. Then we will show that for each $\mathcal{B} \in \mathcal{R}^{k_{1} + k_{2}}$, 
\begin{equation}
\begin{split}\label{le42}
& \int h(\boldsymbol{X}_{2} , \mathcal{D}_{\boldsymbol{X}_{1}, \boldsymbol{X}_{2}} ) \ \boldsymbol{1}_{\{(\boldsymbol{X}_{1}, \boldsymbol{X}_{2}) \in \mathcal{B} \}}  \ d\mathbb{P} \\
& = \int \mathbb{P} ( (\boldsymbol{X}_{1}, \boldsymbol{X}_{2}, \boldsymbol{X}_{3}) \in \mathcal{D}\ \vert \ \boldsymbol{X}_{2}, \boldsymbol{X}_{1})  \ \boldsymbol{1}_{\{(\boldsymbol{X}_{1}, \boldsymbol{X}_{2}) \in \mathcal{B} \}}  \ d\mathbb{P}.
\end{split}
\end{equation}
The left-hand side of (\ref{le42}) is well-defined since $h(\boldsymbol{X}_{2} , \mathcal{D}_{\boldsymbol{X}_{1}, \boldsymbol{X}_{2}} )$ is measurable. To show (\ref{le42}), we again apply similar arguments to those in the proof of Lemma~\ref{le3}. Specifically, let $L$ be the collection of sets in $\mathcal{R}^{k_{1} + k_{2} + k_{3}}$ such that (\ref{le42}) holds. Since for each $\mathcal{D}_{1} \in \mathcal{R}^{k_{1}}$, $\mathcal{D}_{2}\in\mathcal{R}^{k_{2}}$, $\mathcal{D}_{3}\in\mathcal{R}^{k_{3}}$, we have 
\begin{equation*}
\begin{split}
& \mathbb{P}((\boldsymbol{X}_{1}, \boldsymbol{X}_{2}, \boldsymbol{X}_{3}) \in (\mathcal{D}_{1} \times \mathcal{D}_{2} \times \mathcal{D}_{3}) \ \vert \ \boldsymbol{X}_{2}, \boldsymbol{X}_{1}) \\ 
& =  \mathbb{P}(\boldsymbol{X}_{3}\in\mathcal{D}_{3}  \ \vert \ \boldsymbol{X}_{2}, \boldsymbol{X}_{1}) \boldsymbol{1}_{\{\boldsymbol{X}_{1} \in\mathcal{D}_{1}\}} \boldsymbol{1}_{\{\boldsymbol{X}_{2} \in\mathcal{D}_{2}\}}\\
& =  h(\boldsymbol{X}_{2}, \mathcal{D}_{3})\boldsymbol{1}_{\{\boldsymbol{X}_{1} \in\mathcal{D}_{1}\}} \boldsymbol{1}_{\{\boldsymbol{X}_{2} \in\mathcal{D}_{2}\}} \\
& = h(\boldsymbol{X}_{2}, (\mathcal{D}_{1} \times \mathcal{D}_{2} \times \mathcal{D}_{3})_{\boldsymbol{X}_{1}, \boldsymbol{X}_{2}}),
\end{split}
\end{equation*}
it holds that $L$ contains all Borel rectangles $\mathcal{D}\in\mathcal{R}^{k_{1} + k_{2} + k_{3}}$. The remaining arguments follow those in the proof of Lemma~\ref{le3}. Finally, since $h(\boldsymbol{X}_{2}, \mathcal{D}_{\boldsymbol{X}_{1}, \boldsymbol{X}_{2}} )$ is $\sigma(\boldsymbol{X}_{1}, \boldsymbol{X}_{2})$-measurable, by (\ref{le42}) and the definition of conditional expectation, we can obtain the conclusion in part 2) of Lemma~\ref{le4}. This concludes the proof of Lemma~\ref{le4}.

\subsection{Lemma \ref{le1} and its proof} \label{SecA.5}

\begin{lemma} \label{le1}
Let $\{\boldsymbol{U}_{i}\}$ and $\{\boldsymbol{V}_{i}\}$ be sequences of $k_{1}$-dimensional and $k_{2}$-dimensional random vectors, respectively. Assume that  $(\boldsymbol{U}_{i}, \boldsymbol{V}_{i})$'s are identically distributed. Then there exists a transition kernel $g: \mathbb{R}^{k_{1}}\times \mathcal{R}^{k_{2}}\longrightarrow [0, 1]
$ such that for each $i$ and $\mathcal{D}\in \mathcal{R}^{k_{2}}$, $g(\boldsymbol{U}_{i}, \mathcal{D})$ is a version of $\mathbb{P}(\boldsymbol{V}_{i} \in \mathcal{D} \ \vert \ \boldsymbol{U}_{i})$.
\end{lemma}

\noindent \textit{Proof}. For each $(\boldsymbol{U}_{i}, \boldsymbol{V}_{i})$, there exists  a transition kernel $g_{i}: \mathbb{R}^{k_{1}}\times \mathcal{R}^{k_{2}} \longrightarrow[0, 1]
$ such that for each $\mathcal{D}\in\mathcal{R}^{k_{2}}$, $g_{i} (\boldsymbol{U}_{i}, \mathcal{D})$ is a version of $\mathbb{P}(\boldsymbol{V}_{i} \in \mathcal{D} \ \vert \ \boldsymbol{U}_{i})$; see, for example, Theorem 4.1.18 of~\citep{Durrett2019}. By this and the fact that $\mu_{\boldsymbol{U}_{i}} = \mu_{\boldsymbol{U}_{1}}$ for each $i$, we have that for each $\mathcal{A}\in \mathcal{R}^{k_{1}}$ and $\mathcal{B} \in \mathcal{R}^{k_{2}}$, 
\begin{equation}\label{le11}\begin{split}
\mathbb{P}(  (\boldsymbol{U}_{i}, \boldsymbol{V}_{i}) \in \mathcal{A}\times\mathcal{B}) & = \int_{\Omega} \boldsymbol{1}_{\{\boldsymbol{U}_{i} \in \mathcal{A}\}}	\mathbb{P} ( \boldsymbol{V}_{i}\in\mathcal{B}\ \vert \ \boldsymbol{U}_{i}) d\mathbb{P}	\\
& = \int_{\Omega} \boldsymbol{1}_{\{\boldsymbol{U}_{i} \in \mathcal{A}\}} \ g_{i}(\boldsymbol{U}_{i}, \mathcal{B}) d\mathbb{P}	\\
& = \int_{\mathbb{R}^{k_{1}}} \mu_{\boldsymbol{U}_{i}}(dx) \ \boldsymbol{1}_{\{x \in \mathcal{A}\}} \ g_{i}(x, \mathcal{B}) \\
& = \int_{\mathbb{R}^{k_{1}}} \mu_{\boldsymbol{U}_{1}}(dx) \ \boldsymbol{1}_{\{x \in \mathcal{A}\}} \ g_{1}(x, \mathcal{B}) \\
& = \int_{\mathbb{R}^{k_{1}}} \mu_{\boldsymbol{U}_{i}}(dx) \ \boldsymbol{1}_{\{x \in \mathcal{A}\}} \ g_{1}(x, \mathcal{B})\\
& = \int_{\Omega} \ \boldsymbol{1}_{\{\boldsymbol{U}_{i} \in \mathcal{A}\}} \ g_{1}(\boldsymbol{U}_{i}, \mathcal{B})d\mathbb{P},
\end{split}\end{equation}
where $\Omega$ represents the underlying probability space, the third and last equalities above follow from the change of variables formula, and the fourth and fifth equalities above are due to the assumption of identical distributions. Therefore, it follows from (\ref{le11}) and the definition of conditional expectation that $g_{1}(\boldsymbol{U}_{i}, \mathcal{D})$ is a version of $\mathbb{P}(\boldsymbol{V}_{i}\in\mathcal{D} \ \vert \ \boldsymbol{U}_{i} )$ for each $i$ and  $\mathcal{D}\in\mathcal{R}^{k_{2}}$. This completes the proof of Lemma~\ref{le1}.

\subsection{Lemma \ref{sup1} and its proof} \label{SecA.6}

Intuitively, Lemma~\ref{sup1} below extracts the total variation distance from a difference of two integrals. The results are natural, but the arguments are somewhat delicate. We note that if the density functions $f_{X}$ and $f_{Y}$ exist, then it holds that 
\begin{equation*}
\begin{split}
\sup_{\mathcal{D} \in \mathcal{R}}| \int_{\mathcal{D}} f_{X}(z)dz - \int_{\mathcal{D}} f_{Y}(z)dz| & \leq  \sup_{\mathcal{D} \in \mathcal{R}} \int_{\mathcal{D}} |f_{X}(z) - f_{Y}(z)|dz \\ & =\frac{1}{2}\norm{\mu_{X} - \mu_{Y}}_{TV}.
\end{split}
\end{equation*}
However, the same calculation does not apply directly to distributions because the integral $\int |\mu_{X}(dz) - \mu_{Y}(dz)|$ is not well-defined due to the two $dz$s inside the integration.  Lemma~\ref{sup1} provides valid arguments in such situations.

\begin{lemma} \label{sup1}
1)	Let $\mu$ and $\nu$ be two probability measures and $f: \mathbb{R}^{K} \longrightarrow\mathbb{R}$ a measurable function with $0\le f\le 1$. Then it holds that
\begin{equation*}
\sup_{\mathcal{D} \in \mathcal{R}^{K}}\Big| \int_{\mathcal{D}} f (x) \mu (dx)   - \int_{\mathcal{D}} f (x) \nu (dx)   \Big|\le \frac{1}{2}\norm{\mu - \nu}_{TV}.\end{equation*}

2) Let $p(\cdot, \cdot): \mathbb{R}^{K_{1}}\times\mathcal{R}^{K_{2}}\longmapsto [0, 1]$ be a transition kernel and $\mu$ a probability measure on $\mathcal{R}^{K_{2}}$ such that for each $x\in\mathbb{R}^{K_{1}}$, $p(x, \cdot)$ is dominated by $\mu$. Further let $\nu$ be a probability measure on $\mathcal{R}^{K_{1}}$ and $0 \le f\le 1$ a measurable function on $\mathbb{R}^{K_{2}}$. Then it holds that 
\begin{equation}
\begin{split}\label{sup12}
& \sup_{\mathcal{D}\in \mathcal{R}^{K_{1} + K_{2}}}\Big|\int_{\mathcal{D}} \nu(dx_{1}) p(x_{1}, dx_{2}) f(x_{2}) - \int_{\mathcal{D}} \nu(dx_{1}) \mu(dx_{2}) f(x_{2}) \Big| \\ 
& \le \frac{1}{2}\int_{\mathbb{R}^{K_{1}}} \nu(dx_{1})\norm{p(x_{1}, \cdot) - \mu(\cdot)}_{TV}.
\end{split}\end{equation}
\end{lemma}

\noindent \textit{Proof}. We start with proving the first assertion. By the definition of the total variation distance, we can show that there exists some set $\mathcal{A}\in\mathcal{R}^{K}$ such that 
\[ \int_{\mathcal{A}} \mu(dx) - \int_{\mathcal{A}} \nu(dx) = \frac{1}{2}\norm{\mu - \nu}_{TV}. \]
Further it holds that for each  $\mathcal{B}\in\mathcal{R}^{K}$ with $\mathcal{B} \subset \mathcal{A}$, 
\[\int_{\mathcal{B}} \mu(dx) - \int_{\mathcal{B}} \nu(dx) \ge 0.\]
Let us define $\mathcal{D}_{j} \coloneqq \mathcal{A}\cap\{x :\frac{j-1}{M}\le  f(x)< \frac{j}{M}\}$ for $j = 1,\dots, M+1$ with $M$ some positive integer. Denote by $\bar{f}$ a step function such that on $\mathcal{D}_{j}$, it holds that $\bar{f} = \frac{j}{M}$. Then we can deduce that 
\begin{equation}\begin{split}\label{sup5}
&\left|\int_{\cup_{j}\mathcal{D}_{j}} f(x) \mu(dx)  - \int_{\cup_{j}\mathcal{D}_{j}} f(x) \nu(dx) - \left(\int_{\cup_{j}\mathcal{D}_{j}} \bar{f}(x) \mu(dx)  - \int_{\cup_{j}\mathcal{D}_{j}} \bar{f}(x) \nu(dx)\right)\right| \\ 
& \le \int_{\cup_{j}\mathcal{D}_{j}} |f(x) - \bar{f}(x)| \mu(dx)  + \int_{\cup_{j}\mathcal{D}_{j}} |f(x) - \bar{f}(x)| \nu(dx) \\
& \le \frac{1}{M} \left( \int_{\cup_{j}\mathcal{D}_{j}} \mu(dx) + \int_{\cup_{j}\mathcal{D}_{j}} \nu(dx) \right)\\
& \le \frac{2}{M}.
\end{split}\end{equation}

In light of the construction of $\mathcal{D}_{j}$'s and $\bar{f}$ above, we have that  
\begin{equation}
\begin{split}
\label{sup4}
\int_{\cup_{j}\mathcal{D}_{j}} & \bar{f}(x) \mu(dx)  - \int_{\cup_{j}\mathcal{D}_{j}} \bar{f}(x) \nu(dx) = \sum_{j}\int_{\mathcal{D}_{j}} \bar{f}(x) \mu(dx)  - \int_{\mathcal{D}_{j}} \bar{f}(x)\nu(dx)\\
& = \sum_{j}\frac{j}{M}\left(\int_{\mathcal{D}_{j}} \mu(dx) - \int_{\mathcal{D}_{j}} \nu(dx)\right)\\
&\le \sum_{j}\left(\int_{\mathcal{D}_{j}} \mu(dx) - \int_{\mathcal{D}_{j}} \nu(dx)\right) \\
&= \frac{1}{2}\norm{\mu-\nu}_{TV}.
\end{split}
\end{equation}
Then it follows from $\mathcal{A} = \cup_{j}\mathcal{D}_{j}$, (\ref{sup5}), (\ref{sup4}), and the fact that the positive integer $M$ can be arbitrarily large that 
\begin{equation}\label{sup2}
\left|\int_{\mathcal{A} } f(x) \mu(dx)  - \int_{\mathcal{A}} f(x)\nu(dx)\right| \le \frac{1}{2}\norm{\mu-\nu}_{TV}.
\end{equation}
Using similar arguments as above, we can show that 
\begin{equation}\label{sup3}
\sup_{\mathcal{D}\in\mathcal{R}^{K}}\left|\int_{\mathcal{D} } f(x)\mu(dx)  - \int_{\mathcal{D}} f(x)\nu(dx)\right| = \left|\int_{\mathcal{A} } f(x)\mu(dx)  - \int_{\mathcal{A}} f(x)\nu(dx)\right|.\end{equation}
Therefore, combining (\ref{sup2}) and (\ref{sup3}) results in the desired conclusion in part 1) of Lemma~\ref{sup1}.

We now proceed with showing the second assertion. Since $\norm{p(x_{1} , \cdot) - \mu(\cdot)}_{TV}$ is a measurable function in $x_{1}$, the RHS of (\ref{sup12}) is well defined. Such a claim can be established using similar arguments to those in Theorem 5.2.2 of Durrett~\citep{Durrett2019}; for simplicity, we omit the details. By the assumptions, let $f_{1}(x_{1}, x_{2})$ be the Radon–Nikodym derivative such that for each $x_{1}\in\mathbb{R}^{K_{1}}$ and $\mathcal{D} \in \mathcal{R}^{ K_{2}}$, 
\[ \int_{\mathcal{D}} p(x_{1}, dx_{2}) = \int_{\mathcal{D}} f_{1}(x_{1}, x_{2})\mu(dx_{2}). \] 
If $\mu$ is the Lebesgue measure and for each $\mathcal{D}\in\mathcal{R}^{K_{2}}$, $p(\boldsymbol{X}, \mathcal{D})$ is a version of $\mathbb{P}(\boldsymbol{Y} \in \mathcal{D}| \boldsymbol{X})$ for some random mappings $\boldsymbol{Y}$ and $\boldsymbol{X}$ dominated by the Lebesgue measure, such a Radon–Nikodym derivative is usually referred to as the conditional (on the density function of $\boldsymbol{X}$) probability density function; for this, see also Example 4.1.6 in~\citep{Durrett2019}. Thus, for each $\mathcal{D}\in \mathcal{R}^{K_{1} + K_{2}}$, we have that 
\begin{equation}
\begin{split} \label{sup14}
&\Big|\int_{\mathcal{D}} \nu(dx_{1})p(x_{1}, dx_{2}) f(x_{2}) - \int_{\mathcal{D}} \nu(dx_{1}) \mu(dx_{2}) f(x_{2}) \Big| \\
& = \Big|\int_{ \mathcal{D}}  \nu(dx_{1})(f_{1}(x_{1}, x_{2}) - 1 ) \mu(dx_{2})  f(x_{2}) \Big|.
\end{split}\end{equation}

Next let us define $\mathcal{D}^{*}$ as 
\begin{equation}
\label{sup13}
\mathcal{D}^{*}\coloneqq\arg \sup_{\mathcal{D} \in\mathcal{R}^{K_{1} + K_{2}}} \int_{ \mathcal{D}}  \nu(dx_{1})(f_{1}(x_{1}, x_{2}) - 1 ) \mu(dx_{2})  f(x_{2})
\end{equation}
such that for each $\mathcal{A}\subset \mathcal{D}^{*}$, the integration of (\ref{sup13}) over $\mathcal{A}$ is nonnegative. 
Then by (\ref{sup14}), (\ref{sup13}), and the assumption of $0\le f\le 1$, it holds that 
\begin{equation}
\begin{split}\label{sup15}
& \sup_{\mathcal{D}\in \mathcal{R}^{K_{1} + K_{2}}}\Big|\int_{\mathcal{D}} \nu(dx_{1}) p(x_{1}, dx_{2}) f(x_{2}) - \int_{\mathcal{D}} \nu(dx_{1}) \mu(dx_{2}) f(x_{2}) \Big| \\ 
& \le  \int_{ \mathcal{D}^{*}}  \nu(dx_{1})(f_{1}(x_{1}, x_{2}) - 1 ) \mu(dx_{2}) .
\end{split}\end{equation}
Thus, it follows from the definition of $f_{1}$, Lemma~\ref{le3}, and the definition of the total variation norm that 
\begin{equation}
\begin{split}\label{sup16}
& \int_{ \mathcal{D}^{*}}  \nu(dx_{1})(f_{1}(x_{1}, x_{2}) - 1 ) \mu(dx_{2}) \\
& = \int_{ \mathbb{R}^{K_{1}}}  \nu(dx_{1}) (p(x_{1}, (D^{*})_{x_{1}} ) - \mu((D^{*})_{x_{1}} ) )\\
& \le  \int_{ \mathbb{R}^{K_{1}}} \nu(dx_{1}) \frac{1}{2}\norm{p(x_{1}, \cdot) - \mu(\cdot)}_{TV}.
\end{split}
\end{equation}
Here Lemma~\ref{le3} is applicable to the integral with $\mu$ since $\mu$ can be seen as a transition kernel. Therefore, combining (\ref{sup15})  and (\ref{sup16}) 
yields the conclusion in part 2) of Lemma \ref{sup1}. This concludes the proof of Lemma \ref{sup1}.

\subsection{Lemma \ref{le2} and its proof} \label{SecA.7}


\begin{lemma} \label{le2}
Let $\{(\boldsymbol{Y}_{i}, \boldsymbol{X}_{i})\}$ and $\{(\boldsymbol{V}_{i}, \boldsymbol{U}_{i}) \}$ be two sequences of identically distributed random vectors with $\boldsymbol{Y}_{i}$ and $\boldsymbol{X}_{i}$ $k_{1}$-dimensional  and $k_{2}$-dimensional, respectively.  Assume that there exists some positive integer $K$ such that for each $\mathcal{D}_{i}\in\mathcal{R}^{k_{2}}$ with $i = 1,\dots, K$, 
\begin{equation*}\begin{split}
\mathbb{P}(\cap_{i=1}^{K} \{ \boldsymbol{X}_{i} \in \mathcal{D}_{i}\} \ \vert \ \boldsymbol{Y}_{j}, j = 1, \dots, K) &= \Pi_{i=1}^{K} \mathbb{P}(\boldsymbol{X}_{i} \in \mathcal{D}_{i} \ \vert \  \boldsymbol{Y}_{i}),	\\
\mathbb{P}(\cap_{i=1}^{K} \{ \boldsymbol{U}_{i} \in \mathcal{D}_{i}\} \ \vert \ \boldsymbol{V}_{j}, j = 1, \dots, K)&=\Pi_{i=1}^{K} \mathbb{P}(\boldsymbol{U}_{i} \in \mathcal{D}_{i} \ \vert \  \boldsymbol{V}_{i}).
\end{split}	
\end{equation*} 
Let us define $\boldsymbol{Y} \coloneqq (\boldsymbol{Y}_{1}^{\top}, \dots, \boldsymbol{Y}_{K}^{\top})^{\top}$ and $\boldsymbol{V} \coloneqq (\boldsymbol{V}_{1}^{\top}, \dots, \boldsymbol{V}_{K}^{\top})^{\top}$, and  $\boldsymbol{X}$ and $\boldsymbol{U}$ are defined similarly. Then it holds that 
\begin{enumerate}
\item[1)]  There exists a transition kernel $h:\mathbb{R}^{Kk_{1}}  \times \mathcal{R}^{Kk_{2}}\longrightarrow [0,1 ]
$ such that for each  $\mathcal{B}\in\mathcal{R}^{Kk_{2}}$, $h(\boldsymbol{Y}, \mathcal{B})$ and $h(\boldsymbol{V}, \mathcal{B})$ are versions of $\mathbb{P}(\boldsymbol{X} \in \mathcal{B} \ \vert \ \boldsymbol{Y})$ and $\mathbb{P}(\boldsymbol{U} \in \mathcal{B} \ \vert \ \boldsymbol{V})$, respectively.
\item[2)] We have 
\begin{equation*}\begin{split} 
& \sup_{\mathcal{D}\in \mathcal{R}^{K(k_{1} + k_{2})}} \Big|  \mathbb{P} \Big( (\boldsymbol{Y}^{\top}, \boldsymbol{X}^{\top}) \in \mathcal{D} \Big) -  \mathbb{P} \Big( (\boldsymbol{V}^{\top}, \boldsymbol{U}^{\top}) \in \mathcal{D} \Big) \Big| \le \frac{1}{2}\norm{\mu_{\boldsymbol{Y}} - \mu_{\boldsymbol{V}}}_{TV},
\end{split}\end{equation*}
where $\mu_{\boldsymbol{Y}} $ and $\mu_{\boldsymbol{V}}$ denote the distributions of $\boldsymbol{Y} $ and $\boldsymbol{V}$, respectively.
\end{enumerate}
\end{lemma}

\noindent \textit{Proof}. By assumption, it holds that for each $\mathcal{A}\in \mathcal{R}^{Kk_{1}}$ and $B_{i}\in\mathcal{R}^{k_{2}}$ with $i = 1, \dots, K$, 
\begin{equation}\begin{split}\label{le21}
\mathbb{P}((\boldsymbol{Y}^{\top}, \boldsymbol{X}^{\top}) \in \mathcal{A}\times (\bigtimes_{i=1}^{K} B_{i})) & = \mathbb{E} \Big[  \boldsymbol{1}_{\{\boldsymbol{Y} \in \mathcal{A}\}} \ \mathbb{P} \Big( \cap_{i=1}^{K}\{\boldsymbol{X}_{i} \in B_{i} \}\ \vert \ \boldsymbol{Y}\Big) \Big]\\
& = \mathbb{E} \Big[  \boldsymbol{1}_{\{\boldsymbol{Y} \in \mathcal{A}\}} \ \Pi_{i=1}^{K} \mathbb{P} \Big( \boldsymbol{X}_{i} \in B_{i} \ \vert \ \boldsymbol{Y}_{i}\Big) \Big],
\end{split}\end{equation}
where $\bigtimes_{i=1}^{K} B_{i} \coloneqq (B_{1}, \dots, B_{K})$. Similarly, we can show that 
\begin{equation}\begin{split}\label{le22}
\mathbb{P}((\boldsymbol{V}^{\top}, \boldsymbol{U}^{\top}) \in \mathcal{A}\times (\bigtimes_{i=1}^{K} B_{i})) = \mathbb{E} \Big[  \boldsymbol{1}_{\{\boldsymbol{V} \in \mathcal{A}\}} \ \Pi_{i=1}^{K} \mathbb{P} \Big( \boldsymbol{U}_{i} \in B_{i} \ \vert \ \boldsymbol{V}_{i}\Big) \Big].
\end{split}\end{equation}
Since $(\boldsymbol{Y}_{i}^{\top}, \boldsymbol{X}_{i}^{\top})$ and $(\boldsymbol{V}_{i}^{\top}, \boldsymbol{U}_{i}^{\top})$ with $i \geq 1$ are identically distributed, by Lemma~\ref{le1} in Section \ref{SecA.5}, there exists a transition kernel $h_{1}$ such that for each $i$ and $\mathcal{D}\in\mathcal{R}^{k_{2}}$, $h_{1}(\boldsymbol{Y}_{i}, \mathcal{D})$ and $h_{1}( \boldsymbol{V}_{i}, \mathcal{D})$ are versions of $\mathbb{P}(\boldsymbol{X}_{i} \in \mathcal{D} \ \vert \ \boldsymbol{Y}_{i}) $ and $\mathbb{P}(\boldsymbol{U}_{i} \in \mathcal{D} \ \vert \ \boldsymbol{V}_{i}) $, respectively. 

Let us define $h_{2} : \mathbb{R}^{ Kk_{1}} \times \mathcal{R}^{Kk_{2}}\longrightarrow [0,1 ]
$ such that for each $x = (x_{1}^{\top} ,\dots, x_{K}^{\top})^{\top} \in \mathbb{R}^{Kk_{1}}$, $h_{2}(x, \cdot)$ is a probability measure such that for each $\mathcal{D}_{i} \in\mathcal{R}^{k_{2}}$ with $i = 1,\dots, K$,
\begin{equation}\label{le23}
h_{2}(x, \bigtimes_{i=1}^{K} \mathcal{D}_{i}) = \Pi_{i=1}^{K}h_{1}(x, \mathcal{D}_{i}).
\end{equation}
We make a useful claim below.
\begin{claim}\label{c1}
$h_{2}$ is a transition kernel satisfying (\ref{le23}).
\end{claim}
\noindent The proof of Claim~\ref{c1} is provided in Section \ref{SecA.8}. Then it follows from (\ref{le21}), (\ref{le22}), and Claim~\ref{c1} above that for each $\mathcal{B}_{i} \in \mathcal{R}^{k_{2}}$ with $i = 1, \dots, K$, 
\begin{equation}\begin{split}\label{le24}
\mathbb{P}((\boldsymbol{Y}^{\top}, \boldsymbol{X}^{\top}) \in \mathcal{A}\times (\bigtimes_{i=1}^{K} B_{i})) & = \mathbb{E} \Big[  \boldsymbol{1}_{\{\boldsymbol{Y} \in \mathcal{A}\}} \ \Pi_{i=1}^{K} \mathbb{P} \Big( \boldsymbol{X}_{i} \in B_{i} \ \vert \ \boldsymbol{Y}_{i}\Big) \Big] \\
& = \mathbb{E} \Big[  \boldsymbol{1}_{\{\boldsymbol{Y} \in \mathcal{A}\}} \ \Pi_{i=1}^{K} h_{1}\Big(  \boldsymbol{Y}_{i}, B_{i}\Big) \Big]\\
& = \mathbb{E} \Big[  \boldsymbol{1}_{\{\boldsymbol{Y} \in \mathcal{A}\}} \ h_{2}\Big(  \boldsymbol{Y}, \bigtimes_{i=1}^{K} B_{i}\Big) \Big],
\end{split}
\end{equation}
and similarly, 
\begin{equation}\begin{split}\label{le25}
\mathbb{P}((\boldsymbol{V}^{\top}, \boldsymbol{U}^{\top}) \in \mathcal{A}\times (\bigtimes_{i=1}^{K} B_{i}))  = \mathbb{E} \Big[  \boldsymbol{1}_{\{\boldsymbol{V} \in \mathcal{A}\}} \ h_{2}\Big(  \boldsymbol{V}, \bigtimes_{i=1}^{K} B_{i}\Big) \Big].
\end{split}
\end{equation}

By the construction of $h_{2}$, (\ref{le24}), (\ref{le25}), and Lemma~\ref{pilambda} in Section \ref{SecA.2}, it holds that for each  $\mathcal{A}\in\mathcal{R}^{Kk_{1}}$ and $\mathcal{B}\in\mathcal{R}^{Kk_{2}}$,
\begin{equation}\begin{split}\label{le26}
\mathbb{P}((\boldsymbol{Y}^{\top}, \boldsymbol{X}^{\top}) \in \mathcal{A}\times \mathcal{B})) 
& = \mathbb{E} \Big[  \boldsymbol{1}_{\{\boldsymbol{Y} \in \mathcal{A}\}} \ h_{2}\Big(  \boldsymbol{Y},  \mathcal{B} \Big) \Big], \\
\mathbb{P}((\boldsymbol{V}^{\top}, \boldsymbol{U}^{\top}) \in \mathcal{A}\times \mathcal{B})) 
& = \mathbb{E} \Big[  \boldsymbol{1}_{\{\boldsymbol{V} \in \mathcal{A}\}} \ h_{2}\Big(  \boldsymbol{V},  \mathcal{B} \Big) \Big].
\end{split}
\end{equation}
Since $h_{2}$ is a transition kernel, we see that for each $\mathcal{B}\in\mathcal{R}^{Kk_{2}}$, $h_{2}(\boldsymbol{Y}, \mathcal{B})$ is $\sigma(\boldsymbol{Y})$-measurable. Thus, in view of (\ref{le26}) we have that for each $\mathcal{B} \in \mathcal{R}^{Kk_{2}}$, $h_{2}(\boldsymbol{Y}, \mathcal{B})$ is a version of $\mathbb{P}(\boldsymbol{X}\in\mathcal{B} \ \vert \ \boldsymbol{Y})$. A similar result for $\boldsymbol{V}$ and $\boldsymbol{U}$ can also be obtained, which leads to the first assertion.

Finally, by Lemma~\ref{le3} and Lemma~\ref{sup1} in Sections \ref{SecA.3} and \ref{SecA.6}, respectively, we can deduce that for each $\mathcal{D}\in\mathcal{R}^{K(k_{1} + k_{2})}$,		
\begin{equation}
\begin{split}
&\Big| \mathbb{P}((\boldsymbol{Y}^{\top}, \boldsymbol{X}^{\top}) \in \mathcal{D}) - \mathbb{P}((\boldsymbol{V}^{\top}, \boldsymbol{U}^{\top}) \in \mathcal{D})\Big| \\
&  = \Big| \int_{\mathbb{R}^{Kk_{1}}} h_{2} (x, \mathcal{D}) \mu_{\boldsymbol{Y}} (dx)   - \int_{\mathbb{R}^{Kk_{1}}} h_{2} (x, \mathcal{D}) \mu_{\boldsymbol{V}} (dx)   \Big|\\
& \le \frac{1}{2}\norm{\mu_{\boldsymbol{Y}} - \mu_{\boldsymbol{U}}}_{TV},
\end{split}
\end{equation}
which yields the second assertion. This completes the proof of Lemma~\ref{le2}.

\subsection{Lemma \ref{knockoff1} and its proof} \label{SecA.9}

Let $\tilde{\boldsymbol{U}}$ and $\tilde{\boldsymbol{V}}$ be the knockoffs counterparts of $r \times c$ design matrices $\boldsymbol{U}$ and $\boldsymbol{V}$, respectively. The corresponding response vectors are denoted as $\boldsymbol{u}$ and $\boldsymbol{v}$, respectively.  Lemma~\ref{knockoff1} below ensures that the knockoffs matrix construction does not cause additional variation in the distribution in terms of the total variation distance. 

\begin{lemma} \label{knockoff1}
Assume that \textnormal{1)} the rows of $(\boldsymbol{U}, \tilde{\boldsymbol{U}})$ and $(\boldsymbol{V}, \tilde{\boldsymbol{V}})$ are identically distributed, \textnormal{2)} the row vectors of $\tilde{\boldsymbol{U}}$ are independent random vectors conditional on $\boldsymbol{U}$, \textnormal{3)} the $i$th row of $\tilde{\boldsymbol{U}}$ is independent of the other rows of $\boldsymbol{U}$ conditional on the $i$th row of $\boldsymbol{U}$, and \textnormal{4)} $\boldsymbol{u}$ is independent of $\tilde{\boldsymbol{U}}$ conditional on $\boldsymbol{U}$. In addition, we assume the same for $(\boldsymbol{v}, \boldsymbol{V}, \tilde{\boldsymbol{V}})$. Then it holds that 
\[\sup_{\mathcal{D} \in \mathcal{R}^{r(1+2c)}} \left| \mathbb{P} \Big((\boldsymbol{u}, \boldsymbol{U}, \tilde{\boldsymbol{U}} ) \in \mathcal{D} \Big) -  \mathbb{P}\Big((\boldsymbol{v}, \boldsymbol{V}, \tilde{\boldsymbol{V}} ) \in \mathcal{D} \Big)\right| \le \frac{1}{2} \norm{\mu_{\boldsymbol{u, \boldsymbol{U}}} - \mu_{\boldsymbol{v}, \boldsymbol{V}}}_{TV}\]
and
\[\sup_{\mathcal{D} \in \mathcal{R}^{2rc}} \left| \mathbb{P} \Big(( \boldsymbol{U}, \tilde{\boldsymbol{U}} ) \in \mathcal{D} \Big) -  \mathbb{P}\Big(( \boldsymbol{V}, \tilde{\boldsymbol{V}} ) \in \mathcal{D} \Big)\right| \le \frac{1}{2} \norm{\mu_{ \boldsymbol{U}} - \mu_{\boldsymbol{V}}}_{TV} .\]
\end{lemma}

\noindent \textit{Proof}. We start with showing the first assertion. Denote the $i$th rows of $\widetilde{\boldsymbol{U}}$ and $\boldsymbol{U}$ by $\widetilde{\boldsymbol{U}}_{i\bullet}$ and $\boldsymbol{U}_{i\bullet}$, respectively. Then for each $\mathcal{D}_{i} \in \mathcal{R}^{r}$, we have that 
\begin{equation*}
\begin{split}
\mathbb{P}(\cap_{i=1}^{c}\{\widetilde{\boldsymbol{U}}_{i\bullet} \in \mathcal{D}_{i}\}|\boldsymbol{U}) & = \Pi_{i=1}^{c}\mathbb{P}(\widetilde{\boldsymbol{U}}_{i\bullet} \in \mathcal{D}_{i}|\boldsymbol{U}) \\
& = \Pi_{i=1}^{c}\mathbb{P}(\widetilde{\boldsymbol{U}}_{i\bullet} \in \mathcal{D}_{i}| \boldsymbol{U}_{i\bullet}),
\end{split}
\end{equation*}
where the first equality follows from assumption 2) and the second one is due to assumption 3). By this, an application of Lemma~\ref{le2} shows that there exists a transition kernel $h:\mathbb{R}^{rc}\times\mathcal{R}^{rc}\longmapsto[0, 1]$ such that for each $\mathcal{D}\in\mathcal{R}^{rc}$, $h(\boldsymbol{U}, \mathcal{D})$ and $h(\boldsymbol{V}, \mathcal{D})$ are versions of $\mathbb{P}(\tilde{\boldsymbol{U}} \in \mathcal{D} \ \vert \ \boldsymbol{U})$ and $\mathbb{P}(\tilde{\boldsymbol{V}} \in \mathcal{D} \ \vert \ \boldsymbol{V}) $, respectively. 

We will make use of the claim below.
\begin{claim}\label{claim2}
For each $\mathcal{D} \in \mathcal{R}^{r(1+2c)}$, it holds that 
\begin{equation*}
\begin{split}
\mathbb{P} ((\boldsymbol{u}, \boldsymbol{U}, \tilde{\boldsymbol{U}}) \in \mathcal{D} ) & = \int_{\mathbb{R}^{r(1+c)}} h(x_{2}, \mathcal{D}_{x_{1}, x_{2}}) \, \mu_{\boldsymbol{u}, \boldsymbol{U}} (dx_{1} \times dx_{2}),\\
\mathbb{P} ((\boldsymbol{v}, \boldsymbol{V}, \tilde{\boldsymbol{V}}) \in \mathcal{D} ) & = \int_{\mathbb{R}^{r(1+c)}} h(x_{2}, \mathcal{D}_{x_{1}, x_{2}}) \, \mu_{\boldsymbol{v}, \boldsymbol{V}} (dx_{1} \times dx_{2}),
\end{split}
\end{equation*}
where $x_{1}$ and $x_{2}$ denote $r$-dimensional and $(rc)$-dimensional vectors, respectively.
\end{claim}
\noindent The proof of Claim~\ref{claim2} is presented in Section \ref{SecA.10}.  
Then it follows from Claim~\ref{claim2} above, Lemma~\ref{sup1},  and the fact of $0 \leq h\le 1$ that for each $\mathcal{D}\in\mathcal{R}^{r(1+2c)}$,
\begin{equation*}
\begin{split}
& \Big|  \mathbb{P} (( \boldsymbol{u}, \boldsymbol{U}, \tilde{\boldsymbol{U}}) \in \mathcal{D})  -   \mathbb{P} (( \boldsymbol{v}, \boldsymbol{V}, \tilde{\boldsymbol{V}}) \in \mathcal{D}) \Big|\\
& = \Big|  \int_{\mathbb{R}^{r(1+c)}} h(x_{2}, \mathcal{D}_{x_{1}, x_{2} }) \, \mu_{\boldsymbol{u}, \boldsymbol{U} }(dx_{1}\times dx_{2})  - \int_{\mathbb{R}^{r(1+c)}} h(x_{2}, \mathcal{D}_{x_{1}, x_{2} }) \, \mu_{\boldsymbol{v}, \boldsymbol{V} }(dx_{1}\times dx_{2}) \Big|\\
& \le \frac{1}{2} \norm{\mu_{\boldsymbol{u}, \boldsymbol{U}} - \mu_{\boldsymbol{v}, \boldsymbol{V}}}_{TV},
\end{split}
\end{equation*}
which leads to the conclusion in the first assertion. Using similar arguments as above, we can establish the second assertion. This concludes the proof of Lemma~\ref{knockoff1}.

\subsection{Proof of Claim~\ref{c1}} \label{SecA.8}

Note that given $x$, the existence of the probability measure $h_{2}(x, \cdot)$ is guaranteed by an application of Theorem 1.7.1 in~\citep{Durrett2019}.
Let $L$ be the collection of sets such that if $\mathcal{D}\in L$, $h_{2}(x, \mathcal{D})$ is a measurable function of $x$. We will make three observations due to the definition of $h_{2}$. (i) $L$ contains all Borel rectangles since the product of measurable functions is still a measurable function. (ii) For $\mathcal{D}_{1}, \mathcal{D}_{2} \in L$ with $\mathcal{D}_{1}\subset \mathcal{D}_{2}$, it holds that $h_{2} (\cdot, \mathcal{D}_{2} \backslash\mathcal{D}_{1})$ is measurable and hence $\mathcal{D}_{2} \backslash\mathcal{D}_{1}\in L$. (iii) For $\mathcal{D}_{i}\subset \mathcal{D}_{i+1}$, $\mathcal{D}_{i} \in L$, and $\mathcal{D}\coloneqq \cup_{i=1}^{\infty}\mathcal{D}_{i}$, we have 
\[ h_{2}(\cdot,\mathcal{D}) = \sup_{i}h_{2}(\cdot, \mathcal{D}_{i}) \] 
which is measurable, and thus  $\mathcal{D}\in L$. Therefore, it follows from these facts and Lemma~\ref{pilambda} in Section \ref{SecA.2} that for each $\mathcal{D}\in \mathcal{R}^{Kk_{2}}$, $h_{2}(\cdot, \mathcal{D})$ is measurable. This completes the proof of Claim~\ref{c1}.

\subsection{Proof of Claim~\ref{claim2}} \label{SecA.10}

Let us first show the first assertion. For each Borel rectangle $\mathcal{A}_{1}\times \mathcal{A}_{2}\times \mathcal{A}_{3} \in \mathcal{R}^{r}\times \mathcal{R}^{rc}\times \mathcal{R}^{rc}$, it holds that 
\begin{equation*}
\begin{split}
& \mathbb{P} ((\boldsymbol{u}, \boldsymbol{U}, \tilde{\boldsymbol{U}}) \in \mathcal{A}_{1}\times \mathcal{A}_{2}\times \mathcal{A}_{3}) = \mathbb{E} \Big[ \boldsymbol{1}_{\{(\boldsymbol{u}, \boldsymbol{U}) \in \mathcal{A}_{1}\times \mathcal{A}_{2}\}} \ \mathbb{P} ( \tilde{\boldsymbol{U}} \in\mathcal{A}_{3} \ \vert \ \boldsymbol{u}, \boldsymbol{U}) \Big]\\
& = \mathbb{E} \Big[ \boldsymbol{1}_{\{(\boldsymbol{u}, \boldsymbol{U}) \in \mathcal{A}_{1}\times \mathcal{A}_{2}\}} \ \mathbb{P} ( \tilde{\boldsymbol{U}} \in\mathcal{A}_{3} \ \vert \  \boldsymbol{U}) \Big]\\
& = \int_{\mathbb{R}^{r(1+c)}} h(x_{2}, (\mathcal{A}_{1}\times \mathcal{A}_{2} \times \mathcal{A}_{3})_{x_{1}, x_{2}}) \, \mu_{\boldsymbol{u}, \boldsymbol{U}} (dx_{1}\times dx_{2}),
\end{split}
\end{equation*}
where the second equality is due to the assumption of Lemma~\ref{knockoff1} and the last equality is because of the definition of $h$. Let $L$ be a collection of sets in $\mathcal{R}^{r(1+2c)}$ such that if $\mathcal{D}\in L$, 
\[ \mathbb{P}((\boldsymbol{u}, \boldsymbol{U}, \tilde{\boldsymbol{U}}) \in \mathcal{D}) = \int_{\mathbb{R}^{r(1+c)}} h(x_{2}, \mathcal{D}_{x_{1}, x_{2}}) \, \mu_{\boldsymbol{u}, \boldsymbol{U}}(dx_{1}\times dx_{2}). \]
Then we can see that $L$ contains all Borel rectangles which collection is a $\pi$-system.

Let us make a few observations below.
\begin{enumerate}
\item[1)] The set $L$ contains $\mathbb{R}^{r(1+2c)}$.
\item[2)] If $\mathcal{D}_{1}, \mathcal{D}_{2} \in L$ and $\mathcal{D}_{1}\subset \mathcal{D}_{2}$, then some basic measure and integration operations as well as the operation of $\mathcal{D}_{x}$ give that 
\begin{equation*}
\begin{split}
& \mathbb{P}((\boldsymbol{u}, \boldsymbol{U}, \tilde{\boldsymbol{U}}) \in \mathcal{D}_{2}\backslash\mathcal{D}_{1}) = \mathbb{P}((\boldsymbol{u}, \boldsymbol{U}, \tilde{\boldsymbol{U}}) \in \mathcal{D}_{2}) - \mathbb{P}((\boldsymbol{u}, \boldsymbol{U}, \tilde{\boldsymbol{U}}) \in \mathcal{D}_{1}) \\
& = \int_{\mathbb{R}^{r(1+c)}} \Big(h(x_{2} , (\mathcal{D}_{2})_{x_{1}, x_{2} })  -  h(x_{2} , (\mathcal{D}_{1})_{x_{1}, x_{2} })  \Big) \, \mu_{\boldsymbol{u}, \boldsymbol{U}}(dx_{1} \times dx_{2})\\
& = \int_{\mathbb{R}^{r(1+c)}} h(x_{2}, (\mathcal{D}_{2}\backslash\mathcal{D}_{1})_{x_{1}, x_{2}}) \, \mu_{\boldsymbol{u}, \boldsymbol{U}}(dx_{1} \times dx_{2}),
\end{split}
\end{equation*}
which leads to $\mathcal{D}_{2} \backslash\mathcal{D}_{1} \in L$.
\item[3)] If $\mathcal{D}_{n} \in L$ and $\mathcal{D}_{n} \subset \mathcal{D}_{n+1}$, then it follows from the continuity of measure and the monotone convergence theorem that 
\begin{equation*}
\begin{split}
& \mathbb{P}( (\boldsymbol{u} , \boldsymbol{U}, \tilde{\boldsymbol{U}}) \in \cup_{n}\mathcal{D}_{n}  ) = \lim_{n} \mathbb{P}( (\boldsymbol{u} , \boldsymbol{U}, \tilde{\boldsymbol{U}}) \in \mathcal{D}_{n}  ) \\
&  = \lim_{n} \int_{\mathbb{R}^{r(1+c)}} h(x_{2}, (\mathcal{D}_{n} )_{x_{1}, x_{2} }) \, \mu_{\boldsymbol{u}, \boldsymbol{U} }(dx_{1}\times dx_{2}) \\
& = \int_{\mathbb{R}^{r(1+c)}} h(x_{2}, (\cup_{n}\mathcal{D}_{n})_{x_{1}, x_{2}}) \, \mu_{\boldsymbol{u}, \boldsymbol{U} }(dx_{1}\times dx_{2}),
\end{split}
\end{equation*}
which results in $\cup_{n}\mathcal{D}_{n} \in L$.
\end{enumerate}
Therefore, using the aforementioned facts, an application of Lemma~\ref{pilambda} in Section \ref{SecA.2} yields the conclusion in the first assertion. The conclusion in the second assertion can be shown in a similar fashion, which concludes the proof of Claim~\ref{claim2}.

\subsection{Lemma \ref{pilambda}} \label{SecA.2}

\begin{defn*}[$\pi$-system and $\lambda$-system]
A collection of sets $P$ is said to be a $\pi$-system if for any $A, B\in P $, $A\cap B \in P$. A collection $L$ of sets in $\Omega$ is said to be a $\lambda$-system if 
\begin{enumerate}
\item[1)] $\Omega \in L$;
\item[2)] If $A\subset B$ and $A, B \in L $, then $B\backslash A \in L$;
\item[3)] If $A_{n} \in L $ and $A_{n} \subset A_{n+1}$, then $\cup_{n} A_{n} \in L$.
\end{enumerate}
\end{defn*}

\begin{lemma}[$\pi-\lambda$ Theorem in \citep{Durrett2019}] \label{pilambda}
If $P$ is a $\pi$-system and $L$ is a $\lambda$-system that contains $P$, then the smallest $\sigma$-algebra containing $P$ is also contained in $L$.
\end{lemma}


\end{document}